\documentclass[11pt]{article}
\pdfoutput=1 
\usepackage{./aug/jheppub}

\usepackage{graphicx}
\usepackage{amsfonts,amsmath,amssymb}
\usepackage{tikz}
\usepackage{longtable}
\usepackage{verbatim}
\usepackage{array,setspace,mathrsfs,yfonts,dsfont,bbm,colonequals,amscd}
\usepackage{relsize,suffix,mathtools,cancel,bbm}

\newcommand{\be}{\begin{eqnarray}}
\newcommand{\ee}{\end{eqnarray}}
\newcommand{\bea}{\begin{eqnarray}}
\newcommand{\eea}{\end{eqnarray}}

\newcommand{\bn}{\begin{enumerate}}
\newcommand{\en}{\end{enumerate}}

\title{Curious patterns of IR symmetry enhancement}

\preprint{}

\author[a]{Shlomo S. Razamat,\!}
\author[a]{Orr Sela,\!}
\author[b]{and Gabi Zafrir}

\affiliation[a]{Department of Physics, Technion, Haifa, 32000, Israel}
\affiliation[b]{IPMU,  University of Tokyo,  Kashiwa, Chiba 277-8583, Japan}

\emailAdd{razamat@physics.technion.ac.il}
\emailAdd{sorrsela@campus.technion.ac.il}
\emailAdd{gabi.zafrir@ipmu.jp}

\abstract
{We study several cases of IR enhancements of global symmetry in four dimensions. In particular, we consider a sequence of $Spin(n+4)$ supersymmetric gauge theories ($8\geq n\geq 1$) with $n$ vectors and  spinor matter with $32$ components. We show  that the  subgroup of the flavor symmetry of these theories rotating the matter in the spinor representations in the UV, when proper gauge singlet fields are added, enhances to the commutant of $SU(2)$ in $E_{9-n}$. We discuss several other interesting cases of enhanced symmetries and the interplay between symmetry enhancement and self-duality.
We also make some observations about possible interconnections between chiral ring relations and symmetry enhancement. 
Finally, we conjecture relations of the discussed models to compactifications  of certain conformal matter models in six dimensions on tori. The conjecture is based on deriving a relation between five dimensional models with $Spin$ gauge groups and conformal theories in six dimensions.
As a by product of our considerations we discover  a new instance of a simple self-duality of a theory with an $SU(6)$ gauge group. 
}

\begin{document} 

\maketitle
\flushbottom


\section{Introduction}

Quantum field theories exhibit various interesting phenomena at strongly coupled fixed points of renormalization group flows. One of these is a possible enhancement of global symmetry of the UV model in the IR. 
We know of many examples of such enhancements. However, we lack a general understanding of when such an enhancement could happen and what is the general mechanism behind this phenomenon.

The situation in the case of supersymmetric field theories is somewhat more tractable. In particular we have quantitative tools which allow us to access strongly coupled fixed point physics by studying weakly coupled UV descriptions. Some of the supersymmetric quantities one can discuss do not depend on the RG flow \cite{Festuccia:2011ws} and thus should be consistent with the physics along the flow and in particular with the symmetry properties of the fixed point. Typically the tools take the form of different types of indices, that is partition functions with supersymmetric boundary conditions on manifolds which have non contractible cycles, or limits of these. The former partition functions can be thought of as counting certain types of operators in the theory. As the partition functions are not sensitive to the RG scale, the counting captures also the physics of the strongly coupled fixed point. In particular, the counted objects should form the representations of the different symmetry groups. If the symmetry group is bigger in the IR than in the UV, the representations seen in the computation need to be consistent with the enhancement. Moreover,  in some cases one can identify in the computation contributions which can only come from conserved currents. If such a contribution can be found, then it serves as a definite footprint of the enhancement, given that all the assumptions of the computation, such as the identification of the superconformal R symmetry, are true.

In five and three space-time dimensions there are many examples of enhancements of symmetry which were widely studied. A reason for that is that the enhancement often happens because the additional components of the conserved currents come from interesting   non perturbative physics, monopoles in the latter case \cite{Intriligator:1996ex} and instantons in the former \cite{Seiberg:1996bd}. Actually, in these cases the enhancement was predicted even without resorting to more modern partition function techniques by discussing when monopole and instanton operators can produce additional components of the conserved currents. However, the additional components of the conserved currents might not arise from such non perturbative effects but from more mundane local vector operators which become conserved at the fixed points. In fact in four dimensions this is the only way the enhancement can happen. In this paper we will discuss symmetry enhancement in four space-time dimensional theories.

Another interesting phenomenon in quantum field theory is that  two different UV theories might flow to the same fixed point. That is they are dual to each other, or in statistical physics jargon are in the same universality class.  A particularly simple phenomenon of duality is when the two dual models are but slightly different. For example, the models can be given by the same gauge sector differing only by gauge invariant fields. Such dualities are often named self-dualities. Although the two models are similar the map of operators between them can be non obvious. Also, the IR fixed point might be on a conformal manifold of theories, in which case the weakly coupled dual models can flow to different points on that manifold. In other words,  the duality group can also act on the space of conformal couplings. In fact one can construct examples of dualities where the UV theory is the same for the two models of the pair but one can still show that there is a non trivial map of operators as well as an action on the conformal couplings which gives identical physics. An example of this is the S duality of ${\cal N}=4$ SYM, in which the UV theory is actually conformal and there is no flow. Another example is that of ${\cal N}=1$ $SU(2)$ SQCD with four flavors and a quartic superpotential preserving an $SU(4)\times U(1)$ subgroup of the $SU(8)$ symmetry of the UV theory. This theory flows to a fixed point with a large conformal manifold possessing an $SU(4)\times U(1)$ symmetry preserving subspace along which the duality group acts. This is a simple example of Seiberg duality \cite{Seiberg:1994pq}. In fact, $SU(2)$ SQCD with four flavors can be further deformed by the addition of gauge singlet fields to a model preserving only $SU(2) 
\times SU(6)  \times U(1)$ symmetry group, but which is dual to itself under Seiberg duality and can be argued to flow to a fixed point which is isolated. In this case the duality acts only by reorganizing the states of the model and in particular leads to an enhancement of the symmetry to  $E_6\times U(1)$ \cite{Razamat:2017wsk}. The phenomenon of duality and of symmetry enhancement can be thus related in certain examples. We will exploit such a relation here to both discover new enhancements of symmetry starting from dualities and to discover new dualities from observations of enhancements of IR symmetries.

An indirect indication to whether a symmetry of some model can be enhanced is obtained by engineering it as a fixed point of a different flow for which the symmetry is manifest. That is, in some cases, one can claim that the theory of interest flows to the same fixed point as another model which manifests a bigger symmetry  group. A simple example is again given by ${\cal N}=1$ $SU(2)$ SQCD but now with five flavors. The symmetry is $SU(10)$, however the dual model is an $SU(3)$ theory with five flavors, where only $SU(5)\times SU(5)\times   U(1)$ symmetry is manifest. Here, the enhancement to $SU(10)$ only happens at the fixed point. In fact, in recent years, starting with \cite{Gaiotto:2009we}, we have discovered an amazingly rich set of flows which exhibit similar effects. One can engineer fixed points in four dimensions by starting with six dimensional models, put them on a compact two dimensional surface in the presence of background gauge fields for the global symmetries and flow to an effective theory in four dimensions. The symmetry of the theory in four dimensions is determined by the symmetry of the theory in six dimensions and by the details of the background fields.    On the other hand the same models can be argued to appear as fixed points of RG flows of four dimensional gauge theories for which the symmetry is often smaller than the symmetry expected in the IR. To obtain claims of this kind one needs to derive a dictionary between four dimensional constructions and compactifications. Such a dictionary in fact can be made very systematic by realizing that one can understand compactifications on complicated geometries by splitting them into simple geometric blocks and then understanding the compactifications on the blocks and how to combine them in four dimensions. The symmetry of the blocks might be enhanced when they are combined to theories corresponding to compactifications which preserve bigger symmetries. Examples of this type of studies can be found in \cite{Razamat:2016dpl,Bah:2017gph,Kim:2017toz,Kim:2018bpg,Kim:2018lfo} following the works of \cite{Gaiotto:2009we,Benini:2009mz,Bah:2012dg,Gaiotto:2015usa}.  This issue of developing the dictionary between six dimensional and four dimensional models, provides one of the motivations for this paper.

\begin{figure}
 [htbp ]
\center\includegraphics[scale=0.8]{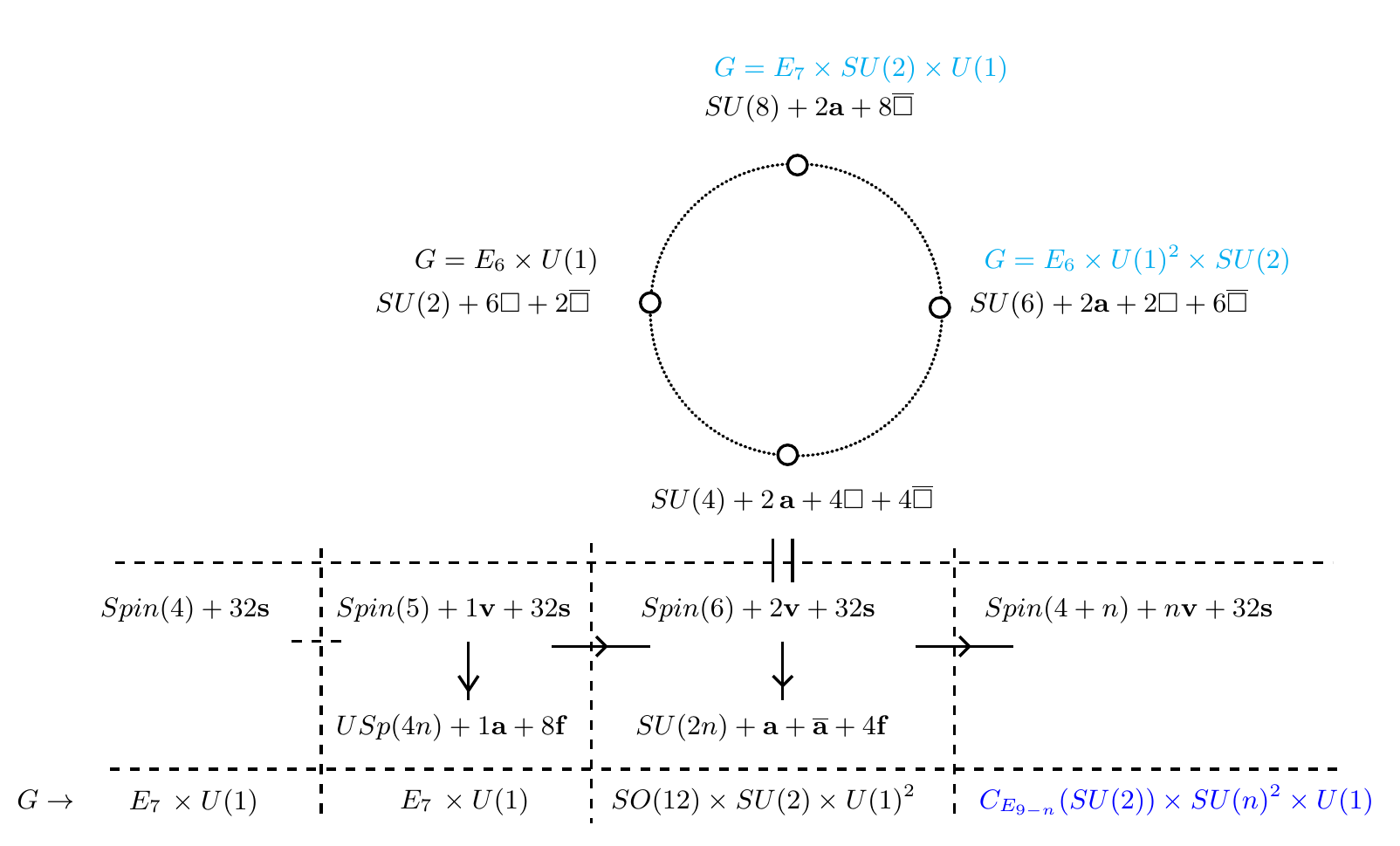}
\caption{The main sequences of theories and symmetry enhancements that we will discuss. The $Spin(4)$ model, the $E_7$ surprise \cite{Dimofte:2012pd}, is equivalent to the $Spin(5)$ model as they reside on the same conformal manifold. Here $C_{E_{9-n}}(SU(2))$ is the centralizer of $SU(2)$ in $E_{9-n}$.  The theories and/or enhancements colored in shades of blue are new to this paper while others appeared in other references. Here $s$ denotes components in the spinor representation, $v$ vector, $a$ is two index antisymmetric, and $f$ flavors which is the number of fundamentals for $USp$ and fundamentals plus anti-fundamentals for $SU$. The flavor symmetry is denoted by $G$.  The symmetry of the $SU(2n)$ sequence with $n>2$ is expected to be $SO(12)\times U(1)\times U(1)^2$.
}
\label{resdy}
\end{figure}

In this work we will study several known self-dualities. Most of these dualities were discussed in \cite{Csaki:1997cu, Karch:1997jp} as generalizations of self-dualities of ${\cal N}=1$ $SU(n)$ SQCD with $2n$ flavors. 
For additional examples see \cite{Ramond:1996ku} and \cite{Distler:1996ub}. In particular we will consider a sequence of self-dual models with $Spin(n+4)$ gauge groups and matter in the spinor and vector representations. We will take the spinors to have thirty two components, which will mean that the number of spinors depends on the gauge group, and $n$ chiral fields transforming as vectors of $Spin(n+4)$. The construction is bounded by $Spin(12)$ as the spinor of $Spin(13)$ has $64$ components. First, by adding certain gauge invariant fields we will construct models which are dual to themselves and that exhibit a rather significant enhancement of symmetry in the IR. 

The tool we will use to make such a statement is the supersymmetric index \cite{Kinney:2005ej, Romelsberger:2005eg,Dolan:2008qi} . When the theory is conformal, the index can be thought of as a sum over different representations of the superconformal group. It is independent of the RG flow \cite{Festuccia:2011ws}, and thus has to be consistent with the properties of the theories anywhere along the flow. In particular the supersymmetric index, which is defined as a partition function on ${\mathbb S}^3\times {\mathbb S}^1$ along the flow, becomes equivalent to the superconformal index at the fixed point. At the fixed point we can make full use of the superconformal symmetry to extract the information about the states. In particular, it has been shown in \cite{Beem:2012yn} that although the index counts states up to cancellations due to recombinations of various short superconformal multiplets to long multiplets, at low orders of the expansion of the index it reliably captures the information about some very important operators. These involve the relevant operators and the marginal operators minus the conserved currents. In particular any negative contribution to the difference between the latter comes from conserved currents. The only assumptions in such computations are that  the model flows to an interacting fixed point without free field, and that we have correctly identified the superconformal R symmetry. As the R symmetry can mix with abelian symmetries, we can use $a$ maximization \cite{Intriligator:2003jj} to determine the superconformal one, which should give the correct result if no accidental $U(1)$ symmetries appear at the fixed point. We will always assume that this is the case, and our results are consistent with the assumption. 

The main result of the paper will be to show that this sequence of theories has a rather regular enhancement of symmetry. The symmetry rotating the spinor fields will enhance to the commutant of $SU(n)\times SU(2)$ in $E_8$, and the $SU(n)$ symmetry rotating the vectors enhances  to $SU(n)\times SU(n)$.  This pattern of enhancement suggests an interpretation of the theories as fixed points of compactifications of six dimensional models. In particular, we conjecture that taking the E-string theory, which has $E_8$ global symmetry, and gauging an $SU(n)$ subgroup with the addition of $2n$ hypermultiplets is the six dimensional model we need to study. When one takes such a model on a torus with flux breaking the $SU(2n)$ symmetry to $SU(n)\times SU(n)\times U(1)$ one obtains a theory in four dimensions with $E_{9-n}$ $\times SU(n)\times SU(n)\times U(1)$ symmetry. We conjecture that the theories we discuss in four dimensions are farther deformations of these models breaking the $SU(2)$ in $E_{9-n}$. We will make this statement very precise for low values of $n$.  We will also discuss a variety of other examples of symmetry enhancements, 
duality, and compactifications. In  Figure \ref{resdy} we summarize the main results regarding symmetry enhancement. We will support our conjectures by discussing first the reduction of a variety of six dimensional conformal theories on a circle to five dimensions.
 
We will also discuss a curious  property of chiral ring relations which many of the models we encounter exhibit. Namely, consider relevant operators ${\cal O}_a$ in representations ${\frak r}_a$ of a global symmetry $G$, such that ${\cal O}_a\left.{\cal O}_b\right|_{{\frak r}_{ab}} \sim 0$ in the chiral ring with ${\frak r}_{ab}$ a representation appearing in the decomposition of ${\frak r}_a\otimes {\frak r}_b$. In all examples we have considered $(\oplus {\frak r}_{ab} )\oplus Adj(G)$ is an adjoint representation of some group $\overline   G$. This group is not in general the symmetry group of the model. In some cases we can construct, by making use of gauge singlet fields, from a given theory a model so that $\overline G$ is it's symmetry. 

The paper has the following outline. In section two we discuss the sequence of four dimensional $Spin(4+n)$ gauge theories and summarize the enhancement of symmetries. In section three we will consider five dimensional models with 
similar gauge groups to the ones discussed in four dimensions and conjecture UV six dimensional completions of such models. 
In section four we use the five dimensional understanding to discuss the relation of the models to compactifications from six dimensions and use this relations to discover more cases of symmetry enhancements and dualities. In section five we discuss an interconnection between chiral ring relations and symmetry enhancements.
Several appendices will contain more technical results.

\

\section{Four dimensional symmetry enhancement in $\boldsymbol{Spin(n+4)}$ models}
\label{fourd}

We consider a sequence of theories with gauge groups ranging from $Spin(12)$ down to $Spin(5)$, with matter content in the spinorial and vector representations. Each such theory has at least one dual theory \cite{Karch:1997jp,Csaki:1997cu} with the same gauge group but with additional gauge singlet fields and with some of the matter in different representations of the global symmetry. We  will in what follows construct models, motivated by the existence of such self-dualities, which have enhanced global symmetries in the IR. For each model in this sequence we will add certain gauge singlet fields and relevant superpotentials to form models with an enhanced symmetry.  With this choice of gauge singlet fields most of the models have a conformal manifold in the IR which is topologically a point and the duality group will then act as the Weyl symmetry of the enhanced global symmetry of the IR theory.
In what follows we present this new sequence of theories along with their enhanced symmetry in the IR.
We will give evidence, in fact a proof in most cases under some assumptions, of such statements. 

The readers unfamiliar with the supersymmetric index terminology should consult the first part of appendix A for a brief summary of the terminology and the technology. Here we just remind the reader that in the expansion of the superconformal index computed with the superconformal R-symmetry in the canonical fugacities $q$ and $p$ \cite{Dolan:2008qi} at order $qp$ one obtains the difference between the marginal operators and the conserved currents \cite{Beem:2012yn}.
In what follows we will quote the result of the computation of this order and deduce the symmetry from this argument. In certain cases we will observe only negative contributions which can come only from the currents, which is a proof following the superconformal representation theory that the symmetry is at least the one corresponding to the currents. The only assumptions are that we have flown to an interacting fixed point without free fields, and that we have identified the R symmetry correctly. In appendix A we give all the details of the computation.

The following models contain various fields which are singlets under the gauge symmetries. We choose these fields in such a manner that  the theories will exhibit the enhancement of symmetry clearly. This happens if one can identify unambiguously the conserved currents in the computation of the index, which typically occurs when the conformal manifold is a point. In most of the models in the sequence one can choose the singlets so that the conformal manifold is a point and the enhancement of the symmetry is maximal. In some cases, the conformal manifold is a point with some choice of singlets, but one can find evidence for a larger enhancement, using a different choice of singlets, which leads to a conformal manifold bigger than a point. In such cases we detail both options for the singlet fields.

\

\subsection*{$\boldsymbol{Spin(5)}$}

\

We start with a $Spin(5)_g$ gauge theory with an octet of spinors and a single vector field. The group is isomorphic to $USp(4)$ and from that perspective we have an octet of fundamental chiral fields and a single field in the two index antisymmetric traceless representation.
This particular case was discussed in much detail in \cite{Razamat:2017hda}.
The matter content is given in the table below. 
\begin{center}
	\begin{tabular}{|c||c|c|c|c|}
		\hline
		Field & $Spin\left(5\right)_{g}$ & $SU\left(8\right)$ & $U\left(1\right)_{a}$ & $U(1)_{\hat r}$\\
		\hline
		$S$ & $\boldsymbol{4}$ & $\boldsymbol{8}$ & 1 & $\frac{1}{2}$\\
		
		$V$ & $\boldsymbol{5}$ & $\boldsymbol{1}$ & -4 & 0\\
		
		$M$ & $\boldsymbol{1}$ & $\boldsymbol{\overline{28}}$ & -2 & 1\\
		
		$\Phi$ & $\boldsymbol{1}$ & $\boldsymbol{1}$ & 8 & 2\\
		\hline
	\end{tabular}
	\label{table:Spin5}
\end{center}
Here $U(1)_{\hat r}$ is the superconformal R symmetry obtained by $a$ maximization \cite{Intriligator:2003jj} and the superpotential is given by,

\begin{equation}
W=MS^{2}+\Phi V^{2}.
\end{equation}
We will in this section denote by $M$ gauge singlet chiral fields which are added to the models so as to make them be completely self dual under the dualities of \cite{Csaki:1997cu}, while fields denoted by $\Phi$ are added to have all operators above the unitarity bound.

In this model, considered in \cite{Razamat:2017hda}, the contribution of the conserved currents to the index at order $pq$ and having zero charge under $U(1)_a$ is given by 
\begin{equation}
-\boldsymbol{133}_{E_{7}}-1 ,
\end{equation}
suggesting that the symmetry $SU\left(8\right)\times U\left(1\right)$ enhances to $E_{7}\times U\left(1\right)$ in the IR.  Here ${\bf 133}_{E_7}$ decomposes as ${\bf 63}_{SU(8)}+{\bf 70}_{SU(8)}$.
Assuming that the theory flows to an interacting SCFT and that we have identified the superconformal R symmetry correctly this is a proof that the symmetry enhances to this group. Note that as discussed in \cite{Razamat:2017hda} the model has marginal operators charged under $U(1)_a$; in particular, on the conformal manifold the theory is equivalent to the $E_7$ surprise model of \cite{Dimofte:2012pd} which is a $Spin(4)$ theory with eight spinors in the representation $(\frac12,0)$ and eight in the $(0,\frac12)$. 
On the conformal manifold the field $V$ acquires a vacuum expectation value breaking $Spin(5)$ to $Spin(4)$.   

We note that this model has a generalization. The group $Spin(5)$ is the same as $USp(4)$ and if one takes $USp(4n)$ models with an antisymmetric field, which generalizes the vector, and eight fundamentals, which generalize the spinor, one also obtains enhancement to $E_7\times U(1)$ as was discussed in \cite{Razamat:2017hda}.  We will not discuss this generalization here. 

\

\subsection*{$\boldsymbol{Spin(6)}$}

\

Let us now consider a $Spin(6)_g$ gauge theory.
The matter content is given in the table below. 
\begin{center}
	\begin{tabular}{|c||c|c|c|c|c|c|c|}
		\hline
		Field & $Spin\left(6\right)_{g}$ & $SU\left(2\right)_{v}$ & $SU\left(4\right)_{c}$ & $SU\left(4\right)_{s}$ & $U\left(1\right)_{b}$ & $U\left(1\right)_{a}$ & $U\left(1\right)_{r}$\\
		\hline 
		$S$ & $\boldsymbol{\overline{4}}$ & $\boldsymbol{1}$ & $\boldsymbol{1}$ & $\boldsymbol{4}$ & 1 & 1 & $\frac{1}{2}$\\
		
		$C$ & $\boldsymbol{4}$ & $\boldsymbol{1}$ & $\boldsymbol{4}$ & $\boldsymbol{1}$ & -1 & 1 & $\frac{1}{2}$\\		
		
		$V$ & $\boldsymbol{6}$ & $\boldsymbol{2}$ & $\boldsymbol{1}$ & $\boldsymbol{1}$ & 0 & -2 & 0\\
		
		$M_0$ & $\boldsymbol{1}$ & $\boldsymbol{1}$ & $\boldsymbol{\overline{4}}$ & $\boldsymbol{\overline{4}}$ & 0 & 2 & 1\\
		
		$M_1$ & $\boldsymbol{1}$ & $\boldsymbol{2}$ & $\boldsymbol{6}$ & $\boldsymbol{1}$ & 2 & 0 & 1\\
		
		$\Phi$ & $\boldsymbol{1}$ & $\boldsymbol{3}$ & $\boldsymbol{1}$ & $\boldsymbol{1}$ & 0 & 4 & 2\\
		\hline
	\end{tabular}
	\label{table:Spin6}
\end{center}
The superpotential is given by, 
\begin{equation}
W=M_{0}CV^{2}S+M_{1}VC^{2}+\Phi V^{2}\,,
\end{equation}
and the superconformal R charge by ${\hat r}=r-0.142q_{a}-0.057q_{b}$. In this model, analyzed in \cite{Razamat:2017wsk} and in appendix A, the contribution to the index at order $pq$  is given by 
\begin{equation*}
-\boldsymbol{15}_{SU(4)_c}-\boldsymbol{15}_{SU(4)_s}-\boldsymbol{6}_{SU(4)_c}\boldsymbol{6}_{SU(4)_s}-\boldsymbol{3}_{SU(2)_v}-2
\end{equation*}
\begin{equation}
=-\boldsymbol{66}_{SO(12)}-\boldsymbol{3}_{SU(2)_v}-2 ,
\end{equation}
suggesting that the symmetry $SU\left(2\right)\times SU\left(4\right)^{2}\times U\left(1\right)^{2} $ enhances to $SU\left(2\right)\times SO\left(12\right)\times U\left(1\right)^{2}$ in the IR. Moreover, the dimension of the conformal manifold vanishes if there are no accidental $U(1)$s, evidence for which we do not see.  

We note that also this sequence has a generalization. Note that $Spin(6)$ is the same as $SU(4)$. Taking $SU(2n)$ with two antisymmetric, which generalize the vector, four fundamentals and four anti fundamentals, which generalize the spinor, we can construct theories which have $SU(4)\times SU(4)$ symmetry enhanced to $SO(12)$ given appropriate gauge singlet fields are coupled. This sequence is somewhat outside our main thread of discussion and we will consider it in appendix C in full detail. In particular we will discuss a new duality related to  symmetry enhancement in this model.

\

\subsection*{$\boldsymbol{Spin(7)}$}

\

We consider next a $Spin(7)_g$ model. 
The matter content is given in the table below. 
\begin{center}
	\begin{tabular}{|c||c|c|c|c|c|}
		\hline
		Field & $Spin\left(7\right)_{g}$ & $SU\left(4\right)_{s}$ & $SU\left(3\right)_{v}$ & $U\left(1\right)_{a}$ & $U\left(1\right)_{r}$\\
		\hline
		$S$ & $\boldsymbol{8}$ & $\boldsymbol{4}$ & $\boldsymbol{1}$ & -3 & $\frac{1}{2}$\\
		
		$V$ & $\boldsymbol{7}$ & $\boldsymbol{1}$ & $\boldsymbol{3}$ & 4 & 0\\
		
		$M_{0}$ & $\boldsymbol{1}$ & $\boldsymbol{6}$ & $\boldsymbol{\overline{3}}$ & 2 & 1\\
		
		$M_{1}$ & $\boldsymbol{1}$ & $\boldsymbol{\overline{10}}$ & $\boldsymbol{1}$ & 6 & 1\\
		
		$\Phi$ & $\boldsymbol{1}$ & $\boldsymbol{1}$ & $\boldsymbol{\overline{6}}$ & -8 & 2\\
		\hline
	\end{tabular}
	\label{table:Spin7}
\end{center}
The superpotential is given by 
\begin{equation}
W=M_{0}S^{2}V+M_{1}S^{2}+\Phi V^{2}
\end{equation}
and the superconformal R charge by ${\hat r}=r+0.039q_{a}$. The contribution to the index at order $pq$  is given by,
\begin{equation*}
-\boldsymbol{15}_{SU(4)_s}-\boldsymbol{20'}_{SU(4)_s}-2\,\boldsymbol{8}_{SU(3)_v}-1
\end{equation*}
\begin{equation}
=-\boldsymbol{35}_{SU(6)}-2\,\boldsymbol{8}_{SU(3)_v}-1\,,
\end{equation}
suggesting that the symmetry $SU\left(4\right)\times SU\left(3\right)\times U\left(1\right)$ enhances to $SU\left(6\right)\times SU\left(3\right)^{2}\times U\left(1\right)$ in the IR. 
Note that we observe the adjoint representation of $SU(3)_v$ twice in the index at order $qp$ with the negative sign. If the basic assumptions are correct this indicates that 
the $SU(3)$ symmetry we see in the UV is a diagonal combination of two $SU(3)$ symmetries in the IR. The $SU(4)_s\sim Spin(6)$ symmetry is enhanced to $SU(6)$ with the vector representation being the (anti)fundamental of $SU(6)$. Note that in this case the rank of the IR symmetry is bigger than the rank of the UV symmetry.
In appendix A we discuss how the conserved currents of the enhanced symmetries are constructed from the operators of the UV theory. This is not a clean statement as there are many recombinations of short multiplets into long ones.  
We note that also here  the dimension of the conformal manifold vanishes as there are no terms with positive sign appearing at order $qp$. 

\

\subsection*{$\boldsymbol{Spin(8)}$}

\

We consider a $Spin(8)_g$ gauge theory with matter in all the three $8$ dimensional representations. We split the discussion into two. In the first part, we consider a model in which the conformal manifold is a point, and in the second part a closely related model with an IR global symmetry that corresponds to the commutant structure mentioned above. The matter content of the first model is given in the table below. 
\begin{center}
	\begin{tabular}{|c||c|c|c|c|c|c|c|}
		\hline
		Field  & $Spin\left(8\right)_{g}$ & $SU\left(2\right)_{s}$ & $SU\left(2\right)_{c}$ & $SU\left(4\right)_{v}$ & $U\left(1\right)_{a}$ & $U\left(1\right)_{b}$ & $U\left(1\right)_{r}$\\
		\hline 
		$S$ & $\boldsymbol{8}_{s}$ & $\boldsymbol{2}$ & $\boldsymbol{1}$ & $\boldsymbol{1}$ & -2 & 0 & $\frac{1}{2}$\\
		
		$C$ & $\boldsymbol{8}_{c}$ & $\boldsymbol{1}$ & $\boldsymbol{2}$ & $\boldsymbol{1}$ & 0 & -2 & $\frac{1}{2}$\\
		
		$V$ & $\boldsymbol{8}_{v}$ & $\boldsymbol{1}$ & $\boldsymbol{1}$ & $\boldsymbol{4}$ & 1 & 1 & 0\\
		
		$M_{0}$ & $\boldsymbol{1}$ & $\boldsymbol{1}$ & $\boldsymbol{1}$ & $\boldsymbol{6}$ & 2 & -2 & 1\\
		
		$M_{1}$ & $\boldsymbol{1}$ & $\boldsymbol{2}$ & $\boldsymbol{2}$ & $\boldsymbol{\overline{4}}$ & 1 & 1 & 1\\
		
		$M_{2}$ & $\boldsymbol{1}$ & $\boldsymbol{3}$ & $\boldsymbol{1}$ & $\boldsymbol{1}$ & 4 & 0 & 1\\
		
		$M_{3}$ & $\boldsymbol{1}$ & $\boldsymbol{1}$ & $\boldsymbol{3}$ & $\boldsymbol{1}$ & 0 & 4 & 1\\
		
		$\Phi$ & $\boldsymbol{1}$ & $\boldsymbol{1}$ & $\boldsymbol{1}$ & $\boldsymbol{\overline{10}}$ & -2 & -2 & 2\\
		\hline
	\end{tabular}
	\label{table:Spin8a}
\end{center}
The superpotential is given by 
\begin{equation}
\label{WSpin8}
W=M_{0}S^{2}V^{2}+M_{1}SCV+M_{2}S^{2}+M_{3}C^{2}+\Phi V^{2}
\end{equation}
and the superconformal R charge by ${\hat r}=r+0.061q_{a}+0.107q_{b}$. The contribution to the index at order $pq$  is given by,
\begin{equation*}
-\boldsymbol{3}_{SU(2)_s}\boldsymbol{3}_{SU(2)_c}-\boldsymbol{3}_{SU(2)_s}-\boldsymbol{3}_{SU(2)_c}-2\,\boldsymbol{15}_{SU(4)_v}-2
\end{equation*}
\begin{equation}
=-\boldsymbol{15}_{\widetilde{SU}(4)}-2\,\boldsymbol{15}_{SU(4)_v}-2\,,
\end{equation}
and the symmetry $SU\left(2\right)^{2}\times SU\left(4\right)\times U\left(1\right)^{2}$ enhances to $SU\left(4\right)^{3}\times U\left(1\right)^{2}$ in the IR. 
Note that the symmetry rotating the matter in the two spinor representations, $SU(2)_s\times SU(2)_c$, enhances to $SU(4)$ with the (anti)fundamental of $SU(4)$ built as $(2,2)$ of the two $SU(2)$ symmetries. The symmetry rotating the matter in the vector representation is, as also in the previous case, enhancing to two copies of $SU(4)$. 
Moreover, the dimension of the conformal manifold vanishes. 

Next, in the second part, we examine a model which is similar to the previous one but without the "flipper" field $M_0$; that is, we remove this field and its corresponding term in the superpotential \eqref{WSpin8}. We obtain the matter content given in the table below. 
\begin{center}
	\begin{tabular}{|c||c|c|c|c|c|c|c|}
		\hline
		Field  & $Spin\left(8\right)_{g}$ & $SU\left(2\right)_{s}$ & $SU\left(2\right)_{c}$ & $SU\left(4\right)_{v}$ & $U\left(1\right)_{a}$ & $U\left(1\right)_{b}$ & $U\left(1\right)_{\hat r}$\\
		\hline 
		$S$ & $\boldsymbol{8}_{s}$ & $\boldsymbol{2}$ & $\boldsymbol{1}$ & $\boldsymbol{1}$ & -2 & 0 & $\frac{1}{3}$\\
		
		$C$ & $\boldsymbol{8}_{c}$ & $\boldsymbol{1}$ & $\boldsymbol{2}$ & $\boldsymbol{1}$ & 0 & -2 & $\frac{1}{3}$\\
		
		$V$ & $\boldsymbol{8}_{v}$ & $\boldsymbol{1}$ & $\boldsymbol{1}$ & $\boldsymbol{4}$ & 1 & 1 & $\frac{1}{6}$\\
		
		$M_{1}$ & $\boldsymbol{1}$ & $\boldsymbol{2}$ & $\boldsymbol{2}$ & $\boldsymbol{\overline{4}}$ & 1 & 1 & $\frac{7}{6}$\\
		
		$M_{2}$ & $\boldsymbol{1}$ & $\boldsymbol{3}$ & $\boldsymbol{1}$ & $\boldsymbol{1}$ & 4 & 0 & $\frac{4}{3}$\\
		
		$M_{3}$ & $\boldsymbol{1}$ & $\boldsymbol{1}$ & $\boldsymbol{3}$ & $\boldsymbol{1}$ & 0 & 4 & $\frac{4}{3}$\\
		
		$\Phi$ & $\boldsymbol{1}$ & $\boldsymbol{1}$ & $\boldsymbol{1}$ & $\boldsymbol{\overline{10}}$ & -2 & -2 & $\frac{5}{3}$\\
		\hline
	\end{tabular}
	\label{table:Spin8b}
\end{center}
The superpotential is now given by 
\begin{equation}
W=M_{1}SCV+M_{2}S^{2}+M_{3}C^{2}+\Phi V^{2}
\end{equation}
and $\hat r$ written in the table is the superconformal R charge. When calculating the index in this case, the representations of $SU(2)_s\times SU(2)_c$ turn out to form representations of $SU(4)$ as before, but in addition to that the index is now consistent with a combination of the two $U(1)$ symmetries enhancing to $SU(2)$. Explicitly, if we define 
\begin{equation}
U(1)_{e}=\frac{1}{4}\left[U(1)_{a}-U(1)_{b}\right]\,,\,\,\,\,U(1)_{h}=\frac{1}{2}\left[U(1)_{a}+U(1)_{b}\right]
\end{equation}
we get that $U(1)_{e}$ enhances to $SU(2)_{e}$. Writing the index in terms of $SU(2)_{e}$ and $\widetilde{SU}(4)\supset SU(2)_{s}\times SU(2)_{c}$, we obtain 
\begin{equation*}
I=1+\boldsymbol{6}_{SU\left(4\right)_{v}}\boldsymbol{2}_{SU\left(2\right)_{e}}\left(pq\right)^{\frac{1}{2}}+2\,\boldsymbol{4}_{\widetilde{SU}(4)}\boldsymbol{\bar{4}}_{SU\left(4\right)_{v}}h\left(pq\right)^{\frac{7}{12}}
\end{equation*}
\begin{equation*}
+\left[\boldsymbol{6}_{\widetilde{SU}(4)}\boldsymbol{2}_{SU\left(2\right)_{e}}h^{2}+h^{-4}\right]\left(pq\right)^{\frac{2}{3}}+\left[\boldsymbol{\overline{10}}_{SU\left(4\right)_{v}}+\boldsymbol{6}_{SU\left(4\right)_{v}}\right]h^{-2}\left(pq\right)^{\frac{5}{6}}
\end{equation*}
\begin{equation}
+\boldsymbol{6}_{SU\left(4\right)_{v}}\boldsymbol{2}_{SU\left(2\right)_{e}}\left(p^{\frac{3}{2}}q^{\frac{1}{2}}+p^{\frac{1}{2}}q^{\frac{3}{2}}\right)+\left[\boldsymbol{20'}_{SU\left(4\right)_{v}}\boldsymbol{3}_{SU\left(2\right)_{e}}-\boldsymbol{15}_{\widetilde{SU}(4)}-\boldsymbol{15}_{SU\left(4\right)_{v}}-1\right]pq+\ldots\,.
\end{equation}
In this case the evidence from the index does not give a proof for the enhancement but rather an indication that the supersymmetric spectrum of states is consistent with such a claim. 
This is because we only see representations of the symmetry appearing but not the conserved currents. If the conjecture is correct the theory  contains marginal operator in the adjoint representation of $SU(2)$ which is canceled in the index with the conserved currents. 

We should also note that in this model we do not observe any evidence for the enhancement of $SU\left(4\right)_{v}$ to $SU\left(4\right)^2$. Particularly, we do not see the additional conserved currents. However, there is no direct contradiction either, as the index can be written in characters of that symmetry and the additional conserved currents can cancel against a marginal operator in the adjoint representation of $SU\left(4\right)_{v}$. One reason why we may expect this enhancement is that we can start from the previous case with $M_0$, where we observe the conserved currents, add a free chiral field and use it to flip $M_0$. This is a relevant deformation compared to the fixed point and so should initiate a flow that is expected to lead to the model without $M_0$. Furthermore, we do not expect this to break the enhanced symmetry as the only other operator with the same R-charge is $C^2 V^2$ and there is no way to combine them to a representation such that the flipping breaks the enhanced symmetry.  

\

\subsection*{$\boldsymbol{Spin(9)}$}

\

Next consider a $Spin(9)_g$ model.
The matter content is given in the table below. 
\begin{center}
	\begin{tabular}{|c||c|c|c|c|c|}
		\hline
		Field & $Spin\left(9\right)_{g}$ & $SU\left(2\right)_{s}$ & $SU\left(5\right)_{v}$ & $U\left(1\right)_{a}$ & $U\left(1\right)_{r}$\\
		\hline 
		$S$ & $\boldsymbol{16}$ & $\boldsymbol{2}$ & $\boldsymbol{1}$ & -5 & $\frac{1}{2}$\\
		
		$V$ & $\boldsymbol{9}$ & $\boldsymbol{1}$ & $\boldsymbol{5}$ & 4 & 0\\
		
		$M_{0}$ & $\boldsymbol{1}$ & $\boldsymbol{1}$ & $\boldsymbol{\overline{10}}$ & 2 & 1\\
		
		$M_{1}$ & $\boldsymbol{1}$ & $\boldsymbol{3}$ & $\boldsymbol{\overline{5}}$ & 6 & 1\\
		
		$M_{2}$ & $\boldsymbol{1}$ & $\boldsymbol{3}$ & $\boldsymbol{1}$ & 10 & 1\\
		
		$\Phi$ & $\boldsymbol{1}$ & $\boldsymbol{1}$ & $\boldsymbol{\overline{15}}$ & -8 & 2\\
		\hline
	\end{tabular}
	\label{table:Spin9}
\end{center}
The superpotential is given by,
\begin{equation}
W=M_{0}S^{2}V^{2}+M_{1}S^{2}V+M_{2}S^{2}+\Phi V^{2}\,,
\end{equation}
and the superconformal R charge by ${\hat r}=r+0.0376q_{a}$. The contribution to the index at order $pq$  is given by
\begin{equation*}
-\boldsymbol{3}_{SU(2)_s}-\boldsymbol{5}_{SU(2)_s}-2\,\boldsymbol{24}_{SU(5)_v}-2
\end{equation*}
\begin{equation}
=-\boldsymbol{8}_{SU(3)}-2\,\boldsymbol{24}_{SU(5)_v}-2 ,
\end{equation}
suggesting that the symmetry $SU\left(2\right)\times SU\left(5\right)\times U\left(1\right)$ enhances to $SU\left(3\right)\times SU\left(5\right)^{2}\times U\left(1\right)^{2}$ in the IR.
Note that $SU(2)_s$ enhances to $SU(3)$ with the adjoint of $SU(2)$ being the (anti)fundamental of $SU(3)$. 
Moreover, the dimension of the conformal manifold vanishes. 

\

\subsection*{$\boldsymbol{Spin(10)}$}

\

The next model in our list is a $Spin(10)_g$ gauge theory.
The matter content is given in the table below. 
\begin{center}
	\begin{tabular}{|c||c|c|c|c|c|}
		\hline
		Field  & $Spin\left(10\right)_{g}$ & $SU\left(2\right)_{s}$ & $SU\left(6\right)_{v}$ & $U\left(1\right)_{a}$ & $U\left(1\right)_{r}$\\
		\hline
		$S$ & $\boldsymbol{16}$ & $\boldsymbol{2}$ & $\boldsymbol{1}$ & 3 & $\frac{1}{2}$\\
		
		$V$ & $\boldsymbol{10}$ & $\boldsymbol{1}$ & $\boldsymbol{6}$ & -2 & 0\\
		
		$M$ & $\boldsymbol{1}$ & $\boldsymbol{3}$ & $\boldsymbol{\overline{6}}$ & -4 & 1\\
		
		$\Phi$ & $\boldsymbol{1}$ & $\boldsymbol{1}$ & $\boldsymbol{\overline{21}}$ & 4 & 2\\
		\hline
	\end{tabular}
	\label{table:Spin10}
\end{center}
The superpotential is given by
\begin{equation}
W=MS^{2}V+\Phi V^{2}\,,
\end{equation}
and the superconformal R charge by ${\hat r}=r-0.076q_{a}$. The contribution to the index at order $pq$  is given by 
\begin{equation}
\boldsymbol{175}_{SU(6)_v}-\boldsymbol{35}_{SU(6)_v}-\boldsymbol{3}_{SU(2)_s}-\boldsymbol{5}_{SU(2)_s}-1\,,
\end{equation} 
and we see that in this case there is a non-vanishing contribution from marginal operators and thus the conformal manifold is not a single point. Analyzing the representations of operators in this theory (see  appendix A), we get that we can write this result in the form 
\begin{equation}
\textrm{Marginals}-\textrm{Conserved currents},
\end{equation}
where 
\begin{equation}
\textrm{Marginals}=\boldsymbol{175}_{SU(6)_v}+\boldsymbol{35}_{SU(6)_v}
\end{equation}
and
\begin{equation}
\textrm{Conserved currents}=2\,\boldsymbol{35}_{SU(6)_v}+\boldsymbol{8}_{SU(3)}+1,
\end{equation}
corresponding to the enhancement of symmetry from $SU\left(2\right)\times SU\left(6\right)\times U\left(1\right)$ to $SU\left(3\right)\times SU\left(6\right)^{2}\times U\left(1\right)$. 
We note that the results are also consistent with the marginals only being in ${\bf 175}_{SU(6)_v}$ and the symmetry rotating the fields in the vector representation not enhancing to two copies. 
We conjecture that the symmetry does enhance because the model is part of a sequence with such a property but we do not have a proof of this statement. The symmetry $SU(2)_s$ does enhance to $SU(3)$ with the adjoint of the former being the (anti)fundamental of $SU(3)$.  Here the results of the index are a proof of the statement modulo the usual assumptions.

\

\subsection*{$\boldsymbol{Spin(11)}$}

\

We consider a $Spin(11)_g$ model with seven chiral fields in the vector representation and a single spinor.
The matter content is given in the table below. 
\begin{center}
	\begin{tabular}{|c||c|c|c|c|}
		\hline
		Field  & $Spin\left(11\right)_{g}$ & $SU\left(7\right)_{v}$ & $U\left(1\right)_{a}$ & $U\left(1\right)_{r}$\\
		\hline 
		$S$ & $\boldsymbol{32}$ & $\boldsymbol{1}$ & -7 & $\frac{1}{2}$\\
		
		$V$ & $\boldsymbol{11}$ & $\boldsymbol{7}$ & 4 & 0\\
		
		$M_{0}$ & $\boldsymbol{1}$ & $\boldsymbol{\overline{21}}$ & 6 & 1\\
		
		$M_{1}$ & $\boldsymbol{1}$ & $\boldsymbol{\overline{7}}$ & 10 & 1\\
		
		$\Phi$ & $\boldsymbol{1}$ & $\boldsymbol{\overline{28}}$ & -8 & 2\\
		\hline
	\end{tabular}
	\label{table:Spin11}
\end{center}
The superpotential is given by 
\begin{equation}
W=M_{0}S^{2}V^{2}+M_{1}S^{2}V+\Phi V^{2}\,,
\end{equation}
and the superconformal R charge by ${\hat r}=r+0.030q_{a}$. The contribution to the index at order $pq$  is given by 
\begin{equation}
-2\,\boldsymbol{48}_{SU(7)_v}-2\,,
\end{equation} 
suggesting that the symmetry $SU\left(7\right)\times U\left(1\right)$ enhances to $SU\left(7\right)^{2}\times U\left(1\right)^{2}$ in the IR. The symmetry rotating the vectors enhances to two copies in the IR.
 Moreover, the dimension of the conformal manifold vanishes.

\

\subsection*{$\boldsymbol{Spin(12)}$}

\

The last model is a $Spin(12)_g$ gauge theory.
The matter content is given in the table below. 
\begin{center}
	\begin{tabular}{|c||c|c|c|c|}
		\hline
		Field  & $Spin\left(12\right)_{g}$ & $SU\left(8\right)_{v}$ & $U\left(1\right)_{a}$ & $U\left(1\right)_{r}$\\
		\hline 
		$S$ & $\boldsymbol{\overline{32}}$ & $\boldsymbol{1}$ & 2 & $\frac{1}{2}$\\
		
		$V$ & $\boldsymbol{12}$ & $\boldsymbol{8}$ & -1 & 0\\
		
		$M$ & $\boldsymbol{1}$ & $\boldsymbol{\overline{28}}$ & -2 & 1\\
		
		$\Phi$ & $\boldsymbol{1}$ & $\boldsymbol{\overline{36}}$ & 2 & 2\\
		\hline
	\end{tabular}
	\label{table:Spin12}
\end{center}
The superpotential is given by
\begin{equation}
W=MS^{2}V^{2}+\Phi V^{2}\,,
\end{equation} 
and the superconformal R charge by ${\hat r}=r-0.110q_{a}$. The contribution to the index at order $pq$  is given by
\begin{equation}
-2\,\boldsymbol{63}_{SU(8)_v}-1\,,
\end{equation} 
suggesting that the symmetry $SU\left(8\right)\times U\left(1\right)$ enhances to $SU\left(8\right)^{2}\times U\left(1\right)$ in the IR. 
Moreover, the dimension of the conformal manifold vanishes.

\

\subsection*{Summary of the symmetry enhancements}

\

We summarize the various symmetry enhancements that occur in the models described above in the table below.

\begin{center}
	\begin{tabular}{|c|c|c|c|c|}
		\hline
		Gauge & UV  symmetry & IR  symmetry & UV rank & IR rank\\
		\hline	\hline  
		$Spin\left(5\right)$ & $SU\left(8\right)\times U\left(1\right)$ & $E_{7}\times U\left(1\right)$ & 8 & 8\\
		\hline 
		$Spin\left(6\right)$ & $SU\left(2\right)\times SU\left(4\right)^{2}\times U\left(1\right)^{2}$  & $SU\left(2\right)\times SO\left(12\right)\times U\left(1\right)^{2}$ & 9 & 9\\
		\hline 
		$Spin\left(7\right)$ & $SU\left(4\right)\times SU\left(3\right)\times U\left(1\right)$ & $SU\left(6\right)\times SU\left(3\right)^{2}\times U\left(1\right)$ & 6 & 10\\
		\hline 
		$Spin\left(8\right)$ & $SU\left(2\right)^{2}\times SU\left(4\right)\times U\left(1\right)^{2}$ & $SU\left(4\right)^{3}\times U\left(1\right)\times (U(1)/\,SU(2))$ & 7 & 11\\
		\hline 
		$Spin\left(9\right)$ & $SU\left(2\right)\times SU\left(5\right)\times U\left(1\right)$ & $SU\left(3\right)\times SU\left(5\right)^{2}\times U\left(1\right)^{2}$ & 6 & 12\\
		\hline 
		$Spin\left(10\right)$ & $SU\left(2\right)\times SU\left(6\right)\times U\left(1\right)$ & $SU\left(3\right)\times SU\left(6\right)^{2}\times U\left(1\right)$ & 7 & 13\\
		\hline 
		$Spin\left(11\right)$ & $SU\left(7\right)\times U\left(1\right)$ & $SU\left(7\right)^{2}\times U\left(1\right)^{2}$ & 7 & 14\\
		\hline 
		$Spin\left(12\right)$ & $SU\left(8\right)\times U\left(1\right)$ & $SU\left(8\right)^{2}\times U\left(1\right)$ & 8 & 15\\
		\hline
	\end{tabular}
	\label{table:Summary}
\end{center}
In the case of $Spin(8)$ we write two options for the enhancement which are determined by the singlets.
We note that the sequence of groups,

$$ E_7,\,\, SO(12),\,\, SU(6),\,\, SU(4)\times SU(2),\,\, SU(3)\times U(1),\,\, SU(3),\,\, U(1),\,\, 1$$
is precisely the commutant of $SU(n)\times SU(2)$ in $E_8$. The commutant of $SU(n)$ in $E_8$ is known as the $E_{9-n}$ algebra.  We will discuss a possible way to understand this enhancement of symmetry by flows of six dimensional theories on a compact space.

\

\section{From four dimensions to five and six}\label{fiveghrt}

We would like to try to find a more fundamental reason as to why such an enhancement should occur. One way to do this is to realize the $4d$ theories as a compactification of a $6d$ model, preserving the enhanced symmetry, which is a subgroup of the symmetry of the $6d$ model. For example, it was argued that the $E_7$ enhancement in the \cite{Dimofte:2012pd} $E_7$ surprise model can be understood as resulting from a compactification of the rank $1$ E-string theory on a torus with fluxes breaking $E_8$ to $E_7 \times U(1)$ \cite{Kim:2017toz}. Specifically, the $E_7$ surprise is a mass deformation of a $4d$ theory that is the above mentioned E-string compactification. 

The $E_7$ surprise can also be regarded as the $Spin(4)$ case in the $Spin$ sequence discussed in the previous section. Thus, it is plausible that there is a generalization of the $6d$ construction also to this case, and one may hope that this can shed light on the observed enhancements. To try and make a concrete proposal, we first recall some aspects of the E-string compactifications, see \cite{Kim:2017toz}. The important observation here is that the $4d$ theory contains an $SU(2)$ gauge theory with $8$ doublet chiral fields, which in turn is related to the fact that the E-string theory when compactified on a circle has a gauge theory description as a $5d$ $\mathcal{N}=1$ $SU(2)$ gauge theory with $8$ doublet hypermultiplet. The reasons regarding this connection were elucidated in \cite{Kim:2017toz}, and further developed in \cite{Kim:2018bpg,Kim:2018lfo}, and we refer the reader to these papers for the details. 

This suggests that, as a starting point in proposing a $6d$ interpretation, we should consider the $5d$ $\mathcal{N}=1$ versions of the $4d$ $Spin$ gauge theories we introduced in the previous section. These have the same matter content as their $4d$ version, but with the $4d$ $\mathcal{N}=1$ matter replaced by $5d$ $\mathcal{N}=1$ matter, so that $4d$ chiral fields become $5d$ hypermultiplets and so forth.    

We are then lead to consider the $5d$ $\mathcal{N}=1$ $Spin(n+4)$ gauge theories with $n$ hypermultiplets in the vector representation and spinor matter with $32$ component whose exact splitting into hypermultiplets depends on $n$. The structure of this family is such that they sit on a Higgs branch flow line generated by a vev to the vector hypermultiplets. The maximal value of $n$ here is $9$, despite the fact that the spinors of $Spin(13)$ are $64$ dimensional, as these are pseudo-real allowing half-hypers. Furthermore for $Spin(12)$, whose spinors are also pseudo-real, we have two possibilities: either a single spinor hyper of a chosen chirality or two spinor half-hypers of opposite chirality\footnote{Similar choices also exist for the $n=4$ case, but here we concentrate only on the $Spin(8)$ parity invariant case as this is the case generated via Higgsing of the $Spin(9)$ case.}. The $Spin(12)$ parity invariant case is the one generated via Higgsing of the $Spin(13)$ case, but it is the other case that appears to be important for the $4d$ story. Nevertheless, we shall include these cases as those naturally fit in the $5d$ to $6d$ story.  

Next, we inquire as to the UV behavior of these theories. Specifically, we remind the reader that $5d$ $\mathcal{N}=1$ gauge theories may be realized by a mass deformation of a $5d$ or $6d$ SCFT, where in the latter case the deformation is given by circle compactification with possible flavor holonomies. This, however, is only true if the matter content is sufficiently small. For the theories we consider, there is evidence that these can be UV completed to $6d$ SCFTs on a circle. This is by itself a promising indication in our search for a $6d$ explanation.

We next briefly review the evidence for the $6d$ UV completion. First, as previously mentioned, these theories sit on a line of Higgs branch flows generated by a vev to the vector hypermultiplets. The end point of the line leads to theories that are known to have a $6d$ lift notably the cases of $n=0$, which lifts to two decoupled rank $1$ E-string theories \cite{Ganor:1996pc}, $n=1$, which lifts to the rank $2$ E-string theory \cite{Ganor:1996pc} and $n=2$, which lifts to the $E_7 \times Spin(7)$ conformal matter SCFT \cite{Zafrir:2015rga}. Thus, it is plausible that the other cases also lift to $6d$ SCFTs. 

This is further supported by analyzing the symmetry enhancement pattern in these theories. Five dimensional gauge theories have non-perturbative instanton excitations that can in some instances provide additional conserved currents that lead to symmetry enhancement at the UV fixed point. It is possible in some cases to determine what these additional currents are using field theory methods, providing a glimpse to the UV global symmetry. In theories where the UV fixed point is a $5d$ SCFT the additional currents always combine with the IR global symmetry to form a finite group. However, in theories where the UV fixed point is a $6d$ SCFT the additional currents combine with the IR global symmetry to form an affine Lie group. Furthermore, the type of affine group is the $6d$ global symmetry, possibly twisted by an outer automorphism. 

For the theories we consider here, such an analysis for the $1$-instanton case was preformed in \cite{Zafrir:2015uaa}, and it was found that the enhancement spectrum is indeed inconsistent with a finite group, but consistent with an affine one. For convenience, and as this will play a role later, we have summarized the resulting symmetries in table \ref{MC}.

Finally, there are various criteria that have been conjectured for the existence of a $5d$ or a $6d$ SCFT fixed points \cite{Jefferson:2017ahm}, and they can also be used. Indeed, when applied to these cases, they too suggest that these $5d$ gauge theories posses a $6d$ SCFT UV completion.   

\begin{table}[h!]
\begin{center}
\begin{tabular}{|c|c|}
  \hline 
  Theory  & Symmetry  \\
\hline
  $Spin(5)+1V+8S$ & $E^{(1)}_{8} \times SU(2)$  \\ 
 \hline
   $Spin(6)+2V+8S$ & $E^{(1)}_{7} \times B^{(1)}_{3}$ \\  
\hline
   $Spin(7)+3V+4S$ & $E^{(2)}_{6}\times A^{(2)}_{5}$  \\ 
\hline
   $Spin(8)+4V+2S+2C$ & $D^{(2)}_{5}\times A^{(2)}_{7}$ \\ 
\hline
   $Spin(9)+5V+2S$ & $A^{(2)}_{9}\times A^{(2)}_{4}$ \\ 
\hline
   $Spin(10)+6V+2S$ & $A^{(2)}_{11}\times A^{(2)}_{2}\times A^{(1)}_{1}$ \\ 
\hline
   $Spin(11)+7V+1S$ & $A^{(2)}_{13}\times A^{(1)}_{1}\times U(1)^{(2)}$ \\  
\hline
   $Spin(12)+8V+\frac{1}{2}S+\frac{1}{2}C$ & $A^{(2)}_{15}\times U(1)^{(2)}$ \\ 
\hline
   $Spin(12)+8V+1S$ & $A^{(2)}_{15}\times A^{(1)}_{1}$ \\ 
\hline
   $Spin(13)+9V+\frac{1}{2}S$ & $A^{(2)}_{17}$ \\ 
\hline
\end{tabular}
 \end{center}
\label{MC}
\caption{The minimal symmetry consistent with the perturbative plus $1$-instanton contribution for the $5d$ $Spin$ gauge theories considered here. The spectrum of the expected additional current is such that it can only be accommodated by an affine group, at least for some factors of the global symmetry. This is interpreted as the theory lifting to a $6d$ SCFT in the infinite coupling limit, whose symmetry is the finite Lie group associated with the affine case. The superscript here denotes whether the affine group is the twisted ($2$) or the untwisted version ($1$). This lifts to whether the compactification of the $6d$ SCFT involves a twist or not. We also use the notation of $U(1)^{(2)}$ for a $U(1)$ group projected out by charge conjugation.}
\end{table}

The remainder of this section will be devoted to conjecturing the $6d$ SCFT fixed points associated with these theories, and then subjecting these conjectures to various consistency checks. This is an interesting problem by itself, and can be tackled independently of the $4d$ story. Furthermore, the answer will assist us in formulating a conjectural explanation for the substantial symmetry enhancement we observed in the previous section. Next, we shall state our conjecture and lay out the supporting evidence in its favor.

We conjecture that this family of theories uplifts to a family of $6d$ SCFTs that can be engineered as follows. Consider taking the rank $1$ E-string theory and gauging an $SU(n)$ subgroup of $E_8$, while also adding $2n$ hypermultiplets in the fundamental representation of $SU(n)$. With this combination, the $SU(n)$ gauge anomaly can be canceled by adding a tensor multiplet as is usual in $6d$ low-energy gauge theory descriptions of $6d$ SCFTs. It is thus plausible that the infinite coupling limit would correspond to the origin of the tensor branch of some $6d$ SCFT. This is further supported as this SCFT has string theory realizations, both in brane constructions \cite{Hanany:1997gh,Brunner:1997gk} and in F-theory compactifications \cite{HMRV}.       

Then our claim is that the $5d$ SCFTs in this family, parametrized as $Spin(n+4)$, lift to this theory where the compactification is done with a twist in the charge conjugation symmetry of the $6d$ SCFT\footnote{To be more precise, we expect the $6d$ SCFT to have a discrete symmetry which descends to the charge conjugation symmetry of the $6d$ gauge theory. We can then incorporate a twist under this discrete symmetry when compactifying the $6d$ SCFT on a circle.}. We next present some evidence for our claim. First we note that the flow pattern is correctly reproduced. The largest group in $5d$ is $Spin(13)$ and this maps to the group theory fact that the largest $SU$ subgroup inside $E_8$ is $SU(9)$. There is a Higgs branch flow in $6d$ that maps an SCFT of this type with some value of $n$ to ones with lower values of $n$. This is seen in the low-energy description as Higgsing down the $SU(n)$ group using the flavor hypermultiplets.

Another strong piece of evidence is that the global symmetry matches, where in cases where the symmetry is complex we indeed see the twisted version in $5d$. The global symmetry of the $6d$ SCFT is $SU(2n)\times E_{9-n}$\footnote{For $n=2$, as the representations of $SU(2)$ are pseudo-real, the $SU(4)$ is enhanced to $Spin(7)$. The low-energy gauge theory in fact shows an $Spin(8)$ global symmetry, but this is argued to be a low-energy enhancement and that only its $Spin(7)$ subgroup is a global symmetry of the SCFT. See \cite{Ohmori:2015pia,HaZ}, for discussions on this in the context of the simpler case without the rank $1$ E-string theory.}. Here $E_{9-n}$ is the commutant of $SU(n)$ inside $E_8$, and it is probably recognizable for those familiar with $5d$ SCFTs as the global symmetry of the $5d$ SCFT whose IR description is $SU(2)+(8-n)F$ \cite{Seiberg:1996bd}. Such people will also know that taking  $n$ to be $8$ there are in fact two possibilities the so called $E_1 = SU(2)$ and $\tilde{E}_1 = U(1)$ theories, and that only $\tilde{E}_1$ can flow to $E_0$. Indeed this also appears in our case though the flow direction is now reversed.

In $6d$ this comes about as there are two inequivalent ways of embedding $SU(8)$ inside $E_8$, where one preserves an $SU(2)$ while the other preserves a $U(1)$. This can be seen by using the $Spin(16)$ subgroup of $E_8$ under which the adjoint decomposes as $\bold{248}\rightarrow \bold{120} + \bold{128}$. We can now embed $SU(8)$ using the $U(1)\times SU(8)$ subgroup of $Spin(16)$, but there are two distinct ways of performing this embedding, that differ by how each spinor decomposes. As only one spinor appears inside the adjoint of $E_8$, this leads to two inequivalent embeddings of $SU(8)$ inside $E_8$. What is even more interesting is that this matches the behavior in $5d$ where for $Spin(12)$ we have two options depending on the chiralities of the spinors. These two options match the two possible embeddings as is apparent from the symmetry enhancement pattern in this theories.      

However, we have seen that the $Spin(13)$ theory can flow to only one of them. This leads us to suspect that only the $\tilde{E}_1$ embedding can be reached from the $n=9$ case. This comes about as embedding the $SU(8)$ inside $E_8$ using its $SU(9)$ maximal subgroup automatically selects this embedding, as can be confirmed using the branching rules. 

Besides the global symmetry and flow pattern, we can also compare the dimension of the Coulomb branch. When we compactify a $6d$ theory with tensor and vector multiplets, we expect the Coulomb branch of the $5d$ theory to be spanned by the tensor multiplets and the vectors on the circle. Its dimension should then be the dimension of the $6d$ tensor branch plus the total rank of the $6d$ gauge groups. In our case we compactify with a charge conjugation twist which projects out some of these directions. Particularly, the Coulomb branch of an $SU(n)$ group is spanned by the operators $Tr(\phi^k)$, for $k=2,3,...,n$. Under the action of charge conjugation the operators with $k$ even are even while those with $k$ odd are odd. Thus, we see that in the compactification, the Coulomb branch directions associated with $k$ odd operators are projected out. Therefore, we conclude that compactifying the $6d$ SCFT in this way we expect a $\left\lfloor \frac{n}{2} \right\rfloor + 2$ dimensional Coulomb branch. This exactly matches the Coulomb branch dimension of the $Spin(n+4)$ $5d$ gauge theory.

Some lower dimensional cases have already been analyzed, particularly the cases of $n=1,2$. In the case of $n=1$ the $5d$ gauge theory is $Spin(5)+1V+8S=USp(4)+1AS+8F$ which is known to lift to the $6d$ rank $2$ E-string \cite{Ganor:1996pc}, which is indeed the $6d$ SCFT we find in this case. For $n=2$ the $5d$ gauge theory is $Spin(6)+2V+8S=SU(4)_0+2AS+8F$ which is known to lift to the $6d$ $E_7 \times Spin(7)$ conformal matter \cite{Zafrir:2015rga}, which is indeed the $6d$ SCFT we find in this case. In both cases charge conjugation acts trivially so the compactification can be done without the twist.  

Finally we can also consider other Higgs branch flows. The $5d$ theory possesses also ones that are associated with giving vevs to the spinor hypermultiplets. These lead to theories outside the family we considered so far. In $6d$ we can Higgs the tensor in the rank $1$ E-string theory, replacing it with $29$ hypers. This process is known to break $E_8$ down to $SU(2)\times E_7$, where the $SU(2)$ is locked to the $SU(2)$ R-symmetry. Of course in order to access the Higgs branch the commutant of $SU(n)$ in $E_8$ must be at least $SU(2)$. Under the unbroken $E_7$, the $29$ hypers transform as $\bold{1}+\frac{1}{2}\bold{56}$. In our case the $E_7$ will be partially gauged by the $SU(n)$ and these will provide additional matter for the $SU(n)$. As we shall now see the remaining flow diagram is also consistent.

First it should be noted that in both $5d$ and $6d$ the Higgs branch at infinite coupling usually differs from the one visible at finite coupling, see for instance \cite{Cremonesi:2015lsa,Ferlito:2017xdq,Hanany:2018uhm,HaZ}. Thus, although in principle the Higgs branches at infinite coupling should agree this does not guarantee that the ones at finite coupling, which are the ones we will be mostly relying on, agree.

With that in mind consider first the $Spin(13)$ SCFT. In this case the  minimal Higgs branch flow Higgses $SU(9)$ down to $SU(8)$, and maps to giving a vev to a vector on the $5d$ side. One cannot Higgs down the rank $1$ E-string theory here because there is no ungauged $SU(2)$ in $E_8$. Alternatively ,on the $5d$ side, it is known that one cannot give a vev just to a half-hyper in the spinor so any Higgs branch flow must involve at least a vev to one vector. 

Going down to $Spin(12)$, we have two theories. Again, one cannot give a vev just to the spinors when we have a half-hyper spinor for each chirality, but such a vev is possible when we have a full hyper in the spinor representation. In the latter case the theory flows to an $SU(6)+16F$ gauge theory, which is known to lift to the $6d$ SCFT associated with the $6d$ gauge theory $USp(8)+16F$\cite{Hayashi:2015fsa,Yonekura:2015ksa}. On the $6d$ side we can only Higgs down the rank $1$ E-string theory when the commutant of $SU(8)$ in $E_8$ is $SU(2)$. This is precisely the case mapped to the full hyper spinor. In that case the $6d$ gauge theory flows to an $SU(8)+1AS+16F$ gauge theory which can be further Higgsed to $USp(8)+16F$. As this theory is real, the charge conjugation twist does not effect it and we find the $6d$ and $5d$ flow patterns consistent.      

The flow patterns for $Spin(11)$ and $Spin(10)$ are similar, both having a Higgs branch flow generated by spinor vevs which leads to an $SU(5)+14F$ gauge theory. This theory is known to lift to the $6d$ SCFT associated with the $6d$ gauge theory $USp(6)+14F$ \cite{Hayashi:2015fsa,Yonekura:2015ksa}. This maps in the $6d$ side to Higgsing down the rank $1$ E-string theory to get $SU(n)+1AS+(n+8)F$ and then further Higgsing down to $USp(6)+14F$ in both cases. One curious thing is that for $n=6$ one can also Higgs down the rank $1$ E-string theory to get $SU(6)+\frac{1}{2}\bold{20}+15F$ instead, though we do not see any good candidate for this flow on the $5d$ side. It is possible that this flow appears only non-perturbatively there.

The $Spin(9)$ and $Spin(8)$ spinor vevs are again very similar, both leading to $Spin(7)+1V+6S$ which we can further Higgs to $SU(4)+12F$. In $6d$ this flow gives $SU(n)+1AS+(n+8)F$ which can be further Higgsed to $USp(4)+12F$, which reduces in $5d$ to the $SU(4)+12F$ gauge theory \cite{Hayashi:2015fsa,Yonekura:2015ksa}. So we see again that the flow appears consistent. Interestingly in both cases we now pass through an intermediate theory. Comparing symmetries and Coulomb branch dimensions we come to the conjecture that the $Spin(7)+1V+6S$ should lift to the twisted compactification of the $6d$ SCFT associated with $SU(4)+1AS+12F$.

Finally, for $Spin(7)$, a vev to a spinor hyper leads to a $G_2+6F$ gauge theory which can be further Higgsed to $SU(3)+10F$. In the $6d$ side this flow maps to Higgsing down the rank $1$ E-string which leads to $SU(3)+12F$ and can be further Higgsed to $SU(2)+10F$. The latter, again, is known to be the $6d$ lift of the $5d$ $SU(3)+12F$ gauge theory \cite{Hayashi:2015fsa,Yonekura:2015ksa}. Interestingly, it has been suggested in \cite{Jefferson:2018irk} that $5d$ $G_2+6F$ gauge theory lifts to a twisted compactification of $SU(3)+12F$, and our results are consistent with that.

\

\section{Symmetry enhancement and compactification from $\boldsymbol{6d}$}
\label{secsuen}

In this section we shall suggest a connection between the observed $4d$ symmetry enhancement phenomena to compactifications of six dimensional theories on a torus. Previously, we have found that the $4d$ $Spin(n+4)$ models have symmetry enhancement consisting, typically, of a $U(1)\times SU(n)\times SU(n)$ part and a remaining symmetry which can be described as the commutant in $E_8$ of $SU(n)\times SU(2)$. 

This suggests the interpretation that the $4d$ $Spin(n+4)$ models can be obtained from a deformation of a compactification of the $6d$ SCFT introduced in the previous section, probably on a torus with fluxes in the global symmetry of the SCFT. We remind the reader that the $6d$ SCFTs in question can be constructed as the UV completion of an $SU(n)$ gauge theory with $2n$ fundamental hypermultiplets connected to a rank $1$ E-string theory via gauging an $SU(n)$ subgroup of $E_8$. The symmetry of the six dimensional theory is then $E_{9-n}\times SU(2n)$. When compactifying down to four dimensions on a surface with flux for some subgroup of the global symmetry, the symmetry can be further broken. We conjecture that a relevant deformation of such a compactification can lead to the sequence we have discussed. The combined effect of the relevant deformation and the flux is to break $SU(2n)$ to $SU(n)^2 \times U(1)$ and $E_{9-n}$ to the commutant of $SU(2)$ in $E_{9-n}$. 

These expectations also hang on some recent understandings regarding the relation between theories in $6d$, $5d$ and $4d$. Particularly, compactifying a $6d$ SCFT on a torus with flux, we can first reduce to $5d$ where in lucky cases the $6d$ SCFT has an effective IR free gauge theory description. Reducing on another circle to $4d$ leads to a $4d$ gauge theory possessing similar matter content as the $5d$ gauge theory. An example of this is the $(2,0)$ theory. Reducing the type $A_{k-1}$ $(2,0)$ model on a circle one obtains the  maximally supersymmetric $SU(n)$ gauge theory, and further reduction to four dimensions leads to ${\cal N}=4$ $SU(n)$ gauge theory. Other examples involve compactifications of $ADE$ conformal matter on the torus \cite{Bah:2017gph,Kim:2017toz, Kim:2018bpg, Kim:2018lfo}.

Thus, from this view point, it would not be surprising if the $4d$ theories are related to compactifications of these $6d$ SCFTs which can be described by $Spin$ gauge theories with vectors and spinors in five dimensions. In section \ref{fiveghrt} we have discussed how the six dimensional models above, when compactified to five dimensions with a twist, are described by such effective IR free gauge theories. One issue that needs to be addressed is the twisting. The $5d$ theories are related to the $6d$ ones under a twisted compactification. However, in the $4d$ theories we observe the untwisted version of the global symmetry. These two observations are not contradictory for the following reason. We can build the torus from two tubes both containing a twist. Then, on one hand as the tubes contain a twist we expect them to have a $Spin$ description, but, on the other hand, the full surface is constructed from two such tubes, and as the twist is ${\mathbb Z}_2$ valued, the two twists should cancel out. In this way we suspect there should be a $Spin$ construction of an untwisted compactification.  

\begin{figure}
\center
\includegraphics[width=0.4\textwidth]{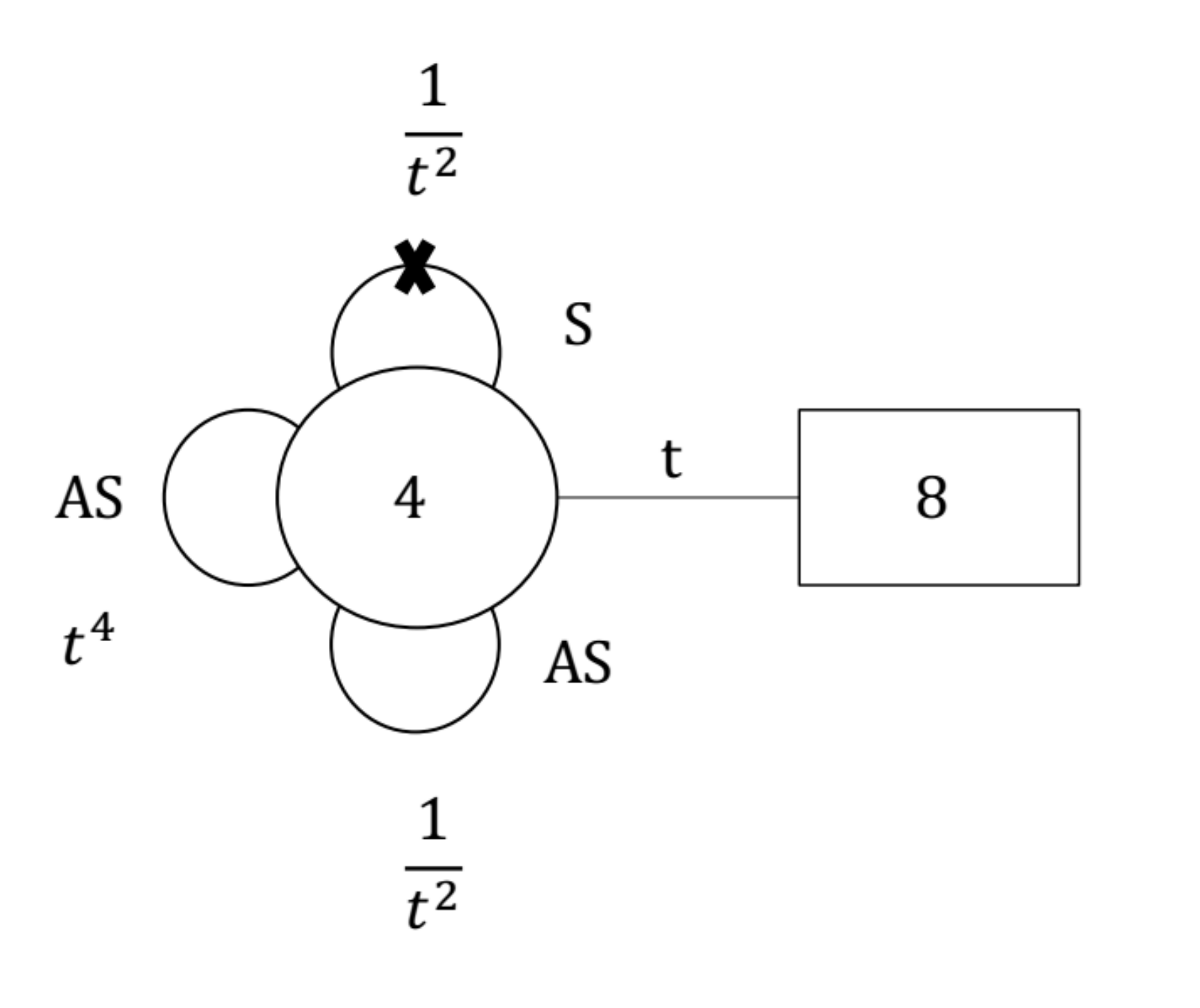} 
\caption{The $4d$ theory conjectured to be the result of the compactification and mass deformation of the rank $2$ E-string. Here the gauge group is $USp(4)$ and $S$ and $AS$ denote the symmetric, that is the adjoint for $USp$ groups, and antisymmetric representations, respectively. Here there is a superpotential coupling the symmetric chiral to the quark bilinear and the middle antisymmetric to the symmetric bilinear. The theory has an $SO(8)$ global symmetry represented by the square as well as a $U(1)$ whose charges we denoted using the fugacity $t$. The black $X$ on the symmetric chiral stands for an additional chiral field which couples linearly to the symmetric bilinear. In fact, due to the charges of the fields, the flip field also couples linearly to the lower antisymmetric bilinear, and so efficiently flips some combination of the two symmetric bilinears. There is also an anomaly free R symmetry under which the symmetric and lower antisymmetric have R charge $0$, the quarks R charge $1$ and the middle antisymmetric and flip fields have R charge $2$.}
\label{ThrUSp4}
\end{figure}

We will now make the conjecture precise for the case of $n=1$ and $n=2$.
We begin with the case of $Spin(5)$. Here the $6d$ SCFT is the rank $2$ E-string theory\footnote{Here there is no gauge group but rather only two tensors, which can be broken on the tensor branch to the rank $1$ E-string and the free $(2,0)$ tensor. This structure is known to be that of a rank $2$ E-string theory as is apparent from both brane constructions and F-theory compactifications.}. The direct compactification of this model to four dimensions is not known, but a mass deformation of it breaking the $SU(2)$ global symmetry is known \cite{Kim:2017toz}. We can consider such a theory corresponding to flux $\frac{1}{2}$ breaking $E_8\rightarrow U(1)\times E_7$. Due to the fractional flux the $E_7$ is broken to $SO(8)$ so the global symmetry of the $4d$ theory is $U(1)\times SO(8)$. See figure \ref{ThrUSp4} with the four dimensional theory. We can now preform an additional deformation by giving a vev to the flip field. This breaks both the $U(1)$ and the R-symmetry which is replaced by a new R-symmetry under which the symmetric and lower antisymmetric have R-charge $1$. Therefore, a mass term for these two fields is consistent with all the symmetries and we expect it to be generated, eliminating these fields from the resulting theory.

We now see that the charged matter content of the resulting theory is the same as the one in the $Spin(5)$ theory we studied though the two differ by the flipping field spectrum. As some operators hit the unitary bound in this theory we expect them to decouple. We do note that due to the superpotential we started with, the theory we get is at a point on the conformal manifold of the $Spin(5)$ theory where $E_7 \times U(1)$ has been broken to $SO(8)$. 

We thus see that the $Spin(5)$ theory can be connected to a compactification of the rank $2$ E-string theory up to decoupling of free fields and flipping of various operators. This still leaves a bit to be desired as the starting point does not posses all of the $6d$ symmetry apparently visible in the end point.

\begin{figure}
\center
\includegraphics[width=0.8\textwidth]{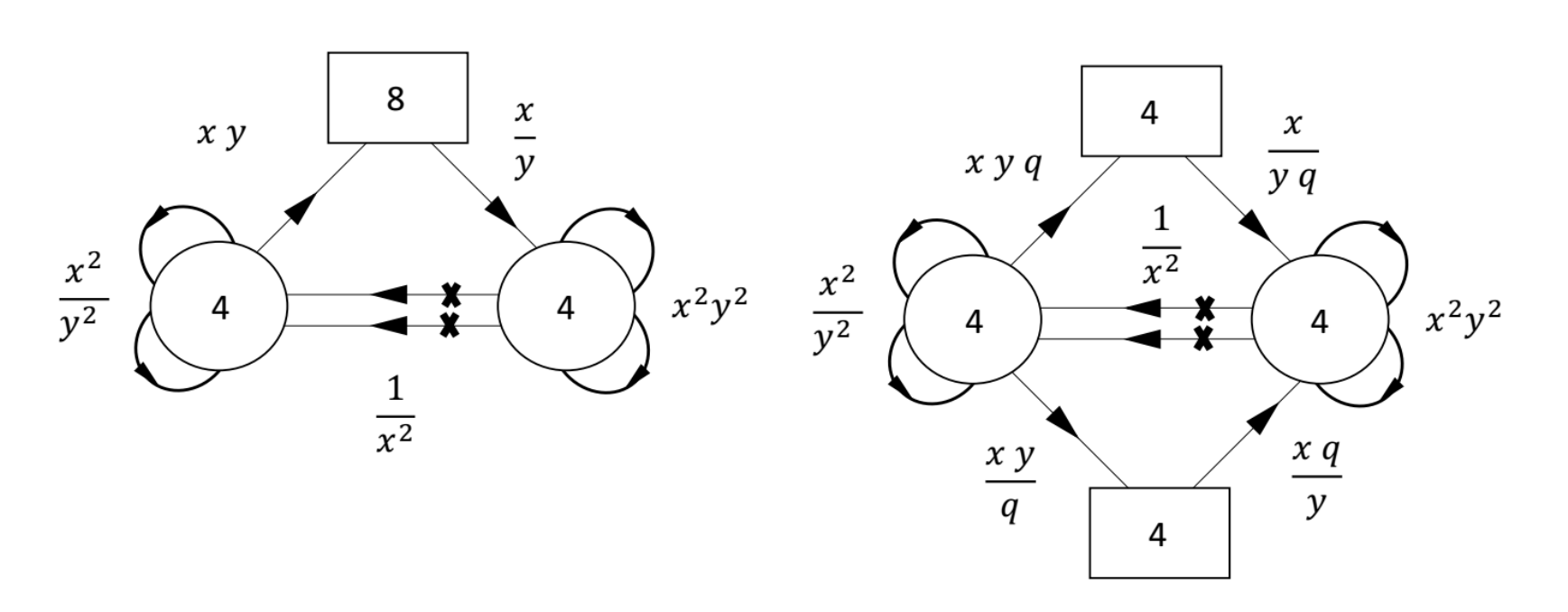} 
\caption{The $4d$ theory conjectured to be the result of the compactification of the $Spin(7)\times E_7$ conformal matter on a torus with flux $1$ in the $Spin(7)$ preserving $U(1)\times SU(2)\times SU(2)$. Here there is a cubic superpotential along the triangle and a quartic one coupling the two antisymmetrics via the bifundamentals. There are also two singlet chiral fields flipping the baryons made from the bifundamentals. The visible global symmetry here is $SU(8)\times SU(2)\times U(1)_x \times U(1)_y$ where the $SU(2)$ rotates the two antisymmetrics. From $6d$, it is expected that $SU(8)\rightarrow E_7$ and $U(1)_y \rightarrow SU(2)$. There is also a $U(1)_R$ R-symmetry under which the bifundamentals have R-charge $0$, the flippers have R-charge $2$, and the rest have R-charge $1$.}
\label{SOe7Quiver}
\end{figure}

Another case where we understand the details is the case of $Spin(6)$. The six dimensional theory is the $Spin(7)\times E_7$ conformal matter,  which is the same as an $SU(2)$ gauging of the rank $1$ E-string theory with additional four fundamental hypermultiplets. The $E_7$ is the commutant of $SU(2)$ in $E_8$ and $Spin(7)$ is the symmetry rotating the four hypermultiplets. Figure \ref{SOe7Quiver} shows the $4d$ theory resulting from the compactification of the $Spin(7)\times E_7$ conformal matter  on a torus with flux $1$ in $Spin(7)$ preserving $U(1)\times SU(2)\times SU(2)$ \cite{ZafTA}. Next consider separating the $8$ flavors to a pair of $4$. This splits the $SU(8)$ to $U(1)\times SU(4)\times SU(4)$. As the $SU(8)$ enhances to $E_7$, the $U(1)$ should enhance to $SU(2)$, and the $SU(4)\times SU(4)$ should enhance to $SO(12)$ which together form the maximal subgroup of $E_7$.

\begin{figure}
\center
\includegraphics[width=0.4\textwidth]{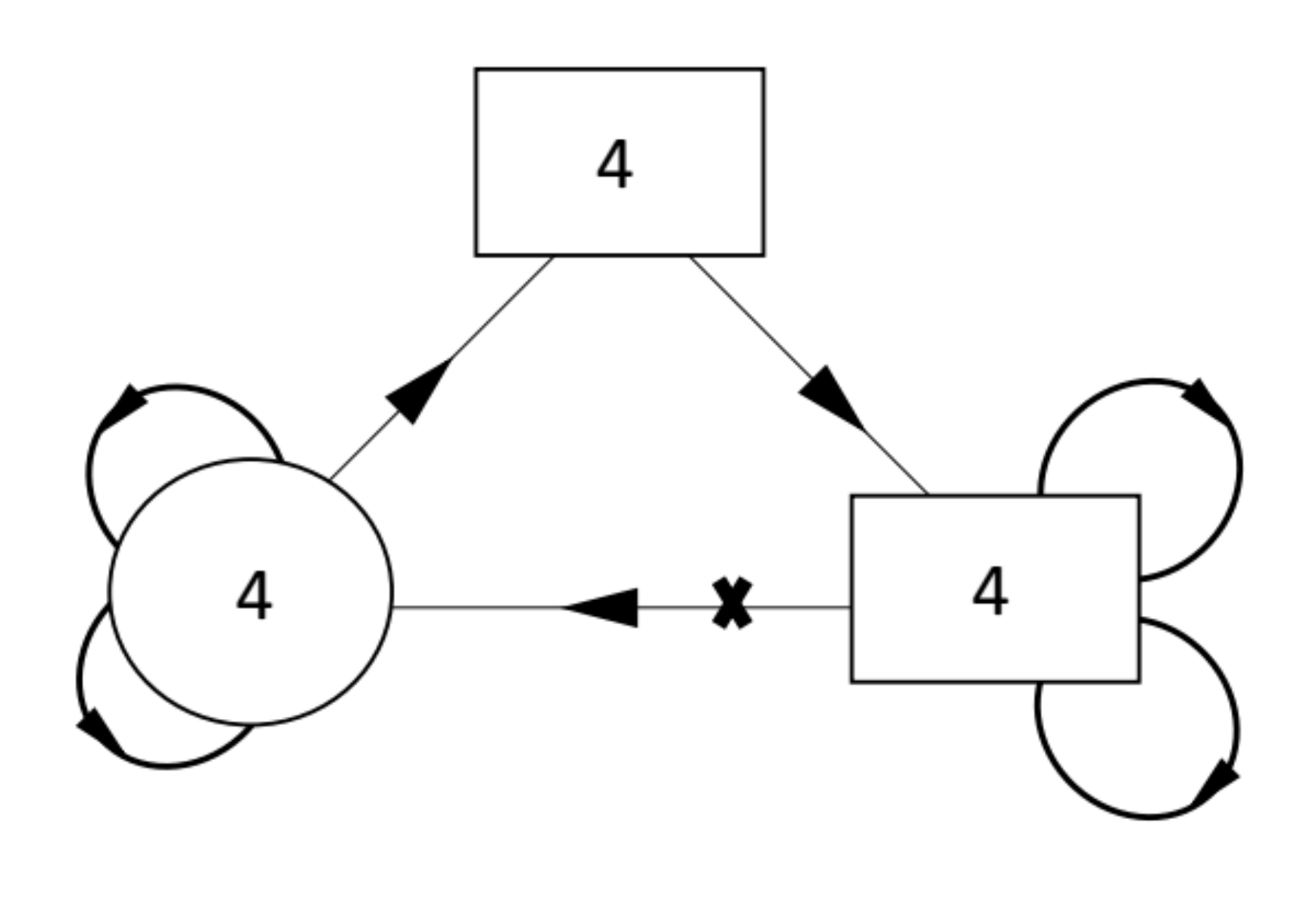} 
\caption{The $4d$ theory one gets after giving a vev to the baryon associated with the bottom right $SU(4)\times SU(4)$ bifundamental. This is just a $Spin(6)$ gauge theory with two chiral fields in the vector representation, $4$ in the spinor and another $4$ in the conjugate spinor. Additionally there are various chiral fields flipping the gauge invariant operators: $SSV, S\bar{S}, S^4$. There is also a decoupled field coming from one of the previous flippers.}
\label{QuiverRed}
\end{figure}

We next consider giving a vev to a baryon. This breaks the $U(1)$ and breaks one of the $SU(4)$ gauge symmetries which is then identified with one of the $SU(4)$ global symmetries. This is expected to preserve a $U(1)\times SU(2)\times SO(12)$ global symmetry, which is expected to be the global symmetry of the resulting theory. The resulting theory is shown in figure \ref{QuiverRed}. We indeed see that it is the theory we proposed up to a different choice of flipping fields.
  
\

\subsection*{More symmetry enhancement and duality}

The way we obtained the $Spin(6)$ theory with enhanced symmetry can be generalized in various ways.
Instead of starting with the $Spin(7)\times E_7$ conformal matter in six dimensions we can start from closely related models which have the symmetry $E_7\times SU(2)\times SU(2)\times SU(2)$. These are obtained by taking a collection of M$5$-branes near the end of the world M$9$-brane and on a ${\mathbb C}/{\mathbb Z}_2$ singularity. This type of models also depends on the choice of action of the ${\mathbb Z}_2$ orbifold group on the $E_8$ symmetry of the M$9$-brane, where there are three distinct choices \cite{HMRV}. Here we concentrate on the choice preserving the $E_7\times SU(2)$ subgroup of $E_8$.

The resulting $6d$ SCFTs can be described on the tensor branch as a linear $SU(2)$ quiver gauge theory, connected by bifundamental hypermultiplets, with two fundamental hypermultiplets for both of the edge $SU(2)$ groups. One of the edge $SU(2)$ groups is further connected to a rank $1$ E-string theory via gauging. We shall denote the number of $SU(2)$ groups in the quiver by $n-1$. Then the case of $n=1$ is the rank $1$ E-string theory, the case of $n=2$ is the $Spin(7)\times E_7$ conformal matter and the higher $n$ are the generalizations. For generic $n$ the theory has an $E_7$ global symmetry coming from the commutant of $SU(2)$ gauge in $E_8$ as well as an $SU(2)^3$ factor coming from the various fundamental and bifundamental rotation groups\footnote{Similarly to the $Spin(7)\times E_7$ conformal matter case, the relationship between the global symmetries of the $6d$ SCFT and the $6d$ gauge theory is subtle, see \cite{Mekareeya:2017jgc} for the details.}. The three $SU(2)$ groups are not symmetric, and we shall refer to one of them, roughly associated with rotations of the fundamental flavors at the edge of the quiver not connected to the rank $1$ E-string, as $SU(2)_E$. 

\begin{figure}
\center
\includegraphics[width=0.8\textwidth]{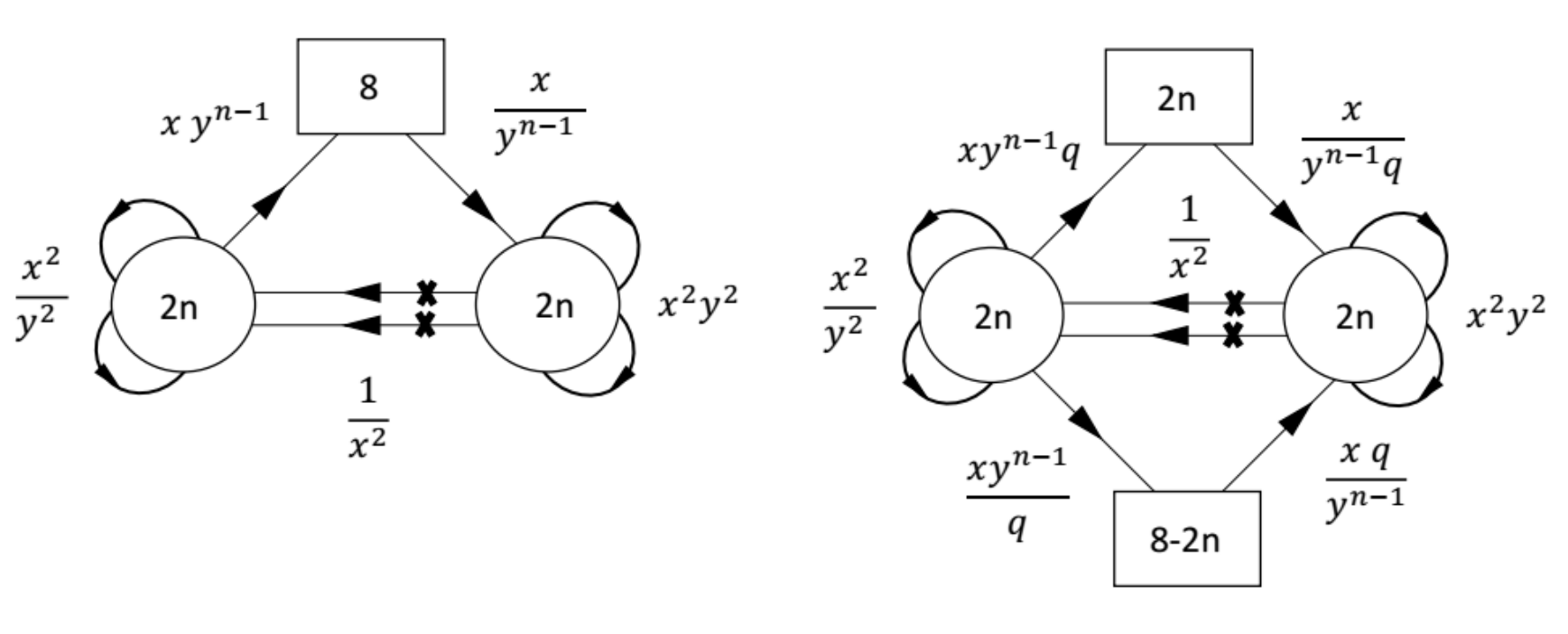} 
\caption{The $4d$ theory conjectured to be the result of the compactification of the class of theories considered here on a torus with flux $1$ in $SU(2)_E$. Here again there is a cubic superpotential along the triangle and a quartic one coupling the two antisymmetrics via the bifundamentals. There are also two singlet chiral fields flipping the baryons made from the bifundamentals. The visible global symmetry here is $SU(8)\times SU(2)\times U(1)_x \times U(1)_y$ where the $SU(2)$ rotates the two antisymmetrics. From $6d$, it is expected that $SU(8)\rightarrow E_7$ and $U(1)_y \rightarrow SU(2)$. There is also a $U(1)_R$ R-symmetry under which the bifundamentals have R-charge $0$, the flippers have R-charge $2$, and the rest have R-charge $1$.}
\label{MoreDuals}
\end{figure}

When reduced on a circle to $5d$, this class of theories has an affective $5d$ description as an $SU(2n)$ gauge theory with two antisymmetric hypermultiplets, $8$ fundamental hypermultiplets and Chern-Simons level $0$. This can be used to conjecture the $4d$ theories resulting from torus compactifications with fluxes of this class of $6d$ SCFTs \cite{ZafTA}. Specifically, the gauge theory description upon compactifying the models on a torus with flux breaking $SU(2)_E$ to $U(1)$ is shown in figure \ref{MoreDuals}. This is very similar to what we had above (see Figure 3) except the gauge groups are now $SU(2n)$.  In particular it is then natural to split the eight flavors into $8-2n$ and $2n$ and give vacuum expectation value to the baryons generalizing what we did before. This procedure will Higgs one of the two gauge groups and we will be left with an $SU(2n)$ gauge theory with $2n$ fundamentals, $8-2n$ antifundamentals, additional fields in the antisymmetric representation and singlet fields. Note that this procedure makes sense only for $n=1,2,3,4$. 
The $SU(8)$ symmetry we break by the procedure enhances to $E_7$ and thus the symmetry we expect is the commutant of the $U(1)$ in $E_7$ associated with the split, which is then in order of increasing $n$: $E_6$, $SO(12)$, $E_6$, and $E_7$.  The first case was discussed in \cite{Razamat:2017wsk}, the second is the case we discussed here, and finally the fourth case was discussed in \cite{Csaki:1997cu} in the context of self duality. The third case to the best of our knowledge was not considered till now.  

We will next consider each case in detail discussing the theories and the enhancement of symmetry. 
 In addition, we present a new self-duality associated with the $n=3$ case.  The models obtained from six dimensions have a particular set of gauge singlet fields. In the models below we alter this set in such a way so that the enhanced symmetry will remain but the conformal manifold will reduce to a point and thus the symmetry of the theory will  appear in the index computation through the contribution of the conserved currents.

\

\subsection*{$\boldsymbol{n=1}$}

\

We have an $SU(2)$ gauge theory with matter content as follows, 
\begin{center}
	\begin{tabular}{|c||c|c|c|c|c|}
		\hline
		Field & $SU(2)_g$ & $SU(6)_A$ & $SU(2)_B$ & $U(1)_h$ & $U(1)_{\hat r}$\\
		\hline
		$Q_A$ & $\boldsymbol{2}$ & $\boldsymbol{6}$ & $\boldsymbol{1}$ & 1 & $\frac{5}{9}$\\
		
		$Q_B$ & $\boldsymbol{2}$ & $\boldsymbol{1}$ & $\boldsymbol{2}$ & -3 & $\frac{1}{3}$\\
		
		$M$ & $\boldsymbol{1}$ & $\boldsymbol{\overline{15}}$ & $\boldsymbol{1}$ & -2 & $\frac{8}{9}$\\
		
		$\Phi$ & $\boldsymbol{1}$ & $\boldsymbol{1}$ & $\boldsymbol{1}$ & 6 & $\frac{4}{3}$\\
		\hline
	\end{tabular}
\end{center}
Here $U(1)_{\hat r}$ is the superconformal R charge and the superpotential is given by 
\begin{equation}
W=MQ_A^2+\Phi Q_B^2\,.
\end{equation}
This model was analyzed in \cite{Razamat:2017wsk}, where it was shown that the index at order $pq$ is given by 
\begin{equation*}
-\boldsymbol{20}_{SU(6)_A}\,\boldsymbol{2}_{SU(2)_B}-\boldsymbol{35}_{SU(6)_A}-\boldsymbol{3}_{SU(2)_B}-1
\end{equation*}
\begin{equation}
=-\boldsymbol{78}_{E_6}-1
\end{equation}
and that the UV symmetry $SU(6) \times SU(2) \times U(1)$ enhances to $E_6 \times U(1)$ in the IR. 

\

\subsection*{$\boldsymbol{n=2}$}

\

The model we obtain is the $SU(4)$ (or $Spin(6)$) gauge theory considered in section \ref{fourd} in table \ref{table:Spin6}, which we reproduce here for convenience, 
\begin{center}
	\begin{tabular}{|c||c|c|c|c|c|c|c|}
		\hline
		Field & $Spin\left(6\right)_{g}$ & $SU\left(2\right)_{v}$ & $SU\left(4\right)_{c}$ & $SU\left(4\right)_{s}$ & $U\left(1\right)_{b}$ & $U\left(1\right)_{a}$ & $U\left(1\right)_{r}$\\
		\hline 
		$S$ & $\boldsymbol{\overline{4}}$ & $\boldsymbol{1}$ & $\boldsymbol{1}$ & $\boldsymbol{4}$ & 1 & 1 & $\frac{1}{2}$\\
		
		$C$ & $\boldsymbol{4}$ & $\boldsymbol{1}$ & $\boldsymbol{4}$ & $\boldsymbol{1}$ & -1 & 1 & $\frac{1}{2}$\\		
		
		$V$ & $\boldsymbol{6}$ & $\boldsymbol{2}$ & $\boldsymbol{1}$ & $\boldsymbol{1}$ & 0 & -2 & 0\\
		
		$M_0$ & $\boldsymbol{1}$ & $\boldsymbol{1}$ & $\boldsymbol{\overline{4}}$ & $\boldsymbol{\overline{4}}$ & 0 & 2 & 1\\
		
		$M_1$ & $\boldsymbol{1}$ & $\boldsymbol{2}$ & $\boldsymbol{6}$ & $\boldsymbol{1}$ & 2 & 0 & 1\\
		
		$\Phi$ & $\boldsymbol{1}$ & $\boldsymbol{3}$ & $\boldsymbol{1}$ & $\boldsymbol{1}$ & 0 & 4 & 2\\
		\hline
	\end{tabular}
\end{center}
The superpotential is given by 
\begin{equation}
W=M_{0}CV^{2}S+M_{1}VC^{2}+\Phi V^{2}
\end{equation}
and the superconformal R charge by ${\hat r}=r-0.142q_{a}-0.057q_{b}$. As discussed in section \ref{fourd} and derived in the appendix, the index at order $pq$ is given by 
\begin{equation*}
-\boldsymbol{15}_{SU(4)_c}-\boldsymbol{15}_{SU(4)_s}-\boldsymbol{6}_{SU(4)_c}\boldsymbol{6}_{SU(4)_s}-\boldsymbol{3}_{SU(2)_v}-2
\end{equation*}
\begin{equation}
=-\boldsymbol{66}_{SO(12)}-\boldsymbol{3}_{SU(2)_v}-2
\end{equation}
and the UV symmetry $SU\left(2\right)\times SU\left(4\right)^{2}\times U\left(1\right)^{2}$ enhances to $SU\left(2\right)\times SO\left(12\right)\times U\left(1\right)^{2}$ in the IR. 

\

\subsection*{$\boldsymbol{n=3}$}

\

We have an $SU(6)$ gauge theory with matter content as follows, 
\begin{center}
	\begin{tabular}{|c||c|c|c|c|c|c|c|}
		\hline
		Field & $SU\left(6\right)_{g}$ & $SU\left(2\right)_{1}$ & $SU\left(2\right)_{2}$ & $SU\left(6\right)$ & $U\left(1\right)_{a}$ & $U\left(1\right)_{b}$ & $U\left(1\right)_{r}$ \\
		\hline 
		$A$ & $\boldsymbol{15}$ & $\boldsymbol{2}$ & $\boldsymbol{1}$ & $\boldsymbol{1}$ & -3 & -1 & 0\\
		
		$\bar{Q}$ & $\boldsymbol{\overline{6}}$ & $\boldsymbol{1}$ & $\boldsymbol{1}$ & $\boldsymbol{6}$ & 4 & 0 & $\frac{1}{2}$\\
		
		$F$ & $\boldsymbol{6}$ & $\boldsymbol{1}$ & $\boldsymbol{2}$ & $\boldsymbol{1}$ & 0 & 4 & $\frac{1}{2}$\\
		
		$M_{0}$ & $\boldsymbol{1}$ & $\boldsymbol{1}$ & $\boldsymbol{2}$ & $\boldsymbol{\overline{6}}$ & -4 & -4 & 1\\
		
		$M_{1}$ & $\boldsymbol{1}$ & $\boldsymbol{2}$ & $\boldsymbol{1}$ & $\boldsymbol{\overline{15}}$ & -5 & 1 & 1\\
		
		$\Phi$ & $\boldsymbol{1}$ & $\boldsymbol{4}$ & $\boldsymbol{1}$ & $\boldsymbol{1}$ & 9 & 3 & 2\\
		\hline
	\end{tabular}
	\label{tabN6}
\end{center}
The superpotential is given by 
\begin{equation}
W=M_{0}F\bar{Q}+M_{1}A\bar{Q}^{2}+\Phi A^{3}
\end{equation}
and the superconformal R charge by $\hat{r}=r-0.0175q_{a}-0.0481q_{b}$. The index at order $pq$ equals (see the appendix for more details)
\begin{equation*}
-\boldsymbol{3}_{SU(2)_1}-\boldsymbol{20}_{SU(6)}\,\boldsymbol{2}_{SU(2)_2}-\boldsymbol{35}_{SU(6)}-\boldsymbol{3}_{SU(2)_2}-2
\end{equation*}
\begin{equation}
=-\boldsymbol{3}_{SU(2)_1}-\boldsymbol{78}_{E_6}-2
\end{equation}
and the UV symmetry $SU(2)^2 \times SU(6) \times U(1)^2$ enhances to $SU(2) \times E_6 \times U(1)^2$ in the IR. 

\

\subsection*{$\boldsymbol{n=4}$}

\
This model was discussed in \cite{Csaki:1997cu}.
We have an $SU(8)$ gauge theory with matter content as follows, 
\begin{center}
	\begin{tabular}{|c||c|c|c|c|c|}
		\hline
		Field & $SU\left(8\right)_{g}$ & $SU\left(2\right)$ & $SU\left(8\right)$ & $U\left(1\right)_{a}$ & $U\left(1\right)_{r}$\\
		\hline 
		$A$ & $\boldsymbol{28}$ & $\boldsymbol{2}$ & $\boldsymbol{1}$ & 2 & 0\\
		
		$\bar{Q}$ & $\boldsymbol{\overline{8}}$ & $\boldsymbol{1}$ & $\boldsymbol{8}$ & -3 & $\frac{1}{2}$\\
		
		$M$ & $\boldsymbol{1}$ & $\boldsymbol{2}$ & $\boldsymbol{\overline{28}}$ & -4 & 1\\
		
		$\Phi$ & $\boldsymbol{1}$ & $\boldsymbol{5}$ & $\boldsymbol{1}$ & -8 & 2\\
		\hline
	\end{tabular}
\end{center}
The superpotential is given by 
\begin{equation}
W=MA^{5}\bar{Q}^{2}+\Phi A^{4}
\end{equation}
and the superconformal R charge by $\hat{r}=r+0.0693q_{a}$. The index at order $pq$ equals (see the appendix for more details)
\begin{equation*}
-\boldsymbol{70}_{SU(8)}-\boldsymbol{63}_{SU(8)}-\boldsymbol{3}_{SU(2)}-1
\end{equation*}
\begin{equation}
=-\boldsymbol{133}_{E_7}-\boldsymbol{3}_{SU(2)}-1
\end{equation}
and the UV symmetry $SU(8) \times SU(2) \times U(1)$ enhances to $E_7 \times SU(2) \times U(1)$ in the IR. 

\

\

\subsection*{$\boldsymbol{SU(6)}$ self-duality}

\

As discussed above, the symmetry enhancements in the $n=1,\,2,\,4$ theories are related to self-dualities, and we expect the same to be true in the case of the $n=3$ model. We find a self-duality as follows. We begin by decomposing the $SU(6)$ flavor symmetry to $SU(2)_3 \times SU(4) \times U(1)_c$ and write the first duality frame as the matter content of Table (\ref{tabN6}) without the gauge singlet fields and with no superpotential,
\begin{center}
	\begin{tabular}{|c||c|c|c|c|c|c|c|c|c|}
		\hline
		Field & $SU(6)_{g}$ & $SU(2)_{1}$ & $SU(2)_{2}$ & $SU(2)_{3}$ & $SU(4)$ & $U(1)_{c}$ & $U(1)_{a}$ & $U(1)_{b}$ & $U(1)_{r}$ \\
		\hline 
		$A$ & $\boldsymbol{15}$ & $\boldsymbol{2}$ & $\boldsymbol{1}$ & $\boldsymbol{1}$ & $\boldsymbol{1}$ & 0 & -3 & -1 & 0\\
		
		$\bar{Q}_{a}$ & $\boldsymbol{\overline{6}}$ & $\boldsymbol{1}$ & $\boldsymbol{1}$ & $\boldsymbol{2}$ & $\boldsymbol{1}$ & 2 & 4 & 0 & $\frac{1}{2}$\\
		
		$\bar{Q}_{b}$ & $\boldsymbol{\overline{6}}$ & $\boldsymbol{1}$ & $\boldsymbol{1}$ & $\boldsymbol{1}$ & $\boldsymbol{4}$ & -1 & 4 & 0 & $\frac{1}{2}$\\
		
		$F$ & $\boldsymbol{6}$ & $\boldsymbol{1}$ & $\boldsymbol{2}$ & $\boldsymbol{1}$ & $\boldsymbol{1}$ & 0 & 0 & 4 & $\frac{1}{2}$\\
		\hline
	\end{tabular}
\end{center}
Then, the second duality frame is given by the following matter content, 
\begin{center}
	\begin{tabular}{|c||c|c|c|c|c|c|c|c|c|}
		\hline
		Field & $SU(6)_{g}$ & $SU(2)_{1}$ & $SU(2)_{2}$ & $SU(2)_{3}$ & $SU(4)$ & $U(1)_{c}$ & $U(1)_{a}$ & $U(1)_{b}$ & $U(1)_{r}$ \\
		\hline 
		$a$ & $\boldsymbol{\overline{15}}$ & $\boldsymbol{2}$ & $\boldsymbol{1}$ & $\boldsymbol{1}$ & $\boldsymbol{1}$ & 0 & -3 & -1 & 0\\
		
		$\bar{q}_{a}$ & $\boldsymbol{\overline{6}}$ & $\boldsymbol{1}$ & $\boldsymbol{1}$ & $\boldsymbol{2}$ & $\boldsymbol{1}$ & 0 & 0 & 4 & $\frac{1}{2}$\\
		
		$\bar{q}_{b}$ & $\boldsymbol{6}$ & $\boldsymbol{1}$ & $\boldsymbol{1}$ & $\boldsymbol{1}$ & $\boldsymbol{\overline{4}}$ & -1 & 4 & 0 & $\frac{1}{2}$\\
		
		$f$ & $\boldsymbol{6}$ & $\boldsymbol{1}$ & $\boldsymbol{2}$ & $\boldsymbol{1}$ & $\boldsymbol{1}$ & 2 & 4 & 0 & $\frac{1}{2}$\\
		
		$m_{0}$ & $\boldsymbol{1}$ & $\boldsymbol{1}$ & $\boldsymbol{1}$ & $\boldsymbol{2}$ & $\boldsymbol{4}$ & 1 & -4 & -4 & 1\\
		
		$m_{1}$ & $\boldsymbol{1}$ & $\boldsymbol{2}$ & $\boldsymbol{2}$ & $\boldsymbol{1}$ & $\boldsymbol{4}$ & -1 & -5 & 1 & 1\\
		
		$m_{2}$ & $\boldsymbol{1}$ & $\boldsymbol{1}$ & $\boldsymbol{2}$ & $\boldsymbol{1}$ & $\boldsymbol{4}$ & -1 & 4 & 4 & 1\\
		
		$m_{3}$ & $\boldsymbol{1}$ & $\boldsymbol{2}$ & $\boldsymbol{1}$ & $\boldsymbol{2}$ & $\boldsymbol{4}$ & 1 & 5 & -1 & 1\\
		\hline
	\end{tabular}
\end{center}
along with the superpotential 
\begin{equation}
W=m_{0}\bar{q}_{a}\bar{q}_{b}+m_{1}a\bar{q}_{b}f+m_{2}a^{4}\bar{q}_{b}f+m_{3}a^{3}\bar{q}_{a}\bar{q}_{b}.
\end{equation}

Note that the representations of the matter content of the two dual theories under the gauge groups are different. The 't Hooft anomalies of the two models match and the indices agree.

\

\section{Ring relations and hidden symmetries}

So far we have discussed an interplay between symmetry enhancement and duality. There is another interesting field theoretical effect related to enhancement of symmetry. Supersymmetric theories in general  have chiral operators forming an algebraic  ring. This ring is specified by generators and relations, see \cite{Kutasov:1995ss} for a nice exposition. In this section we explore some interconnections between the ring relations and enhancements of symmetry. In particular we will show that a theory with chiral ring relations of a certain type entails a theory with an enhanced symmetry. We will also make an experimental observation that the representations under the global symmetry of the chiral ring relations of a certain type seem to always complement the adjoint representation of the symmetry to the adjoint representation of a bigger group, which is not in general a symmetry group of the model. 
In more detail, we will discuss chiral ring relations in theories with at least four supercharges in four  dimensions appearing at R charge $\leq 2$. By studying many examples we observe that the representations of the relations under the global symmetry group $G$ extends the adjoint representation of $G$ to an adjoint representation of a bigger group $\overline G$. We will give examples related to the $Spin(n+4)$ sequence of models we discussed here and also discuss some different cases.

\

\subsection*{The basic setup for the claims}  \label{model}

We examine a model, which we will denote ${\cal T}_{\cal O}$, with chiral relevant operators  ${\cal O}_i$ such that ${\cal O}_i{\cal O}_j$ are marginal, where $i$ and $j$ can be either equal or different.  We assume that the model has at least ${\cal N}=1$ supersymmetry, an R symmetry $U(1)_r$, and a global non R symmetry $G$ under which the operators ${\cal O}_i$ are in the representations ${\mathfrak R}_i$. 
Now, these operators can satisfy chiral ring relations,
\begin{equation}
\label{relgen}
\left.{\cal O}_i{\cal O}_j\right|_{{\mathfrak r}_{ij}} \sim 0\,,
\end{equation}
where ${\mathfrak r}_{ij}$ are some representations of $G$ which appear in the decomposition of
 the products ${{\mathfrak R}_i}\times{{\mathfrak R}_j}$. In general, the set of representations ${\mathfrak r}_{ij}$ can be either non trivial or  empty. We will show next that
\begin{equation}
\label{sum}
\sum_{ij} {\mathfrak r}_{ij}+\textrm{Adj}(G)
\end{equation}
forms in many examples the adjoint representation of some group that we denote by $\overline G$. In addition, in the cases where $i\neq j$ and ${\cal O}_i$ and ${\cal O}_j$ are oppositely charged under all the abelian factors of $G$, this observation can be proven, under certain assumptions, and the group $\overline G$ turns out to serve as the global symmetry of another model  related to ${\cal T}_{\cal O}$ by the addition of some gauge singlets and a superpotential. Note that the sum $\sum_{ij} {\mathfrak r}_{ij}$ might be in the adjoint representation of some group by itself or it can extend $\textrm{Adj}(G)$ to the adjoint representation of a bigger group for which $G$ is a non trivial subgroup.

\

We first discuss the case mentioned above for which we have a proved argument, and then continue with nontrivial examples that illustrate the observations here. 

\

\subsection*{Mapping relations into conserved currents}  \label{mapprel}

 We begin by considering models in which all the marginal operators are of the form ${\cal O}_i^{(1)}{\cal O}_i^{(2)}$, where ${\cal O}_i^{(1)}$ and ${\cal O}_i^{(2)}$ are different relevant operators oppositely charged under all the abelian factors of $G$.
Some of the assumptions can be relaxed as we will discuss later. For each such model we define a new one in which there are no marginal operators and in which the relations \eqref{relgen} combine in the IR with the conserved currents of ${\cal T}_{\cal O}$ to form conserved currents of a larger symmetry group. In other words, in the new model the symmetry group $G$ enhances in the IR to $\overline G$. To see this, we define the new model ${\cal T}_F$ by adding to ${\cal T}_{\cal O}$ the superpotential  
\begin{equation}
W=\sum_{i}F_i{\cal O}_i^{(2)}\,,
\end{equation}
where $F_i$ are  gauge singlet fields we add to the model. This superpotential sets ${\cal O}_i^{(2)}$ to zero in the chiral ring and as a result there are no marginal operators in ${\cal T}_F$. Note that since the operators $F_i$ are charged under the abelian symmetries, the superconformal symmetry of ${\cal T}_F$ is not the same as of ${\cal T}_{\cal O}$; nonetheless, we assume that all the operators are above the unitarity bound. 

We assume that in ${\cal T}_{\cal O}$,

\be
\left.{\cal O}^{(1)}_i{\cal O}^{(2)}_i\right|_{{\frak r}_i}\sim 0\,.
\ee
Next, in order to find the IR symmetry of ${\cal T}_F$, we analyze the supersymmetric index of this model at order $pq$ and deduce the spectrum of marginal operators and the conserved currents. The various contributions are as follows. Each operator ${\cal O}_i^{(1)}{\cal O}_i^{(2)}$ contributes $\chi({{\mathfrak R}_i^{(1)}}\times{{\mathfrak R}_i^{(2)}})-\chi({\mathfrak r}_i)$ and each $F_i{\cal O}_i^{(2)}$ contributes $\chi(\overline{{\mathfrak R}_i^{(2)}}\times{{\mathfrak R}_i^{(2)}})$. The operators ${\cal O}_i^{(1)} \overline\psi_{F_i}$ and $F_i\overline\psi_{F_i}$ contribute $-\chi({{\mathfrak R}_i^{(1)}}\times{{\mathfrak R}_i^{(2)}})$ and $-\chi(\overline{{\mathfrak R}_i^{(2)}}\times{{\mathfrak R}_i^{(2)}})$, respectively. In addition, we have the contribution of the UV conserved currents $-\chi(\textrm{Adj}(G))$, and so in total we obtain
\begin{equation}
-\sum_{i}\chi\left({\mathfrak r}_i\right)-\chi\left(\textrm{Adj}\left(G\right)\right)\,.
\end{equation}
We see, as stated above, that in the model ${\cal T}_F$ the dimension of the conformal manifold vanishes and that there is an enhancement of the symmetry from $G$ in the UV to a larger group $\overline G$ in the IR (if ${\mathfrak r}_i$ are not empty). In particular, we see that in the model ${\cal T}_{\cal O}$ the sum of the representations of the relations extends the adjoint representation of the global symmetry $G$ to the adjoint representation of some other, larger group.

We can now apply the same argument for theories that contain in addition to the marginal operators which are of the form ${\cal O}_i^{(1)}{\cal O}_i^{(2)}$ other, more general marginal operators, as long as the sum of their representations do not include the sum $\sum_{i}{\mathfrak r}_i$. Alternatively, the argument can be made for each relation corresponding to an operator of the form ${\cal O}_i^{(1)}{\cal O}_i^{(2)}$ as long as this representation does not appear among the representations of the other marginal operators. 

We finally comment that the argument is also applicable to theories in which ${\cal O}_i^{(1)}$ and ${\cal O}_i^{(2)}$ are not oppositely charged under all the abelian factors of $G$ (such that ${\cal O}_i^{(1)}{\cal O}_i^{(2)}$ is charged) as long as these $U(1)$s do not mix in the expression for the superconformal R charge and that it remains this way after the flipping (e.g. if the operators that are flipped are uncharged under these $U(1)$s).

\

We now turn to analyze examples that illustrate our claim.  

\

\subsection*{Examples}

\

\subsubsection*{$\boldsymbol{SU(2)}$ with eight fundamentals}  \label{su2}

We consider an $SU(2)$ gauge theory with eight chiral fields $Q_i$ in the fundamental representation, that is $SU(2)$ $N_f=4$ SQCD. In the infrared, this theory flows to a superconformal field theory in which the gauge invariant composite operator $Q_i Q_j$ (where the gauge indices are suppressed and contracted with $\varepsilon^{\alpha\beta}$) has an R-charge which equals to 1 and all the marginal operators are of the form $(Q_i Q_j)(Q_l Q_n)$. Moreover, these marginal operators satisfy a relation and therefore this theory serves as a nontrivial example and we denote the operator $Q_i Q_j$ by $\cal O$. 

Let us analyze explicitly the relation satisfied by ${\cal O}^2$ and the relevant representations that appear in the model. The global symmetry group is $G=SU(8)$, under which the chiral fields $Q_i$ transform in the fundamental representation $\bf 8$ and the operators ${\cal O}=Q_i Q_j$ transform in the representation $\bf 28$. Due to the compositeness of ${\cal O}$, we have the following relation, 
\begin{equation}
\label{relsu2}
\left.{\cal O}^2\right|_{{\mathfrak r}={\bf 70}} \sim 0\,,
\end{equation}
and the marginal operators ${\cal O}^2$ transform in the irreducible representation $\bf 336$ of $SU(8)$. We see that the relation \eqref{relsu2} is not in the adjoint representation of some group, hence according to our discussion it should extend the adjoint representation of the global symmetry group $\textrm{Adj}\left(G\right)=\bf 63$ to the adjoint representation of some other, bigger group. Indeed, adding the representation of the relation, we get 
\begin{equation}
\textrm{Adj}\left(G\right)+{\mathfrak r}={\bf 63}_{SU(8)}+{\bf 70}_{SU(8)} ={\bf 133}_{E_7}=\textrm{Adj}\left(E_7\right)=\textrm{Adj}\left({\overline G}\right).
\end{equation}
We obtained the bigger group $\overline G=E_7$, for which $SU(8)$ is a non trivial subgroup.

\

\subsubsection*{$\boldsymbol{SU(N_c)}$ with $\boldsymbol{2N_c}$ flavors, $\boldsymbol{N_c>2}$}

In contrast to the previous case ({\it i.e.} where $N_c=2$), now we cannot treat the baryons and mesons on an equal footing. Since the IR R-charge of the quarks $Q_i$ and anti-quarks $\tilde{Q}_{\tilde{j}}$ ($i,{\tilde{j}}=1,\ldots,2N_c$) is equal to $1/2$, now only the mesons $Q_i \tilde{Q}_{\tilde{j}}$ have R-charge 1 and correspond to the relevant operators ${\cal O}_{i{\tilde{j}}}$ that build the marginal ones (the baryons $(Q_i)^{N_c}$ and anti-baryons $(\tilde{Q}_{\tilde{j}})^{N_c}$ have R-charge $>1$). These mesons ${\cal O}_{i{\tilde{j}}}$ transform in the representation $({\bf 2N_c},{\bf 2N_c})_0$ of the flavor symmetry $SU(2N_c)\times\widetilde{SU}(2N_c)\times U(1)$, and the marginal operators ${\cal O}^2$ are in the representation 
\begin{equation}
\label{repsu2n}
(Sym^2({\bf 2N_c}),Sym^2({\bf 2N_c}))_0+(ASym^2({\bf 2N_c}),ASym^2({\bf 2N_c}))_0.
\end{equation}
None of these operators satisfy a relation since $N_c>2$ and so all the representations in \eqref{repsu2n} are present. Therefore, ${\mathfrak r}$ is the empty set and we have $\overline G=G$. Note that in the case $N_c=2$ the second representation in \eqref{repsu2n} (which is $(\bf 6,\bf 6)_0$) is "missing" since it corresponds to the following product of (gauge singlet) operators, $(Q_i)^2 (\tilde{Q}_{\tilde{j}})^2$, in contrast to the fact that the gauge indices are contracted between $Q_i$ and $\tilde{Q}_{\tilde{j}}$ in each ${\cal O}_{i{\tilde{j}}}$. 

\

\subsubsection*{$\boldsymbol{E_7}$ surprise}

In this case, discussed in detail in \cite{Dimofte:2012pd}, two copies of the $SU(2)$ gauge theory with four flavors are taken such that the chiral fields $Q_i^{(1)}$ of the first copy transform in the fundamental representation $\bf 8$ of its global symmetry group $SU(8)_1$, while those of the second copy $Q^{(2)i}$ are in the anti-fundamental representation $\bf\overline 8$ of its group $SU(8)_2$. This product theory is then deformed by the exactly marginal operator 
\begin{equation}
\label{deform}
W=\lambda \sum_{i,j} {\cal O}^{(1)}\cdot\, {\cal O}^{(2)},
\end{equation}
where  ${\cal O}^{(1)}=Q^{(1)}_i Q^{(1)}_j$, and analogously for ${\cal O}^{(2)}$.  The superpotential identified the two $SU(8)$ symmetries, that is preserves a diagonal combination of the two symmetries.
The flavor symmetry of the deformed theory is $SU(8)_d$, the diagonal part of $SU(8)_1 \times SU(8)_2$. The exactly marginal operator in \eqref{deform} that preserves this $SU(8)_d$ is unique and therefore $\lambda$ parametrizes a line of fixed points with $SU(8)_d$ flavor symmetry. As was shown in \cite{Dimofte:2012pd}, there is a special point on this line of fixed points at which the flavor symmetry enhances to $E_7$. (In fact, it enhances to $E_7\times U(1)$ as discussed in \cite{Razamat:2017hda}.) We would like to apply our general argument to two kinds of points on this line, one is the special point with the enhanced $E_7$ symmetry and the other is a generic point where the symmetry is $SU(8)_d$.

We begin by considering the marginal operators and the corresponding relations at a generic point on the line of fixed points. The operators ${\cal O}^{(1)}$ and ${\cal O}^{(2)}$ are in the representations $\bf 28$ and $\bf\overline{28}$ of $SU(8)_d$, respectively, while the marginal operators of the forms $({\cal O}^{(1)})^2$ and $({\cal O}^{(2)})^2$ transform in the representations $\bf 336$ and $\bf\overline{336}$, and satisfy the relations 
\begin{equation}
\left.\left({\cal O}^{(k)}\right)^2\right|_{{\mathfrak r}_k} \sim 0\,,
\end{equation}
where $k=1,2$ and ${\mathfrak r}_1={\mathfrak r}_2={\bf 70}$. In addition, we have 721 marginal operators of the form ${\cal O}^{(1)} {\cal O}^{(2)}$, where 63 out of the product ${\bf 28}\times{\bf\overline{28}}={\bf 1}+{\bf 63}+{\bf 720}$ combined with the off-diagonal currents of $SU(8)_1 \times SU(8)_2$ that were broken by the deformation \eqref{deform}. We therefore have
\begin{equation}
\left.{\cal O}^{(1)} {\cal O}^{(2)}\right|_{{\mathfrak r}_3={\bf 63}} \sim 0
\end{equation}
in the chiral ring. Note that while the relations ${\mathfrak r}_1$ and ${\mathfrak r}_2$ are kinematical, ${\mathfrak r}_3$ is dynamical. Now, since ${\mathfrak r}_1+{\mathfrak r}_2+{\mathfrak r}_3$ does not form adjoint representation of some group, we expect it to extend the adjoint representation of the flavor symmetry $SU(8)_d$ to the adjoint representation of some other, bigger group. We get 
\begin{equation}
\label{sumrepe7}
\textrm{Adj}\left(SU(8)_d\right)+{\mathfrak r}_1+{\mathfrak r}_2+{\mathfrak r}_3={\bf 133}_{E_7}+{\bf 133}_{E_7}=\textrm{Adj}\left({\overline G}\right),
\end{equation}
where $\overline G= E_7\times E_7$. 

Next, we consider the special point on the line of fixed points where the flavor symmetry $SU(8)$ enhances to $E_7$. At this point, there are 70 new conserved flavor symmetry currents that come with 70 new marginal operators. Therefore, there are fewer relations and the sum of their representations is 
\begin{equation}
\sum {\mathfrak r} = {\bf 133}_{E_7}=\textrm{Adj}\left(E_7\right).
\end{equation}
We see that in this case the relations form an adjoint representation by themselves. The sum of  the representations of the relations and the adjoint of the symmetry group $E_7$ yields of course the same result as in \eqref{sumrepe7}. 

\

\subsubsection*{$\boldsymbol{Spin(n+4)}$ and conformal matter sequence }

Most of the theories in the sequences considered in sections \ref{fourd} and \ref{secsuen} are examples of ${\cal T}_{F}$. For each such theory, one can easily find the related model ${\cal T}_{\cal O}$ in which there are marginal operators and relations. These relations, in turn, extend the adjoint representation of the symmetry group of ${\cal T}_{\cal O}$ to the adjoint of the larger group that serves as the IR symmetry in the model ${\cal T}_F$.
We now turn to analyze two ${\cal T}_{\cal O}$ models related to two theories from the $Spin(n+4)$ sequence, such that these models do not satisfy all the assumptions of the proof (in particular, some of the relations involve operators that are charged under $U(1)$ factors of the symmetry group) we gave but we still observe that the relations combine with the conserved currents to form the adjoint representation of a bigger group. 

\

\subsubsection*{$\boldsymbol{SU(4)}$ model}

We look at an $SU(4)$ gauge theory with 2 antisymmetric tensors, 4 flavors and a gauge singlet field. Explicitly, the matter content is given as follows, 
\begin{center}
	\begin{tabular}{|c||c|c|c|c|c|c|c|}
		\hline
		Field & $SU(4)$ & $SU(4)_L$ & $SU(4)_R$ & $SU(2)$ & $U(1)_a$ & $U(1)_b$ & $U(1)_r$\\
		\hline 
		$Q$ & $\bf{4}$ & $\bf{4}$ & $\bf{1}$ & $\bf{1}$ & -1 & 1 & $\frac{1}{2}$\\
		
		$\widetilde Q$ & $\bf{\overline{4}}$ & $\bf{1}$ & $\bf{4}$ & $\bf{1}$ & -1 & -1 & $\frac{1}{2}$\\		
		
		$X$ & $\bf{6}$ & $\bf{1}$ & $\bf{1}$ & $\bf{2}$ & 2 & 0 & 0\\
		
		$\Phi$ & $\bf{1}$ & $\bf{1}$ & $\bf{1}$ & $\bf{3}$ & -4 & 0 & 2\\
		\hline
	\end{tabular}
\end{center}
The superpotential is given by
\begin{equation}
W=\Phi X^{2}\,,
\end{equation}
the superconformal R-charge by $\hat{r}=r+0.12845 q_{a}$ and all the operators are above the unitarity bound. 

We begin as usual by identifying the relevant operators that constitute marginal ones. They are listed in the following table,
\begin{center}
	\begin{tabular}{|c||c|c|c|c|c|c|c|}
		\hline
		Operator & $SU(4)$ & $SU(4)_L$ & $SU(4)_R$ & $SU(2)$ & $U(1)_a$ & $U(1)_b$ & $U(1)_r$\\
		\hline 
		${\cal O}_1=Q\widetilde Q$ & $\bf{1}$ & $\bf{4}$ & $\bf{4}$ & $\bf{1}$ & -2 & 0 & 1\\
		
		${\cal O}_2=X^2Q\widetilde Q$ & $\bf{1}$ & $\bf{4}$ & $\bf{4}$ & $\bf{1}$ & 2 & 0 & 1\\		
		
		${\cal O}_3=XQ^2$ & $\bf{1}$ & $\bf{6}$ & $\bf{1}$ & $\bf{2}$ & 0 & 2 & 1\\
		
		${\cal O}_4=X{\widetilde Q}^2$ & $\bf{1}$ & $\bf{1}$ & $\bf{6}$ & $\bf{2}$ & 0 & -2 & 1\\
		\hline
	\end{tabular}
\end{center}
The marginal operators are of the forms ${\cal O}_1 {\cal O}_2$, $({\cal O}_3)^2$, $({\cal O}_4)^2$ and ${\cal O}_3 {\cal O}_4$, but there are also relations. First, by examining the representations of the operators of the forms ${\cal O}_1 {\cal O}_2$ and ${\cal O}_3 {\cal O}_4$ it turns out that all the operators corresponding to the various representations of ${\cal O}_1 {\cal O}_2$ obtained by naively multiplying those of ${\cal O}_1$ and ${\cal O}_2$ are present, but this is not the case for ${\cal O}_3 {\cal O}_4$. Explicitly, If we denote the representations under the global symmetries by 
\begin{equation}
\left(SU\left(4\right)_{L},SU\left(4\right)_{R},SU\left(2\right)\right)_{U\left(1\right)_{a},U\left(1\right)_{b}}\,,
\end{equation}
then ${\cal O}_1$ and ${\cal O}_2$ are in the representations $\left(\bf{4},\bf{4},\bf{1}\right)_{-2,0}$ and $\left(\bf{4},\bf{4},\bf{1}\right)_{2,0}$, respectively, and by multiplying them we get the representations $\left(\bf{6},\bf{6},\bf{1}\right)_{0,0}$, $\left(\bf{10},\bf{6},\bf{1}\right)_{0,0}$, $\left(\bf{6},\bf{10},\bf{1}\right)_{0,0}$ and $\left(\bf{10},\bf{10},\bf{1}\right)_{0,0}$ that all correspond to different operators of the form ${\cal O}_1 {\cal O}_2$ (in which $X^2$ is in the adjoint representation of the gauge group). In the case of ${\cal O}_3 {\cal O}_4$, however, we have the operators ${\cal O}_3$ and ${\cal O}_4$ in the representations $\left(\bf{6},\bf{1},\bf{2}\right)_{0,2}$ and $\left(\bf{1},\bf{6},\bf{2}\right)_{0,-2}$, respectively, and by naively multiplying them we get the representations $\left(\bf{6},\bf{6},\bf{1}\right)_{0,0}$ and $\left(\bf{6},\bf{6},\bf{3}\right)_{0,0}$. Now, as one can deduce from analyzing the possible representations of operators of the form $X^{2}Q^{2}\widetilde{Q}^{2}$, the representation $\left(\bf{6},\bf{6},\bf{1}\right)_{0,0}$ is missing from ${\cal O}_3 {\cal O}_4$. Therefore, the only representation that we get is $\left(\bf{6},\bf{6},\bf{3}\right)_{0,0}$ (in which $X^2$ is in the representation $\bf{20'}$ of the gauge group) and we have the kinematical relation 
\begin{equation}
\left.{\cal O}_{3}{\cal O}_{4}\right|_{{\mathfrak r}_1=({\bf 6},{\bf 6},{\bf 1})_{0,0}} \sim 0\,.
\end{equation}
Second, the marginal operators $({\cal O}_3)^2$ and $({\cal O}_4)^2$ corresponding to the representations $({\bf 1},{\bf 1},{\bf 3})_{0,4}$ and $({\bf 1},{\bf 1},{\bf 3})_{0,-4}$, respectively, contain $X^2$ as a singlet of the gauge group, in contrast to the fact that $X$ is contracted with $Q^2$ in ${\cal O}_3$ and with ${\widetilde Q}^2$ in ${\cal O}_4$. Therefore, these representations are missing from $({\cal O}_3)^2$ and $({\cal O}_4)^2$ and we have the kinematical relations
\begin{equation}
\left.({\cal O}_3)^2\right|_{{\mathfrak r}_2=({\bf 1},{\bf 1},{\bf 3})_{0,4}} \sim 0
\end{equation}
and
\begin{equation}
\left.({\cal O}_4)^2\right|_{{\mathfrak r}_3=({\bf 1},{\bf 1},{\bf 3})_{0,-4}} \sim 0\,.
\end{equation}
Now, since ${\mathfrak r}_1+{\mathfrak r}_2+{\mathfrak r}_3$ does not form the adjoint representation of some group, we expect it to extend the adjoint representation of the global symmetry $G$ to the adjoint representation of some other, bigger group. We have 
\begin{equation}
\textrm{Adj}\left(G\right)=({\bf 15},{\bf 1},{\bf 1})_{0,0}+({\bf 1},{\bf 15},{\bf 1})_{0,0}+({\bf 1},{\bf 1},{\bf 3})_{0,0}+2\,,
\end{equation}
\begin{equation}
{\mathfrak r}_1+{\mathfrak r}_2+{\mathfrak r}_3=({\bf 6},{\bf 6},{\bf 1})_{0,0}+({\bf 1},{\bf 1},{\bf 3})_{0,4}+({\bf 1},{\bf 1},{\bf 3})_{0,-4}
\end{equation}
and so
\begin{equation}
\textrm{Adj}\left(G\right)+{\mathfrak r}_1+{\mathfrak r}_2+{\mathfrak r}_3={\bf 66}_{SO (12)}+{\bf 10}_{SO (5)}+1=\textrm{Adj}\left({\overline G}\right),
\end{equation}
where $\overline G = SO(12)\times SO(5)\times U(1)$.

\

\subsubsection*{$\boldsymbol{Spin(8)}$ model}

We consider a $Spin(8)$ gauge theory with matter content as follows, 
\begin{center}
	\begin{tabular}{|c||c|c|c|c|c|c|c|}
		\hline
		Field  & $Spin\left(8\right)_{g}$ & $SU\left(2\right)_{s}$ & $SU\left(2\right)_{c}$ & $SU\left(4\right)_{v}$ & $U\left(1\right)_{a}$ & $U\left(1\right)_{b}$ & $U\left(1\right)_{r}$\\
		\hline 
		$S$ & $\boldsymbol{8}_{s}$ & $\boldsymbol{2}$ & $\boldsymbol{1}$ & $\boldsymbol{1}$ & -2 & 0 & $\frac{1}{2}$\\
		
		$C$ & $\boldsymbol{8}_{c}$ & $\boldsymbol{1}$ & $\boldsymbol{2}$ & $\boldsymbol{1}$ & 0 & -2 & $\frac{1}{2}$\\
		
		$V$ & $\boldsymbol{8}_{v}$ & $\boldsymbol{1}$ & $\boldsymbol{1}$ & $\boldsymbol{4}$ & 1 & 1 & 0\\
		
		$\Phi_1$ & $\boldsymbol{1}$ & $\boldsymbol{3}$ & $\boldsymbol{1}$ & $\boldsymbol{1}$ & 4 & 0 & 1\\
		
		$\Phi_2$ & $\boldsymbol{1}$ & $\boldsymbol{1}$ & $\boldsymbol{3}$ & $\boldsymbol{1}$ & 0 & 4 & 1\\
		
		$\Phi_3$ & $\boldsymbol{1}$ & $\boldsymbol{1}$ & $\boldsymbol{1}$ & $\boldsymbol{\overline{10}}$ & -2 & -2 & 2\\
		\hline
	\end{tabular}
\end{center}
The superpotential is given by
\begin{equation}
W=\Phi_{1}S^{2}+\Phi_{2}C^{2}+\Phi_{3}V^{2}\,,
\end{equation}
all the operators are above the unitarity bound and the superconformal R-charge is ${\hat r}=r+0.107(q_a+q_b)$.

Denoting the representations under the global symmetries by 
\begin{equation}
\left(SU (2)_s,SU(2)_c,SU(4)_v\right)_{U\left(1\right)_{a},U\left(1\right)_{b}}\,,
\end{equation}
the relevant operators that constitute marginal ones are of the forms 
\begin{equation}
{\cal O}_1=S^2V^2 \,\, , \,\, {\cal O}_2=C^2V^2 \,\, , \,\, {\cal O}_3=SCV \,\, , \,\, {\cal O}_4=SCV^3
\end{equation}
and transform in the representations $\left({\bf 1},{\bf 1},{\bf 6}\right)_{-2,2}$, $\left({\bf 1},{\bf 1},{\bf 6}\right)_{2,-2}$, $\left({\bf 2},{\bf 2},{\bf 4}\right)_{-1,-1}$ and $\left({\bf 2},{\bf 2},{\bf \overline 4}\right)_{1,1}$, respectively. The marginal operators have the forms $({\cal O}_1)^2$, $({\cal O}_2)^2$, ${\cal O}_1 {\cal O}_2$ and ${\cal O}_3 {\cal O}_4$, and we find nontrivial relations when considering the  first two and the last one. Let us begin with the first two and look for example at $({\cal O}_1)^2$. Taking naively the second symmetric power of $\left({\bf 1},{\bf 1},{\bf 6}\right)_{-2,2}$, we get the representations $\left({\bf 1},{\bf 1},{\bf 1}\right)_{-4,4}$ and $\left({\bf 1},{\bf 1},{\bf 20'}\right)_{-4,4}$. However, the operator corresponding to the representation $\left({\bf 1},{\bf 1},{\bf 1}\right)_{-4,4}$ contains one $S^2$ as a singlet of the gauge group, in contrast to the fact that both of these two factors of $S^2$ are contracted with $V^2$ in the two copies of ${\cal O}_1$. Therefore, this representation is missing from $({\cal O}_1)^2$ and we have the kinematical relation
\begin{equation}
\left.({\cal O}_1)^2\right|_{{\mathfrak r}_1=\left({\bf 1},{\bf 1},{\bf 1}\right)_{-4,4}} \sim 0\,.
\end{equation}
Similarly, the operator corresponding to the representation $\left({\bf 1},{\bf 1},{\bf 1}\right)_{4,-4}$ of $({\cal O}_2)^2$ contains one of the $C^2$ as a singlet of the gauge group, and we obtain the kinematical relation
\begin{equation}
\left.({\cal O}_2)^2\right|_{{\mathfrak r}_2=\left({\bf 1},{\bf 1},{\bf 1}\right)_{4,-4}} \sim 0\,.
\end{equation}

Next, let us examine the relations associated with ${\cal O}_3 {\cal O}_4$. Multiplying naively the representations of ${\cal O}_3$ and ${\cal O}_4$, we get the representations $\left({\bf 3},{\bf 3},{\bf 1}\right)_{0,0}$ and $\left({\bf 1},{\bf 1},{\bf 15}\right)_{0,0}$ along with several others. However, $\left({\bf 3},{\bf 3},{\bf 1}\right)_{0,0}$ corresponds to either $S^2$ or $C^2$ being singlets of the gauge group and is therefore missing. In addition, also the representation $\left({\bf 1},{\bf 1},{\bf 15}\right)_{0,0}$ does not appear in the product ${\cal O}_3 {\cal O}_4$. As a result, we have the relation
\begin{equation}
\left.{\cal O}_3 {\cal O}_4\right|_{{\mathfrak r}_3=\left({\bf 3},{\bf 3},{\bf 1}\right)_{0,0}+\left({\bf 1},{\bf 1},{\bf 15}\right)_{0,0}} \sim 0\,.
\end{equation}

As before, since ${\mathfrak r}_1+{\mathfrak r}_2+{\mathfrak r}_3$ does not form the adjoint representation of some group, we expect it to extend the adjoint representation of the global symmetry $G$ to the adjoint representation of some other, bigger group. We have
\begin{equation*}
\textrm{Adj}\left(G\right)=({\bf 3},{\bf 1},{\bf 1})_{0,0}+({\bf 1},{\bf 3},{\bf 1})_{0,0}+({\bf 1},{\bf 1},{\bf 15})_{0,0}+2\,,
\end{equation*}
\begin{equation*}
{\mathfrak r}_1+{\mathfrak r}_2+{\mathfrak r}_3=\left({\bf 1},{\bf 1},{\bf 1}\right)_{-4,4}+\left({\bf 1},{\bf 1},{\bf 1}\right)_{4,-4}+\left({\bf 3},{\bf 3},{\bf 1}\right)_{0,0}+\left({\bf 1},{\bf 1},{\bf 15}\right)_{0,0}
\end{equation*}
and so 
\begin{equation}
\textrm{Adj}\left(G\right)+{\mathfrak r}_1+{\mathfrak r}_2+{\mathfrak r}_3={\bf 15}_{\widetilde{SU}(4)}+2\,{\bf 15}_{SU(4)_v}+{\bf 3}_{\widetilde{SU}(2)}+1=\textrm{Adj}\left({\overline G}\right),
\end{equation}
where $\overline G = SU(4)^3\times {\widetilde{SU}(2)}\times U(1)$.

\

\subsubsection*{Quiver example}

We now turn to examine a model with more than one gauge group. This theory, analyzed in \cite{Bah:2017gph}, is given by the following matter content, 
\begin{center}
	\begin{tabular}{|c||c|c|c|c|c|c|}
		\hline
		Field  & $SU\left(2\right)_{g_{1}}$ & $SU\left(2\right)_{g_{2}}$ & $\widetilde{SU}\left(2\right)$ & $U\left(1\right)_{a}$ & $U\left(1\right)_{b}$ & $U\left(1\right)_{r}$\\
		\hline 
		$Q_{1}$ & $\boldsymbol{2}$ & $\boldsymbol{2}$ & $\boldsymbol{1}$ & 1 & 1 & $\frac{1}{2}$\\
		
		$Q_{2}$ & $\boldsymbol{2}$ & $\boldsymbol{2}$ & $\boldsymbol{1}$ & -1 & 1 & $\frac{1}{2}$\\
		
		$\widetilde{Q}$ & $\boldsymbol{2}$ & $\boldsymbol{2}$ & $\boldsymbol{2}$ & 0 & -1 & $\frac{1}{2}$\\
		
		$\Phi_{1}$ & $\boldsymbol{1}$ & $\boldsymbol{1}$ & $\boldsymbol{1}$ & -2 & -2 & 1\\
		
		$\Phi_{2}$ & $\boldsymbol{1}$ & $\boldsymbol{1}$ & $\boldsymbol{1}$ & 2 & -2 & 1\\
		\hline
	\end{tabular}
\end{center}
The superpotential is given by
\begin{equation}
W=\widetilde{Q}^{2}Q_{1}Q_{2}+\Phi_{1}Q_{1}^{2}+\Phi_{2}Q_{2}^{2}\,,
\end{equation}
all the operators are above the unitarity bound and the superconformal R-charge is ${\hat r}=r+0.027q_{b}$. This model is Klebanov-Witten \cite{Klebanov:1998hh} with gauge singlets added.

In order to identify the relations in this case, we consider the superconformal index. As was discussed in \cite{Bah:2017gph}, in the IR the UV symmetry $\widetilde{SU}(2)\times U(1)_{a}\times U(1)_{b}$ is expected to enhance to $USp(4)\times U(1)_{b}$. The reason is that this model is engineered as a compactification on a torus with flux of class ${\cal S}_2$ with $N=2$, that is two M5 branes probing ${\mathbb Z}_2$ singularity. The flux is such that the $SO(7)$ symmetry of the six dimensional model is broken to $SO(5)\times U(1)$ and that is the reason we expect this symmetry. In $4d$ it appears to be further extended to $USp(4)\times U(1)_{b}$, which might be accidental from the $6d$ viewpoint. Note that the fact that the symmetry of the fixed point here is enhanced is  new for this example as compared to the other examples we have studied.
Then the index is computed in \cite{Bah:2017gph} and given by 
\begin{equation*}
I=1+\left(b^{2}+b^{-2}\boldsymbol{5}_{USp(4)}+\boldsymbol{4}_{USp(4)}\right)\left(pq\right)^{\frac{1}{2}}+b^{-2}\left(\boldsymbol{5}_{USp(4)}-1\right)\left(p^{\frac{1}{2}}q^{\frac{3}{2}}+p^{\frac{3}{2}}q^{\frac{1}{2}}\right)
\end{equation*}
\begin{equation}
+\left[b^{4}+\boldsymbol{10}_{SO(5)}+b^{-2}\,\boldsymbol{16}_{SO(5)}+b^{-4}\left(1+\boldsymbol{14}_{SO(5)}\right)\right]pq+\ldots\,.
\end{equation}
It is consistent with an enhancement of the symmetry to $USp(4)\times U(1)$ and from now on we will assume that this is the symmetry of the conformal theory.
Denoting the representations under the UV global symmetry $G$ by 
\begin{equation}
\widetilde{SU}\left(2\right)_{U\left(1\right)_{a},U\left(1\right)_{b}}\,,
\end{equation}
the relevant operators and their representations are as follows, 
\begin{center}
	\begin{tabular}{|c||c|c|c|c|c|c|}
		\hline
		$\mathrm{Operator}$ & $Q_{1}Q_{2}$ & $\Phi_{1}$ & $\Phi_{2}$ & $\widetilde{Q}^{2}$ & $Q_{1}\widetilde{Q}$ & $Q_{2}\widetilde{Q}$\\
		\hline 
		$\mathrm{Representation}$ & $\boldsymbol{1}_{0,2}$ & $\boldsymbol{1}_{-2,-2}$ & $\boldsymbol{1}_{2,-2}$ & $\boldsymbol{3}_{0,-2}$ & $\boldsymbol{2}_{1,0}$ & $\boldsymbol{2}_{-1,0}$\\
		\hline
	\end{tabular}
\end{center}
As can be seen in the expression for the index, the operators $\Phi_{1}$, $\Phi_{2}$ and $\widetilde{Q}^{2}$ form the $\boldsymbol{5}_{-2}$ of the IR symmetry $USp(4)_{U(1)_b}$, and $Q_{1}\widetilde{Q}$ and $Q_{2}\widetilde{Q}$ form the $\boldsymbol{4}_0$. At order $pq$ (with a vanishing $U(1)_{b}$ charge), we have the following contribution: 
\begin{equation}
\boldsymbol{3}_{\widetilde{SU}\left(2\right)}+\boldsymbol{3}_{\widetilde{SU}\left(2\right)}a^{2}+\boldsymbol{3}_{\widetilde{SU}\left(2\right)}a^{-2}+1=\boldsymbol{10}_{USp\left(4\right)}\,,
\end{equation}
and if we take into account the conserved currents $\boldsymbol{10}_{USp(4)}+1$ of the IR symmetry (and recall that the contribution to the index at order $pq$ comes from the marginal operators minus the conserved currents), we find that the contribution of the marginal operators is
\begin{equation}
\label{marquiv}
2\,\boldsymbol{10}_{USp(4)}+1\,.
\end{equation}
On the other hand, from the product of relevant operators $Q_{1}Q_{2}(\Phi_1+\Phi_2+{\widetilde Q}^2)$ we would expect to have a marginal operator in the representation ${\bf 5}_{USp(4)}$, which is not there. We deduce that there is a relation ${\frak r}'={\bf 5}_{USp(4)}$. Adding this relation to the IR currents, we observe that
\be
{\bf 10}_{USp(4)}+{\bf 1}_{USp(4)}+{\bf 5}_{USp(4)}={\bf 15}_{SU(4)}+{\bf 1}_{SU(4)}\,.
\ee 
We obtain that $\overline G=SU(4)\times U(1)$. Here the vector of $so(5)\sim usp(4)$ is embedded in $so(6)\sim su(4)$ as ${\bf 6}_{so(6)}={\bf 5}_{so(5)}+{\bf 1}_{so(5)}$.

Note that we can consider a relevant deformation of the theory turning on a superpotential with the operators charged as $b^{-4} qp$. This is a relevant deformation and it is a singlet of $USp(4)$ obtained from the symmetric square of $b^{-2} {\bf 5}_{USp(4)} (qp)^{\frac12}$. However, it is charged under $U(1)_b$ and therefore this symmetry is broken and we are left only with $USp(4)$. After this deformation, the relevant operators are in the representations
\be
\label{reldef}
{\bf 1}_{USp(4)}+{\bf 5}_{USp(4)}+{\bf 4}_{USp(4)}\,.
\ee 
All these operators have an R charge which equals to one and thus the symmetric square of \eqref{reldef} naively gives marginal operators, modulo chiral ring relations and recombinations with conserved currents. The symmetric square of \eqref{reldef} is
\be
2\;{\bf 1}_{USp(4)}+{\bf 5}_{USp(4)}+2\;{\bf 4}_{USp(4)}+{\bf 10}_{USp(4)}+{\bf 14}_{USp(4)}+{\bf 16}_{USp(4)}\,.
\ee 
Note that one of the singlets should be absent as a marginal operator since it is the superpotential deformation we turned on. That singlet is the one in the decomposition of the symmetric square of ${\bf 5}_{USp(4)}$, and it recombines with the broken current of the abelian symmetry $U(1)_b$. Thus, we have a (dynamical) relation in the singlet representation. We can now look at the marginal operators as they appear in the index and deduce that these are in the representations 
\be
2\;{\bf 1}_{USp(4)}+2\; {\bf 10}_{USp(4)}+{\bf 14}_{USp(4)}+{\bf 16}_{USp(4)}\,.
\ee 
We assume that there is no enhancement of symmetry in the fixed point and thus the current is in the representation ${\bf 10}_{USp(4)}$. From here we see that also the representation ${\bf 5}_{USp(4)}$ and twice the ${\bf 4}_{USp(4)}$ are missing. We deduce that the relations are,
\be
{\frak r} ={\bf 1}_{USp(4)}+2\;{\bf 4}_{USp(4)}+{\bf 5}_{USp(4)}\,.
\ee 
Combining these with the adjoint of $USp(4)$, we get the adjoint of $SU(5)$. The embedding is first $so(5)\sim usp(4)$ in $so(6)\sim su(4)$ as before and then $su(4)$ in $su(5)$ such that ${\bf 5}_{su(5)}={\bf 4}_{su(4)}+{\bf 1}_{su(4)}$. We conclude that $\overline G=SU(5)$.

\

\


\section*{Acknowledgments}
The authors are grateful to   Hee-Cheol Kim, Zohar Komargodski, and Cumrun Vafa for relevant discussions.  GZ is supported in part by World Premier International Research Center Initiative (WPI), MEXT, Japan. The research of SSR and OS  was also supported in part by the Israel Science Foundation under grant \#1696/15 and by I-CORE Program of the Planning and Budgeting Committee.  \vfill\eject


\appendix

\section{Finding the IR symmetries of the 4d $\boldsymbol{Spin(n+4)}$ sequence}
\label{A:AppTables}

In this appendix we present the calculation of the superconformal index at order $pq$ corresponding to the sequence of $Spin(n+4)$ gauge theories described in section \ref{fourd}. We then use it to extract the global symmetry in the IR of the models under consideration, as discussed below. After reviewing the definition of the index, we begin with a general description of the method and continue with its application to the $Spin(n+4)$ sequence of theories. 

\

\subsection*{The superconformal index}

\

The index of a 4d superconformal field theory is defined as the Witten index of the theory in radial quantization. Denoting by $\cal Q$ one of the Poincar\'e supercharges \footnote{More explicitly, we choose here $\cal Q\equiv \widetilde{\mathcal{Q}}_{\dot{-}}$ which in the language of \cite{Rastelli:2016tbz} corresponds to the "right-handed index".}, the index is defined as the following weighted trace over the states of the theory quantized on ${\mathbb S}^3\times {\mathbb R}$ \cite{Kinney:2005ej,Romelsberger:2005eg,Dolan:2008qi,Rastelli:2016tbz}:
\begin{equation}
\label{indapp}
{\cal I}\left(p,q;{u_a}\right)=\mathrm{Tr}_{{\mathbb S}^3}\left[(-1)^F e^{-\beta \delta} p^{j_1+j_2-\frac{r}{2}} q^{j_1-j_2-\frac{r}{2}} \prod_{a}u_{a}^{e_{a}} \right]\,.
\end{equation}
Here $j_1$ and $j_2$ are the Cartan generators of the $SU(2)_1\times SU(2)_2$ isometry group of ${\mathbb S}^3$, $r$ is the $U(1)_r$ charge, and $u_{a}$ and $e_{a}$ are the fugacities and charges of the global symmetries, respectively. Moreover, $\delta$ is defined as the following anti-commutator: 
\begin{equation}
\delta\equiv\left\{ \mathcal{Q},\mathcal{Q}^{\dagger}\right\} =E-2j_{2}-\frac{3}{2}r\,.
\end{equation}
Note that even though the chemical potential $\beta$ appears in the definition \eqref{indapp}, the index is in fact independent of it since the states with $\delta>0$ come in boson/fermion pairs and therefore cancel. Only the states with $\delta=0$, corresponding to short multiplets of the superconformal algebra, contribute.

The most salient feature of the index, which we exploit repeatedly, is its invariance under the RG flow. By calculating the index of an asymptotically free theory at the UV fixed point, we can extract information about the operator spectrum and the global symmetries of the theory at the IR fixed point. Below, we concentrate on calculating the coefficient that appears in the expansion of the index at order $pq$, where $p$ and $q$ are the superconformal fugacities appearing in \eqref{indapp}. From this coefficient we will be able to identify the enhanced symmetry of the theory in the IR, as explained below. 

\

\subsection*{The method}

\

To evaluate the index of a gauge theory we need to list all the operators one can build from the fields and then project on gauge invariants. This is often done by writing a matrix integral, see for example \cite{Dolan:2008qi, Aharony:2003sx}. The integrand corresponds to the generating function of all the operators and the integral projects on gauge invariant states. The theories we consider have a relatively high rank and a lot of matter making the evaluation of such integrals time consuming. Instead, we will just use representation theory to deduce the needed terms in the index computation.
We begin every calculation with listing the operators that contribute at order $pq$ to the index. In order to do that, we first note that for a chiral multiplet $\chi$ only the scalar component $\chi$ and the anti-spinor $\bar{\psi}^{\chi}$ contribute to the index \cite{Rastelli:2016tbz}: The scalar contributes $\left(pq\right)^{\frac{r}{2}}$ and the anti-spinor $-\left(pq\right)^{\frac{2-r}{2}}$ with $r$ being the R charge, along with contributions resulting from their charges and representations under the global symmetries. Note that if some $U(1)$ factors of the global symmetry mix with the R symmetry (such that the corresponding mixing coefficients are not some simple rational numbers), then the operators that we build out of the scalars and anti-fermions should have vanishing charges under these $U(1)$s so that their contribution to the index will indeed be at order $pq$. In addition, the only contribution from the vector multiplet that concerns us comes from the gaugino $\lambda$, and since it is not charged under the (non R) global symmetry and its R charge equals to 1, the corresponding contribution to the index at order $pq$ will always be 1 and will result from the operator $\lambda \lambda$.

Next, after listing the operators that contribute at order $pq$ to the index, we turn to find their representations under the global symmetry. It is important to note that in this list of operators only the forms of the operators are written ({\it i.e.} the contraction of gauge indices is not specified) and so it often happens that more than one (gauge invariant) operator corresponds to each form. We then write a table with all the operators that one can build corresponding to each form. After summing the characters of the representations of all these operators with the correct signs, we finally obtain the index at order $pq$. 

At this stage, in order to extract the IR symmetry of the model we use the following property of the index \cite{Beem:2012yn}: The contribution at order $pq$ comes from the marginal operators minus the conserved currents. In all the theories considered below except for the case of $Spin(10)$ (and in one of the models in the $Spin(8)$ case), we get only negative contributions at order $pq$ that correspond to the IR conserved currents and to the absence of marginal operators (so that the dimensions of the conformal manifolds vanish). In the cases of $Spin(10)$ and $Spin(8)$, we also find a positive contribution at order $pq$ and so we first need to identify the marginal operators before obtaining the conserved currents. 

It is important to note that in all the theories analyzed below, the superpotential is a relevant deformation. It is checked in each case by first considering the model without the superpotential and then coupling gauge singlet fields through new superpotential terms to the operators that violate the unitarity bound ({\it i.e.} we flip these operators). This way, we obtain a model in which all the operators satisfy the unitarity bound and the corresponding superconformal R charge is the correct one. The gauge singlet fields that do not appear in the new superpotential are of course free (and have IR R charge $2/3$). Then, it is easy to verify that the original superpotential terms (that are not in the new superpotential) are relevant operators in this model. We show it below explicitly for the $Spin(6)$ theory.

We now turn to the application of this procedure to the $Spin(n+4)$ sequence, starting from $n=2$ (the $n=1$ case was discussed in \cite{Razamat:2017hda}). In the simple cases (up to $n=4$) the results were confirmed by a calculation of the index in Mathematica. 

\

\subsection*{$\boldsymbol{Spin(6)}$}

\

We start by describing again the model discussed in section \ref{fourd}. The matter content is given in table \ref{table:Spin6}, which we reproduce here for convenience, 
\begin{center}
	\begin{tabular}{|c||c|c|c|c|c|c|c|}
		\hline
		Field & $Spin\left(6\right)_{g}$ & $SU\left(2\right)_{v}$ & $SU\left(4\right)_{c}$ & $SU\left(4\right)_{s}$ & $U\left(1\right)_{b}$ & $U\left(1\right)_{a}$ & $U\left(1\right)_{r}$\\
		\hline 
		$S$ & $\boldsymbol{\overline{4}}$ & $\boldsymbol{1}$ & $\boldsymbol{1}$ & $\boldsymbol{4}$ & 1 & 1 & $\frac{1}{2}$\\
		
		$C$ & $\boldsymbol{4}$ & $\boldsymbol{1}$ & $\boldsymbol{4}$ & $\boldsymbol{1}$ & -1 & 1 & $\frac{1}{2}$\\		
		
		$V$ & $\boldsymbol{6}$ & $\boldsymbol{2}$ & $\boldsymbol{1}$ & $\boldsymbol{1}$ & 0 & -2 & 0\\
		
		$M_0$ & $\boldsymbol{1}$ & $\boldsymbol{1}$ & $\boldsymbol{\overline{4}}$ & $\boldsymbol{\overline{4}}$ & 0 & 2 & 1\\
		
		$M_1$ & $\boldsymbol{1}$ & $\boldsymbol{2}$ & $\boldsymbol{6}$ & $\boldsymbol{1}$ & 2 & 0 & 1\\
		
		$\Phi$ & $\boldsymbol{1}$ & $\boldsymbol{3}$ & $\boldsymbol{1}$ & $\boldsymbol{1}$ & 0 & 4 & 2\\
		\hline
	\end{tabular}
\end{center}
The superpotential is given by 
\begin{equation}
W=M_{0}CV^{2}S+M_{1}VC^{2}+\Phi V^{2} \label{SupSpin6}
\end{equation}
and the superconformal R charge by ${\hat r}=r-0.142q_{a}-0.057q_{b}$. In order to check that (\ref{SupSpin6}) is a relevant deformation, we begin by considering the model without a superpotential. When doing this, one finds that the operator $V^2$ violates the unitarity bound; therefore, it should be flipped and we couple the field $\Phi$ to this operator by adding the superpotential $W=\Phi V^{2}$.\footnote{See \cite{Benvenuti:2017kud, Benvenuti:2017lle} for some recent applications of the flipping procedure.} Now all the operators satisfy the unitarity bound and the superconformal R charge in this model is given by ${\hat r}=r+0.128q_{a}$. The ${\hat r}$ charges of the operators $M_{0}CV^{2}S$ and $M_{1}VC^{2}$ are $1.92$ and $5/3$, respectively, and since they are smaller than 2, we find that these operators are relevant. 

We can now turn to calculate the index at order $pq$. We see that $U\left(1\right)_{b}$ and $U\left(1\right)_{a}$ mix with $U\left(1\right)_{r}$ in the expression for the IR R symmetry, and so, because the mixing is through irrational coefficients, the operators that contribute to the index at order $pq$ should have vanishing charges under $U\left(1\right)_{b}$ and $U\left(1\right)_{a}$. These operators are of the forms
\begin{equation*}
\lambda\lambda,\,\,\,\,\bar{\psi}^{V}V,\,\,\,\,\bar{\psi}^{C}C,\,\,\,\,\bar{\psi}^{S}S,\,\,\,\,\bar{\psi}^{M_{0}}M_{0},\,\,\,\,\bar{\psi}^{M_{1}}M_{1},\,\,\,\,\bar{\psi}^{\Phi}\Phi,\,\,\,\,C^{2}VM_{1},\,\,\,\,C^{2}S^{2}V^{2},
\end{equation*}
\begin{equation}
C^{2}S^{2}\bar{\psi}^{\Phi},\,\,\,\,S^{2}\bar{\psi}^{M_{1}}V,\,\,\,\,CSV^{2}M_{0},\,\,\,\,CS\bar{\psi}^{\Phi}M_{0},\,\,\,\,CS\bar{\psi}^{M_{0}},\,\,\,\,\Phi V^{2},\,\,\,\,M_{0}^{2}V^{2},\,\,\,\,M_{0}^{2}\bar{\psi}^{\Phi}.
\label{listSpin6}
\end{equation}

Next, denoting the irreducible representations (and the corresponding characters) of the nonabelian groups in the theory by 
\begin{equation}
\left(Spin\left(6\right)_{g},SU\left(2\right)_{v},SU\left(4\right)_{c},SU\left(4\right)_{s}\right), 
\end{equation}
we find the following representations of the gauge singlets corresponding
to the operators in \eqref{listSpin6}, 
\begin{center}
	\begin{longtable}{|c|c|c|}
		\hline
		$\mathrm{Operator}$ & $\left(-1\right)^{F}$ & $\mathrm{Representations}\,\,(R)$\\
		\hline \hline 
		$\lambda\lambda$ & + & $\left(\boldsymbol{1},\boldsymbol{1},\boldsymbol{1},\boldsymbol{1}\right)$\\
		\hline 
		$\bar{\psi}^{V}V$ & - & $\left(\boldsymbol{1},\boldsymbol{1},\boldsymbol{1},\boldsymbol{1}\right),\,\,\left(\boldsymbol{1},\boldsymbol{3},\boldsymbol{1},\boldsymbol{1}\right)$\\
		\hline 
		$\bar{\psi}^{C}C$ & - & $\left(\boldsymbol{1},\boldsymbol{1},\boldsymbol{1},\boldsymbol{1}\right),\,\,\left(\boldsymbol{1},\boldsymbol{1},\boldsymbol{15},\boldsymbol{1}\right)$\\
		\hline 
		$\bar{\psi}^{S}S$ & - & $\left(\boldsymbol{1},\boldsymbol{1},\boldsymbol{1},\boldsymbol{1}\right),\,\,\left(\boldsymbol{1},\boldsymbol{1},\boldsymbol{1},\boldsymbol{15}\right)$\\
		\hline 
		$\bar{\psi}^{M_{0}}M_{0}$ & - & $\left(\boldsymbol{1},\boldsymbol{1},\boldsymbol{1},\boldsymbol{1}\right),\,\,\left(\boldsymbol{1},\boldsymbol{1},\boldsymbol{1},\boldsymbol{15}\right),\,\,\left(\boldsymbol{1},\boldsymbol{1},\boldsymbol{15},\boldsymbol{1}\right),\,\,\left(\boldsymbol{1},\boldsymbol{1},\boldsymbol{15},\boldsymbol{15}\right)$\\
		\hline 
		$\bar{\psi}^{M_{1}}M_{1}$ & - & $\begin{array}{c}
		\left(\boldsymbol{1},\boldsymbol{1},\boldsymbol{1},\boldsymbol{1}\right),\,\,\left(\boldsymbol{1},\boldsymbol{1},\boldsymbol{15},\boldsymbol{1}\right),\,\,\left(\boldsymbol{1},\boldsymbol{1},\boldsymbol{20'},\boldsymbol{1}\right),\\
		\left(\boldsymbol{1},\boldsymbol{3},\boldsymbol{1},\boldsymbol{1}\right),\,\,\left(\boldsymbol{1},\boldsymbol{3},\boldsymbol{15},\boldsymbol{1}\right),\,\,\left(\boldsymbol{1},\boldsymbol{3},\boldsymbol{20'},\boldsymbol{1}\right)
		\end{array}$\\
		\hline 
		$\bar{\psi}^{\Phi}\Phi$ & - & $\left(\boldsymbol{1},\boldsymbol{1},\boldsymbol{1},\boldsymbol{1}\right),\,\,\left(\boldsymbol{1},\boldsymbol{3},\boldsymbol{1},\boldsymbol{1}\right),\,\,\left(\boldsymbol{1},\boldsymbol{5},\boldsymbol{1},\boldsymbol{1}\right)$\\
		\hline 
		$C^{2}VM_{1}$ & + & $\begin{array}{c}
		\left(\boldsymbol{1},\boldsymbol{1},\boldsymbol{1},\boldsymbol{1}\right),\,\,\left(\boldsymbol{1},\boldsymbol{1},\boldsymbol{15},\boldsymbol{1}\right),\,\,\left(\boldsymbol{1},\boldsymbol{1},\boldsymbol{20'},\boldsymbol{1}\right)\\
		\left(\boldsymbol{1},\boldsymbol{3},\boldsymbol{1},\boldsymbol{1}\right),\,\,\left(\boldsymbol{1},\boldsymbol{3},\boldsymbol{15},\boldsymbol{1}\right),\,\,\left(\boldsymbol{1},\boldsymbol{3},\boldsymbol{20'},\boldsymbol{1}\right)
		\end{array}$\\
		\hline 
		$C^{2}S^{2}V^{2}$ & + & $\begin{array}{c}
		\left(\boldsymbol{1},\boldsymbol{3},\boldsymbol{10},\boldsymbol{10}\right),\,\,2\left(\boldsymbol{1},\boldsymbol{3},\boldsymbol{6},\boldsymbol{6}\right),\,\,\left(\boldsymbol{1},\boldsymbol{1},\boldsymbol{10},\boldsymbol{10}\right),\\
		\left(\boldsymbol{1},\boldsymbol{1},\boldsymbol{10},\boldsymbol{6}\right),\,\,\left(\boldsymbol{1},\boldsymbol{1},\boldsymbol{6},\boldsymbol{10}\right),\,\,\left(\boldsymbol{1},\boldsymbol{1},\boldsymbol{6},\boldsymbol{6}\right)
		\end{array}$\\
		\hline 
		$C^{2}S^{2}\bar{\psi}^{\Phi}$ & - & $\left(\boldsymbol{1},\boldsymbol{3},\boldsymbol{6},\boldsymbol{6}\right),\,\,\left(\boldsymbol{1},\boldsymbol{3},\boldsymbol{10},\boldsymbol{10}\right)$\\
		\hline 
		$S^{2}\bar{\psi}^{M_{1}}V$ & - & $\left(\boldsymbol{1},\boldsymbol{1},\boldsymbol{6},\boldsymbol{6}\right),\,\,\left(\boldsymbol{1},\boldsymbol{3},\boldsymbol{6},\boldsymbol{6}\right)$\\
		\hline 
		$CSV^{2}M_{0}$ & + & $\begin{array}{c}
		\left(\boldsymbol{1},\boldsymbol{3},\boldsymbol{1},\boldsymbol{1}\right),\,\,\left(\boldsymbol{1},\boldsymbol{3},\boldsymbol{1},\boldsymbol{15}\right),\,\,\left(\boldsymbol{1},\boldsymbol{3},\boldsymbol{15},\boldsymbol{1}\right),\,\,\left(\boldsymbol{1},\boldsymbol{3},\boldsymbol{15},\boldsymbol{15}\right),\\
		\left(\boldsymbol{1},\boldsymbol{1},\boldsymbol{1},\boldsymbol{1}\right),\,\,\left(\boldsymbol{1},\boldsymbol{1},\boldsymbol{1},\boldsymbol{15}\right),\,\,\left(\boldsymbol{1},\boldsymbol{1},\boldsymbol{15},\boldsymbol{1}\right),\,\,\left(\boldsymbol{1},\boldsymbol{1},\boldsymbol{15},\boldsymbol{15}\right)
		\end{array}$\\
		\hline 
		$CS\bar{\psi}^{\Phi}M_{0}$ & - & $\left(\boldsymbol{1},\boldsymbol{3},\boldsymbol{1},\boldsymbol{1}\right),\,\,\left(\boldsymbol{1},\boldsymbol{3},\boldsymbol{1},\boldsymbol{15}\right),\,\,\left(\boldsymbol{1},\boldsymbol{3},\boldsymbol{15},\boldsymbol{1}\right),\,\,\left(\boldsymbol{1},\boldsymbol{3},\boldsymbol{15},\boldsymbol{15}\right)$\\
		\hline 
		$CS\bar{\psi}^{M_0}$ & - & $\left(\boldsymbol{1},\boldsymbol{1},\boldsymbol{6},\boldsymbol{6}\right),\,\,\left(\boldsymbol{1},\boldsymbol{1},\boldsymbol{10},\boldsymbol{6}\right),\,\,\left(\boldsymbol{1},\boldsymbol{1},\boldsymbol{6},\boldsymbol{10}\right),\,\,\left(\boldsymbol{1},\boldsymbol{1},\boldsymbol{10},\boldsymbol{10}\right)$\\
		\hline 
		$\Phi V^{2}$ & + & $\left(\boldsymbol{1},\boldsymbol{1},\boldsymbol{1},\boldsymbol{1}\right),\,\,\left(\boldsymbol{1},\boldsymbol{3},\boldsymbol{1},\boldsymbol{1}\right),\,\,\left(\boldsymbol{1},\boldsymbol{5},\boldsymbol{1},\boldsymbol{1}\right)$\\
		\hline 
		$M_{0}^{2}V^{2}$ & + & $\left(\boldsymbol{1},\boldsymbol{3},\boldsymbol{6},\boldsymbol{6}\right),\,\,\left(\boldsymbol{1},\boldsymbol{3},\boldsymbol{\overline{10}},\boldsymbol{\overline{10}}\right)$\\
		\hline 
		$M_{0}^{2}\bar{\psi}^{\Phi}$ & - & $\left(\boldsymbol{1},\boldsymbol{3},\boldsymbol{6},\boldsymbol{6}\right),\,\,\left(\boldsymbol{1},\boldsymbol{3},\boldsymbol{\overline{10}},\boldsymbol{\overline{10}}\right)$\\
		\hline
	\end{longtable}
\end{center}
Summing (with signs) the above characters, we obtain 
\begin{equation*}
\sum R\left(-1\right)^{F}=-\left(\boldsymbol{1},\boldsymbol{1},\boldsymbol{15},\boldsymbol{1}\right)-\left(\boldsymbol{1},\boldsymbol{1},\boldsymbol{1},\boldsymbol{15}\right)-\left(\boldsymbol{1},\boldsymbol{1},\boldsymbol{6},\boldsymbol{6}\right)-\left(\boldsymbol{1},\boldsymbol{3},\boldsymbol{1},\boldsymbol{1}\right)-2\left(\boldsymbol{1},\boldsymbol{1},\boldsymbol{1},\boldsymbol{1}\right)
\end{equation*}
\begin{equation}
=-\boldsymbol{66}_{SO(12)}-\boldsymbol{3}_{SU(2)_v}-2
\end{equation}
which corresponds to the enhancement of the flavor symmetry $SU\left(2\right)\times SU\left(4\right)^{2}\times U\left(1\right)^{2}$ to $SU\left(2\right)\times SO\left(12\right)\times U\left(1\right)^{2}$ in the IR. Moreover, the dimension of the conformal manifold vanishes. 

We next continue to analyze the other models in the sequence in a similar way.

\

\subsection*{$\boldsymbol{Spin(7)}$}

\

The matter content is given in the following table, 
\begin{center}
	\begin{tabular}{|c||c|c|c|c|c|}
		\hline
		Field & $Spin\left(7\right)_{g}$ & $SU\left(4\right)_{s}$ & $SU\left(3\right)_{v}$ & $U\left(1\right)_{a}$ & $U\left(1\right)_{r}$\\
		\hline
		$S$ & $\boldsymbol{8}$ & $\boldsymbol{4}$ & $\boldsymbol{1}$ & -3 & $\frac{1}{2}$\\
		
		$V$ & $\boldsymbol{7}$ & $\boldsymbol{1}$ & $\boldsymbol{3}$ & 4 & 0\\
		
		$M_{0}$ & $\boldsymbol{1}$ & $\boldsymbol{6}$ & $\boldsymbol{\overline{3}}$ & 2 & 1\\
		
		$M_{1}$ & $\boldsymbol{1}$ & $\boldsymbol{\overline{10}}$ & $\boldsymbol{1}$ & 6 & 1\\
		
		$\Phi$ & $\boldsymbol{1}$ & $\boldsymbol{1}$ & $\boldsymbol{\overline{6}}$ & -8 & 2\\
		\hline
	\end{tabular}
\end{center}
The superpotential is given by 
\begin{equation}
W=M_{0}S^{2}V+M_{1}S^{2}+\Phi V^{2}
\end{equation}
and the superconformal R charge by ${\hat r}=r+0.039q_{a}$. We see that $U\left(1\right)_{a}$ mixes with $U\left(1\right)_{r}$ in the expression for the IR R symmetry, and so the operators that contribute to the index at order $pq$ should have a vanishing charge under $U\left(1\right)_{a}$. These operators are of the forms
\begin{equation*}
\lambda\lambda,\,\,\,\,\bar{\psi}^{S}S,\,\,\,\,\bar{\psi}^{V}V,\,\,\,\,\bar{\psi}^{M_{0}}M_{0},\,\,\,\,\bar{\psi}^{M_{1}}M_{1},\,\,\,\,\bar{\psi}^{\Phi}\Phi,\,\,\,\,S^{2}VM_{0},\,\,\,\,S^{2}M_{1},\,\,\,\,\Phi V^{2},
\end{equation*}
\begin{equation*}
S^{4}V^{3},\,\,\,\,S^{2}V^{2}\bar{\psi}^{M_{0}},\,\,\,\,V\left(\bar{\psi}^{M_{0}}\right)^{2},\,\,\,\,V^{2}\bar{\psi}^{M_{1}}\bar{\psi}^{M_{0}},\,\,\,\,S^{2}\bar{\psi}^{\Phi}\bar{\psi}^{M_{0}},\,\,\,\,\bar{\psi}^{\Phi}\bar{\psi}^{M_{1}}\bar{\psi}^{M_{0}},
\end{equation*}
\begin{equation}
S^{2}V^{3}\bar{\psi}^{M_{1}},\,\,\,\,V^{3}\left(\bar{\psi}^{M_{1}}\right)^{2},\,\,\,\,S^{2}V\bar{\psi}^{\Phi}\bar{\psi}^{M_{1}},\,\,\,\,V\bar{\psi}^{\Phi}\left(\bar{\psi}^{M_{1}}\right)^{2},\,\,\,\,S^{4}V\bar{\psi}^{\Phi}.
\label{listSpin7}
\end{equation}

Next, denoting the irreducible representations (and the corresponding characters) of the nonabelian groups in the theory by 
\begin{equation}
\left(Spin\left(7\right)_{g},SU\left(4\right)_{s},SU\left(3\right)_{v}\right),
\end{equation}
we find the following representations of the gauge singlets corresponding
to the operators in \eqref{listSpin7}, 
\begin{center}
	\begin{longtable}{|c|c|c|}
		\hline
		$\mathrm{Operator}$ & $\left(-1\right)^{F}$ & $\mathrm{Representations}\,\,(R)$\\
		\hline \hline 
		$\lambda\lambda$ & + & $\left(\boldsymbol{1},\boldsymbol{1},\boldsymbol{1}\right)$\\
		\hline 
		$\bar{\psi}^{S}S$ & - & $\left(\boldsymbol{1},\boldsymbol{1},\boldsymbol{1}\right),\,\,\left(\boldsymbol{1},\boldsymbol{15},\boldsymbol{1}\right)$\\
		\hline 
		$\bar{\psi}^{V}V$ & - & $\left(\boldsymbol{1},\boldsymbol{1},\boldsymbol{1}\right),\,\,\left(\boldsymbol{1},\boldsymbol{1},\boldsymbol{8}\right)$\\
		\hline 
		$\bar{\psi}^{M_{0}}M_{0}$ & - & $\left(\boldsymbol{1},\boldsymbol{1},\boldsymbol{1}\right),\,\,\left(\boldsymbol{1},\boldsymbol{15},\boldsymbol{1}\right),\,\,\left(\boldsymbol{1},\boldsymbol{20'},\boldsymbol{1}\right),\,\,\left(\boldsymbol{1},\boldsymbol{1},\boldsymbol{8}\right),\,\,\left(\boldsymbol{1},\boldsymbol{15},\boldsymbol{8}\right),\,\,\left(\boldsymbol{1},\boldsymbol{20'},\boldsymbol{8}\right)$\\
		\hline 
		$\bar{\psi}^{M_{1}}M_{1}$ & - & $\left(\boldsymbol{1},\boldsymbol{1},\boldsymbol{1}\right),\,\,\left(\boldsymbol{1},\boldsymbol{15},\boldsymbol{1}\right),\,\,\left(\boldsymbol{1},\boldsymbol{84},\boldsymbol{1}\right)$\\
		\hline 
		$\bar{\psi}^{\Phi}\Phi$ & - & $\left(\boldsymbol{1},\boldsymbol{1},\boldsymbol{1}\right),\,\,\left(\boldsymbol{1},\boldsymbol{1},\boldsymbol{8}\right),\,\,\left(\boldsymbol{1},\boldsymbol{1},\boldsymbol{27}\right)$\\
		\hline 
		$S^{2}VM_{0}$ & + & $\left(\boldsymbol{1},\boldsymbol{1},\boldsymbol{1}\right),\,\,\left(\boldsymbol{1},\boldsymbol{15},\boldsymbol{1}\right),\,\,\left(\boldsymbol{1},\boldsymbol{20'},\boldsymbol{1}\right),\,\,\left(\boldsymbol{1},\boldsymbol{1},\boldsymbol{8}\right),\,\,\left(\boldsymbol{1},\boldsymbol{15},\boldsymbol{8}\right),\,\,\left(\boldsymbol{1},\boldsymbol{20'},\boldsymbol{8}\right)$\\
		\hline 
		$S^{2}M_{1}$ & + & $\left(\boldsymbol{1},\boldsymbol{1},\boldsymbol{1}\right),\,\,\left(\boldsymbol{1},\boldsymbol{15},\boldsymbol{1}\right),\,\,\left(\boldsymbol{1},\boldsymbol{84},\boldsymbol{1}\right)$\\
		\hline 
		$\Phi V^{2}$ & + & $\left(\boldsymbol{1},\boldsymbol{1},\boldsymbol{1}\right),\,\,\left(\boldsymbol{1},\boldsymbol{1},\boldsymbol{8}\right),\,\,\left(\boldsymbol{1},\boldsymbol{1},\boldsymbol{27}\right)$\\
		\hline 
		$S^{4}V^{3}$ & + & $\begin{array}{c}
		\left(\boldsymbol{1},\boldsymbol{1},\boldsymbol{8}\right),\,\,2\left(\boldsymbol{1},\boldsymbol{15},\boldsymbol{8}\right),\,\,\left(\boldsymbol{1},\boldsymbol{45},\boldsymbol{8}\right),\,\,\left(\boldsymbol{1},\boldsymbol{1},\boldsymbol{10}\right),\,\,\left(\boldsymbol{1},\boldsymbol{15},\boldsymbol{10}\right),\,\,\left(\boldsymbol{1},\boldsymbol{45},\boldsymbol{10}\right),\\
		\left(\boldsymbol{1},\boldsymbol{1},\boldsymbol{1}\right),\,\,\left(\boldsymbol{1},\boldsymbol{15},\boldsymbol{1}\right),\,\,\left(\boldsymbol{1},\boldsymbol{20'},\boldsymbol{1}\right),\,\,\left(\boldsymbol{1},\boldsymbol{35},\boldsymbol{1}\right),\,\,\left(\boldsymbol{1},\boldsymbol{45},\boldsymbol{1}\right),\,\,\left(\boldsymbol{1},\boldsymbol{20'},\boldsymbol{8}\right)
		\end{array}$\\
		\hline 
		$S^{2}V^{2}\bar{\psi}^{M_{0}}$ & - & $\begin{array}{c}
		\left(\boldsymbol{1},\boldsymbol{15},\boldsymbol{8}\right),\,\,\left(\boldsymbol{1},\boldsymbol{15},\boldsymbol{10}\right),\,\,\left(\boldsymbol{1},\boldsymbol{45},\boldsymbol{8}\right),\,\,\left(\boldsymbol{1},\boldsymbol{45},\boldsymbol{10}\right),\\
		\left(\boldsymbol{1},\boldsymbol{1},\boldsymbol{1}\right),\,\,\left(\boldsymbol{1},\boldsymbol{1},\boldsymbol{8}\right),\,\,\left(\boldsymbol{1},\boldsymbol{15},\boldsymbol{1}\right),\,\,\left(\boldsymbol{1},\boldsymbol{15},\boldsymbol{8}\right),\,\,\left(\boldsymbol{1},\boldsymbol{20'},\boldsymbol{1}\right),\,\,\left(\boldsymbol{1},\boldsymbol{20'},\boldsymbol{8}\right)
		\end{array}$\\
		\hline 
		$V\left(\bar{\psi}^{M_{0}}\right)^{2}$ & + & $\textrm{No gauge singlets}$\\
		\hline 
		$V^{2}\bar{\psi}^{M_{1}}\bar{\psi}^{M_{0}}$ & + & $\left(\boldsymbol{1},\boldsymbol{15},\boldsymbol{8}\right),\,\,\left(\boldsymbol{1},\boldsymbol{15},\boldsymbol{10}\right),\,\,\left(\boldsymbol{1},\boldsymbol{45},\boldsymbol{8}\right),\,\,\left(\boldsymbol{1},\boldsymbol{45},\boldsymbol{10}\right)$\\
		\hline 
		$S^{2}\bar{\psi}^{\Phi}\bar{\psi}^{M_{0}}$ & + & $\left(\boldsymbol{1},\boldsymbol{15},\boldsymbol{8}\right),\,\,\left(\boldsymbol{1},\boldsymbol{15},\boldsymbol{10}\right),\,\,\left(\boldsymbol{1},\boldsymbol{45},\boldsymbol{8}\right),\,\,\left(\boldsymbol{1},\boldsymbol{45},\boldsymbol{10}\right)$\\
		\hline 
		$\bar{\psi}^{\Phi}\bar{\psi}^{M_{1}}\bar{\psi}^{M_{0}}$ & - & $\left(\boldsymbol{1},\boldsymbol{15},\boldsymbol{8}\right),\,\,\left(\boldsymbol{1},\boldsymbol{15},\boldsymbol{10}\right),\,\,\left(\boldsymbol{1},\boldsymbol{45},\boldsymbol{8}\right),\,\,\left(\boldsymbol{1},\boldsymbol{45},\boldsymbol{10}\right)$\\
		\hline 
		$S^{2}V^{3}\bar{\psi}^{M_{1}}$ & - & $\begin{array}{c}
		\left(\boldsymbol{1},\boldsymbol{15},\boldsymbol{8}\right),\,\,\left(\boldsymbol{1},\boldsymbol{15},\boldsymbol{10}\right),\,\,\left(\boldsymbol{1},\boldsymbol{45},\boldsymbol{8}\right),\,\,\left(\boldsymbol{1},\boldsymbol{45},\boldsymbol{10}\right)\\
		\left(\boldsymbol{1},\boldsymbol{20'},\boldsymbol{1}\right),\,\,\left(\boldsymbol{1},\boldsymbol{35},\boldsymbol{1}\right),\,\,\left(\boldsymbol{1},\boldsymbol{45},\boldsymbol{1}\right)
		\end{array}$\\
		\hline 
		$V^{3}\left(\bar{\psi}^{M_{1}}\right)^{2}$ & + & $\textrm{No gauge singlets}$\\
		\hline 
		$S^{2}V\bar{\psi}^{\Phi}\bar{\psi}^{M_{1}}$ & + & $\left(\boldsymbol{1},\boldsymbol{15},\boldsymbol{8}\right),\,\,\left(\boldsymbol{1},\boldsymbol{15},\boldsymbol{10}\right),\,\,\left(\boldsymbol{1},\boldsymbol{45},\boldsymbol{8}\right),\,\,\left(\boldsymbol{1},\boldsymbol{45},\boldsymbol{10}\right)$\\
		\hline 
		$V\bar{\psi}^{\Phi}\left(\bar{\psi}^{M_{1}}\right)^{2}$ & - & $\textrm{No gauge singlets}$\\
		\hline 
		$S^{4}V\bar{\psi}^{\Phi}$ & - & $\begin{array}{c}
		\left(\boldsymbol{1},\boldsymbol{1},\boldsymbol{8}\right),\,\,\left(\boldsymbol{1},\boldsymbol{15},\boldsymbol{8}\right),\,\,\left(\boldsymbol{1},\boldsymbol{45},\boldsymbol{8}\right)\\
		\left(\boldsymbol{1},\boldsymbol{1},\boldsymbol{10}\right),\,\,\left(\boldsymbol{1},\boldsymbol{15},\boldsymbol{10}\right),\,\,\left(\boldsymbol{1},\boldsymbol{45},\boldsymbol{10}\right)
		\end{array}$\\
		\hline
	\end{longtable}
\end{center}
Summing (with signs) the above characters, we obtain  
\begin{equation*}
\sum R\left(-1\right)^{F}=-\left(\boldsymbol{1},\boldsymbol{15},\boldsymbol{1}\right)-\left(\boldsymbol{1},\boldsymbol{20'},\boldsymbol{1}\right)-2\left(\boldsymbol{1},\boldsymbol{1},\boldsymbol{8}\right)-\left(\boldsymbol{1},\boldsymbol{1},\boldsymbol{1}\right)
\end{equation*}
\begin{equation}
=-\boldsymbol{35}_{SU(6)}-2\,\boldsymbol{8}_{SU(3)_v}-1
\end{equation}
which corresponds to the enhancement of the flavor symmetry $SU\left(4\right)\times SU\left(3\right)\times U\left(1\right)$ to $SU\left(6\right)\times SU\left(3\right)^{2}\times U\left(1\right)$ in the IR. Moreover, the dimension of the conformal manifold vanishes. 

\

\subsection*{$\boldsymbol{Spin(8)}$}

\

As in section \ref{fourd}, we split the discussion into two. In the first part, we consider a model in which the conformal manifold is a point, and in the second part a closely related model with an IR global symmetry that corresponds to the commutant structure discussed in section \ref{fourd}. The matter content of the first model is given in the following table, 
\begin{center}
	\begin{tabular}{|c||c|c|c|c|c|c|c|}
		\hline
		Field  & $Spin\left(8\right)_{g}$ & $SU\left(2\right)_{s}$ & $SU\left(2\right)_{c}$ & $SU\left(4\right)_{v}$ & $U\left(1\right)_{a}$ & $U\left(1\right)_{b}$ & $U\left(1\right)_{r}$\\
		\hline 
		$S$ & $\boldsymbol{8}_{s}$ & $\boldsymbol{2}$ & $\boldsymbol{1}$ & $\boldsymbol{1}$ & -2 & 0 & $\frac{1}{2}$\\
		
		$C$ & $\boldsymbol{8}_{c}$ & $\boldsymbol{1}$ & $\boldsymbol{2}$ & $\boldsymbol{1}$ & 0 & -2 & $\frac{1}{2}$\\
		
		$V$ & $\boldsymbol{8}_{v}$ & $\boldsymbol{1}$ & $\boldsymbol{1}$ & $\boldsymbol{4}$ & 1 & 1 & 0\\
		
		$M_{0}$ & $\boldsymbol{1}$ & $\boldsymbol{1}$ & $\boldsymbol{1}$ & $\boldsymbol{6}$ & 2 & -2 & 1\\
		
		$M_{1}$ & $\boldsymbol{1}$ & $\boldsymbol{2}$ & $\boldsymbol{2}$ & $\boldsymbol{\overline{4}}$ & 1 & 1 & 1\\
		
		$M_{2}$ & $\boldsymbol{1}$ & $\boldsymbol{3}$ & $\boldsymbol{1}$ & $\boldsymbol{1}$ & 4 & 0 & 1\\
		
		$M_{3}$ & $\boldsymbol{1}$ & $\boldsymbol{1}$ & $\boldsymbol{3}$ & $\boldsymbol{1}$ & 0 & 4 & 1\\
		
		$\Phi$ & $\boldsymbol{1}$ & $\boldsymbol{1}$ & $\boldsymbol{1}$ & $\boldsymbol{\overline{10}}$ & -2 & -2 & 2\\
		\hline
	\end{tabular}
\end{center}
The superpotential is given by 
\begin{equation}
\label{AppWSpin8}
W=M_{0}S^{2}V^{2}+M_{1}SCV+M_{2}S^{2}+M_{3}C^{2}+\Phi V^{2}
\end{equation}
and the superconformal R charge by ${\hat r}=r+0.061q_{a}+0.107q_{b}$. We see that $U\left(1\right)_{b}$ and $U\left(1\right)_{a}$ mix with $U\left(1\right)_{r}$ in the expression for the IR R symmetry, and so the operators that contribute to the index at order $pq$ should have vanishing charges under $U\left(1\right)_{b}$ and $U\left(1\right)_{a}$. These operators are of the forms
\begin{equation*}
\lambda\lambda,\,\,\,\,\bar{\psi}^{S}S,\,\,\,\,\bar{\psi}^{C}C,\,\,\,\,\bar{\psi}^{V}V,\,\,\,\,\bar{\psi}^{M_{0}}M_{0},\,\,\,\,\bar{\psi}^{M_{1}}M_{1},\,\,\,\,\bar{\psi}^{M_{2}}M_{2},\,\,\,\,\bar{\psi}^{M_{3}}M_{3},\,\,\,\,\bar{\psi}^{\Phi}\Phi,
\end{equation*}
\begin{equation*}
M_{0}S^{2}V^{2},\,\,\,\,M_{1}SCV,\,\,\,\,M_{2}S^{2},\,\,\,\,M_{3}C^{2},\,\,\,\,\Phi V^{2},\,\,\,\,M_{0}\bar{\psi}^{M_{2}}V^{2},\,\,\,\,M_{0}\bar{\psi}^{M_{2}}\bar{\psi}^{\Phi},\,\,\,\,M_{0}S^{2}\bar{\psi}^{\Phi},
\end{equation*}
\begin{equation*}
S^{2}C^{2}V^{4},\,\,\,\,C^{2}V^{2}\bar{\psi}^{M_{0}},\,\,\,\,V^{2}\bar{\psi}^{M_{3}}\bar{\psi}^{M_{0}},\,\,\,\,\bar{\psi}^{\Phi}\bar{\psi}^{M_{3}}\bar{\psi}^{M_{0}},\,\,\,\,C^{2}\bar{\psi}^{\Phi}\bar{\psi}^{M_{0}},
\end{equation*}
\begin{equation*}
SCV^{3}\bar{\psi}^{M_{1}},\,\,\,\,V^{2}\left(\bar{\psi}^{M_{1}}\right)^{2},\,\,\,\,\bar{\psi}^{\Phi}\left(\bar{\psi}^{M_{1}}\right)^{2},\,\,\,\,SCV\bar{\psi}^{\Phi}\bar{\psi}^{M_{1}},\,\,\,\,C^{2}V^{4}\bar{\psi}^{M_{2}},
\end{equation*}
\begin{equation*}
V^{4}\bar{\psi}^{M_{3}}\bar{\psi}^{M_{2}},\,\,\,\,V^{2}\bar{\psi}^{\Phi}\bar{\psi}^{M_{3}}\bar{\psi}^{M_{2}},\,\,\,\,\left(\bar{\psi}^{\Phi}\right)^{2}\bar{\psi}^{M_{3}}\bar{\psi}^{M_{2}},\,\,\,\,C^{2}V^{2}\bar{\psi}^{\Phi}\bar{\psi}^{M_{2}},\,\,\,\,C^{2}\left(\bar{\psi}^{\Phi}\right)^{2}\bar{\psi}^{M_{2}},
\end{equation*}
\begin{equation}
S^{2}V^{4}\bar{\psi}^{M_{3}},\,\,\,\,S^{2}V^{2}\bar{\psi}^{\Phi}\bar{\psi}^{M_{3}},\,\,\,\,S^{2}\left(\bar{\psi}^{\Phi}\right)^{2}\bar{\psi}^{M_{3}},\,\,\,\,S^{2}C^{2}V^{2}\bar{\psi}^{\Phi},\,\,\,\,S^{2}C^{2}\left(\bar{\psi}^{\Phi}\right)^{2}.
\label{listSpin8}
\end{equation}

Next, denoting the irreducible representations (and the corresponding characters) of the nonabelian groups in the theory by 
\begin{equation}
\left(Spin\left(8\right)_{g},SU\left(2\right)_{s},SU\left(2\right)_{c},SU\left(4\right)_{v}\right),
\end{equation}
we find the following representations of the gauge singlets corresponding
to the operators in \eqref{listSpin8}, 
\begin{center}
	\begin{longtable}{|c|c|c|}
		\hline
		$\mathrm{Operator}$ & $\left(-1\right)^{F}$ & $\mathrm{Representations}\,\,(R)$\\
		\hline \hline 
		$\lambda\lambda$ & + & $\left(\boldsymbol{1},\boldsymbol{1},\boldsymbol{1},\boldsymbol{1}\right)$\tabularnewline
		\hline 
		$\bar{\psi}^{S}S$ & - & $\left(\boldsymbol{1},\boldsymbol{1},\boldsymbol{1},\boldsymbol{1}\right),\,\,\left(\boldsymbol{1},\boldsymbol{3},\boldsymbol{1},\boldsymbol{1}\right)$\tabularnewline
		\hline 
		$\bar{\psi}^{C}C$ & - & $\left(\boldsymbol{1},\boldsymbol{1},\boldsymbol{1},\boldsymbol{1}\right),\,\,\left(\boldsymbol{1},\boldsymbol{1},\boldsymbol{3},\boldsymbol{1}\right)$\tabularnewline
		\hline 
		$\bar{\psi}^{V}V$ & - & $\left(\boldsymbol{1},\boldsymbol{1},\boldsymbol{1},\boldsymbol{1}\right),\,\,\left(\boldsymbol{1},\boldsymbol{1},\boldsymbol{1},\boldsymbol{15}\right)$\tabularnewline
		\hline 
		$\bar{\psi}^{M_{0}}M_{0}$ & - & $\left(\boldsymbol{1},\boldsymbol{1},\boldsymbol{1},\boldsymbol{1}\right),\,\,\left(\boldsymbol{1},\boldsymbol{1},\boldsymbol{1},\boldsymbol{15}\right),\,\,\left(\boldsymbol{1},\boldsymbol{1},\boldsymbol{1},\boldsymbol{20'}\right)$\tabularnewline
		\hline 
		$\bar{\psi}^{M_{1}}M_{1}$ & - & $\begin{array}{c}
		\left(\boldsymbol{1},\boldsymbol{1},\boldsymbol{1},\boldsymbol{1}\right),\,\,\left(\boldsymbol{1},\boldsymbol{1},\boldsymbol{1},\boldsymbol{15}\right),\,\,\left(\boldsymbol{1},\boldsymbol{1},\boldsymbol{3},\boldsymbol{1}\right),\,\,\left(\boldsymbol{1},\boldsymbol{1},\boldsymbol{3},\boldsymbol{15}\right),\,\,\left(\boldsymbol{1},\boldsymbol{3},\boldsymbol{1},\boldsymbol{1}\right),\,\,\left(\boldsymbol{1},\boldsymbol{3},\boldsymbol{1},\boldsymbol{15}\right),\\
		\left(\boldsymbol{1},\boldsymbol{3},\boldsymbol{3},\boldsymbol{1}\right),\,\,\left(\boldsymbol{1},\boldsymbol{3},\boldsymbol{3},\boldsymbol{15}\right)
		\end{array}$\tabularnewline
		\hline 
		$\bar{\psi}^{M_{2}}M_{2}$ & - & $\left(\boldsymbol{1},\boldsymbol{1},\boldsymbol{1},\boldsymbol{1}\right),\,\,\left(\boldsymbol{1},\boldsymbol{3},\boldsymbol{1},\boldsymbol{1}\right),\,\,\left(\boldsymbol{1},\boldsymbol{5},\boldsymbol{1},\boldsymbol{1}\right)$\tabularnewline
		\hline 
		$\bar{\psi}^{M_{3}}M_{3}$ & - & $\left(\boldsymbol{1},\boldsymbol{1},\boldsymbol{1},\boldsymbol{1}\right),\,\,\left(\boldsymbol{1},\boldsymbol{1},\boldsymbol{3},\boldsymbol{1}\right),\,\,\left(\boldsymbol{1},\boldsymbol{1},\boldsymbol{5},\boldsymbol{1}\right)$\tabularnewline
		\hline 
		$\bar{\psi}^{\Phi}\Phi$ & - & $\left(\boldsymbol{1},\boldsymbol{1},\boldsymbol{1},\boldsymbol{1}\right),\,\,\left(\boldsymbol{1},\boldsymbol{1},\boldsymbol{1},\boldsymbol{15}\right),\,\,\left(\boldsymbol{1},\boldsymbol{1},\boldsymbol{1},\boldsymbol{84}\right)$\tabularnewline
		\hline 
		$\Phi V^{2}$ & + & $\left(\boldsymbol{1},\boldsymbol{1},\boldsymbol{1},\boldsymbol{1}\right),\,\,\left(\boldsymbol{1},\boldsymbol{1},\boldsymbol{1},\boldsymbol{15}\right),\,\,\left(\boldsymbol{1},\boldsymbol{1},\boldsymbol{1},\boldsymbol{84}\right)$\tabularnewline
		\hline 
		$M_{3}C^{2}$ & + & $\left(\boldsymbol{1},\boldsymbol{1},\boldsymbol{1},\boldsymbol{1}\right),\,\,\left(\boldsymbol{1},\boldsymbol{1},\boldsymbol{3},\boldsymbol{1}\right),\,\,\left(\boldsymbol{1},\boldsymbol{1},\boldsymbol{5},\boldsymbol{1}\right)$\tabularnewline
		\hline 
		$M_{2}S^{2}$ & + & $\left(\boldsymbol{1},\boldsymbol{1},\boldsymbol{1},\boldsymbol{1}\right),\,\,\left(\boldsymbol{1},\boldsymbol{3},\boldsymbol{1},\boldsymbol{1}\right),\,\,\left(\boldsymbol{1},\boldsymbol{5},\boldsymbol{1},\boldsymbol{1}\right)$\tabularnewline
		\hline 
		$M_{1}SCV$ & + & $\begin{array}{c}
		\left(\boldsymbol{1},\boldsymbol{1},\boldsymbol{1},\boldsymbol{1}\right),\,\,\left(\boldsymbol{1},\boldsymbol{1},\boldsymbol{1},\boldsymbol{15}\right),\,\,\left(\boldsymbol{1},\boldsymbol{1},\boldsymbol{3},\boldsymbol{1}\right),\,\,\left(\boldsymbol{1},\boldsymbol{1},\boldsymbol{3},\boldsymbol{15}\right),\,\,\left(\boldsymbol{1},\boldsymbol{3},\boldsymbol{1},\boldsymbol{1}\right),\,\,\left(\boldsymbol{1},\boldsymbol{3},\boldsymbol{1},\boldsymbol{15}\right),\\
		\left(\boldsymbol{1},\boldsymbol{3},\boldsymbol{3},\boldsymbol{1}\right),\,\,\left(\boldsymbol{1},\boldsymbol{3},\boldsymbol{3},\boldsymbol{15}\right)
		\end{array}$\tabularnewline
		\hline 
		$M_{0}S^{2}V^{2}$ & + & $\left(\boldsymbol{1},\boldsymbol{1},\boldsymbol{1},\boldsymbol{1}\right),\,\,\left(\boldsymbol{1},\boldsymbol{1},\boldsymbol{1},\boldsymbol{15}\right),\,\,\left(\boldsymbol{1},\boldsymbol{1},\boldsymbol{1},\boldsymbol{20'}\right),\,\,\left(\boldsymbol{1},\boldsymbol{3},\boldsymbol{1},\boldsymbol{15}\right),\,\,\left(\boldsymbol{1},\boldsymbol{3},\boldsymbol{1},\boldsymbol{45}\right)$\tabularnewline
		\hline 
		$M_{0}S^{2}\bar{\psi}^{\Phi}$ & - & $\left(\boldsymbol{1},\boldsymbol{3},\boldsymbol{1},\boldsymbol{15}\right),\,\,\left(\boldsymbol{1},\boldsymbol{3},\boldsymbol{1},\boldsymbol{45}\right)$\tabularnewline
		\hline 
		$M_{0}\bar{\psi}^{M_{2}}\bar{\psi}^{\Phi}$ & + & $\left(\boldsymbol{1},\boldsymbol{3},\boldsymbol{1},\boldsymbol{15}\right),\,\,\left(\boldsymbol{1},\boldsymbol{3},\boldsymbol{1},\boldsymbol{45}\right)$\tabularnewline
		\hline 
		$M_{0}\bar{\psi}^{M_{2}}V^{2}$ & - & $\left(\boldsymbol{1},\boldsymbol{3},\boldsymbol{1},\boldsymbol{15}\right),\,\,\left(\boldsymbol{1},\boldsymbol{3},\boldsymbol{1},\boldsymbol{45}\right)$\tabularnewline
		\hline 
		$S^{2}C^{2}V^{4}$ & + & $\begin{array}{c}
		2\left(\boldsymbol{1},\boldsymbol{1},\boldsymbol{1},\boldsymbol{1}\right),\,\,\left(\boldsymbol{1},\boldsymbol{3},\boldsymbol{1},\boldsymbol{1}\right),\,\,\left(\boldsymbol{1},\boldsymbol{1},\boldsymbol{3},\boldsymbol{1}\right),\,\,2\left(\boldsymbol{1},\boldsymbol{3},\boldsymbol{3},\boldsymbol{1}\right),\\
		2\left(\boldsymbol{1},\boldsymbol{1},\boldsymbol{1},\boldsymbol{15}\right),\,\,3\left(\boldsymbol{1},\boldsymbol{3},\boldsymbol{1},\boldsymbol{15}\right),\,\,3\left(\boldsymbol{1},\boldsymbol{1},\boldsymbol{3},\boldsymbol{15}\right),\,\,\left(\boldsymbol{1},\boldsymbol{3},\boldsymbol{3},\boldsymbol{15}\right),\\
		3\left(\boldsymbol{1},\boldsymbol{1},\boldsymbol{1},\boldsymbol{20'}\right),\,\,2\left(\boldsymbol{1},\boldsymbol{3},\boldsymbol{3},\boldsymbol{20'}\right),\,\,2\left(\boldsymbol{1},\boldsymbol{1},\boldsymbol{1},\boldsymbol{35}\right),\,\,2\left(\boldsymbol{1},\boldsymbol{3},\boldsymbol{3},\boldsymbol{35}\right),\\
		2\left(\boldsymbol{1},\boldsymbol{1},\boldsymbol{1},\boldsymbol{45}\right),\,\,2\left(\boldsymbol{1},\boldsymbol{3},\boldsymbol{1},\boldsymbol{45}\right),\,\,2\left(\boldsymbol{1},\boldsymbol{1},\boldsymbol{3},\boldsymbol{45}\right),\,\,\left(\boldsymbol{1},\boldsymbol{3},\boldsymbol{3},\boldsymbol{45}\right)
		\end{array}$\tabularnewline
		\hline 
		$C^{2}\bar{\psi}^{\Phi}\bar{\psi}^{M_{0}}$ & + & $\left(\boldsymbol{1},\boldsymbol{1},\boldsymbol{3},\boldsymbol{15}\right),\,\,\left(\boldsymbol{1},\boldsymbol{1},\boldsymbol{3},\boldsymbol{45}\right)$\tabularnewline
		\hline 
		$\bar{\psi}^{\Phi}\bar{\psi}^{M_{3}}\bar{\psi}^{M_{0}}$ & - & $\left(\boldsymbol{1},\boldsymbol{1},\boldsymbol{3},\boldsymbol{15}\right),\,\,\left(\boldsymbol{1},\boldsymbol{1},\boldsymbol{3},\boldsymbol{45}\right)$\tabularnewline
		\hline 
		$V^{2}\bar{\psi}^{M_{3}}\bar{\psi}^{M_{0}}$ & + & $\left(\boldsymbol{1},\boldsymbol{1},\boldsymbol{3},\boldsymbol{15}\right),\,\,\left(\boldsymbol{1},\boldsymbol{1},\boldsymbol{3},\boldsymbol{45}\right)$\tabularnewline
		\hline 
		$C^{2}V^{2}\bar{\psi}^{M_{0}}$ & - & $\left(\boldsymbol{1},\boldsymbol{1},\boldsymbol{1},\boldsymbol{1}\right),\,\,\left(\boldsymbol{1},\boldsymbol{1},\boldsymbol{1},\boldsymbol{15}\right),\,\,\left(\boldsymbol{1},\boldsymbol{1},\boldsymbol{1},\boldsymbol{20'}\right),\,\,\left(\boldsymbol{1},\boldsymbol{1},\boldsymbol{3},\boldsymbol{15}\right),\,\,\left(\boldsymbol{1},\boldsymbol{1},\boldsymbol{3},\boldsymbol{45}\right)$\tabularnewline
		\hline 
		$SCV\bar{\psi}^{\Phi}\bar{\psi}^{M_{1}}$ & + & $\begin{array}{c}
		\left(\boldsymbol{1},\boldsymbol{1},\boldsymbol{1},\boldsymbol{15}\right),\,\,\left(\boldsymbol{1},\boldsymbol{1},\boldsymbol{1},\boldsymbol{20'}\right),\,\,\left(\boldsymbol{1},\boldsymbol{1},\boldsymbol{1},\boldsymbol{35}\right),\,\,2\left(\boldsymbol{1},\boldsymbol{1},\boldsymbol{1},\boldsymbol{45}\right),\\
		\left(\boldsymbol{1},\boldsymbol{1},\boldsymbol{3},\boldsymbol{15}\right),\,\,\left(\boldsymbol{1},\boldsymbol{1},\boldsymbol{3},\boldsymbol{20'}\right),\,\,\left(\boldsymbol{1},\boldsymbol{1},\boldsymbol{3},\boldsymbol{35}\right),\,\,2\left(\boldsymbol{1},\boldsymbol{1},\boldsymbol{3},\boldsymbol{45}\right),\\
		\left(\boldsymbol{1},\boldsymbol{3},\boldsymbol{1},\boldsymbol{15}\right),\,\,\left(\boldsymbol{1},\boldsymbol{3},\boldsymbol{1},\boldsymbol{20'}\right),\,\,\left(\boldsymbol{1},\boldsymbol{3},\boldsymbol{1},\boldsymbol{35}\right),\,\,2\left(\boldsymbol{1},\boldsymbol{3},\boldsymbol{1},\boldsymbol{45}\right),\\
		\left(\boldsymbol{1},\boldsymbol{3},\boldsymbol{3},\boldsymbol{15}\right),\,\,\left(\boldsymbol{1},\boldsymbol{3},\boldsymbol{3},\boldsymbol{20'}\right),\,\,\left(\boldsymbol{1},\boldsymbol{3},\boldsymbol{3},\boldsymbol{35}\right),\,\,2\left(\boldsymbol{1},\boldsymbol{3},\boldsymbol{3},\boldsymbol{45}\right)
		\end{array}$\tabularnewline
		\hline 
		$\bar{\psi}^{\Phi}\left(\bar{\psi}^{M_{1}}\right)^{2}$ & - & $\begin{array}{c}
		\left(\boldsymbol{1},\boldsymbol{1},\boldsymbol{1},\boldsymbol{15}\right),\,\,\left(\boldsymbol{1},\boldsymbol{1},\boldsymbol{1},\boldsymbol{45}\right),\,\,\left(\boldsymbol{1},\boldsymbol{1},\boldsymbol{3},\boldsymbol{20'}\right),\,\,\left(\boldsymbol{1},\boldsymbol{1},\boldsymbol{3},\boldsymbol{35}\right),\,\,\left(\boldsymbol{1},\boldsymbol{1},\boldsymbol{3},\boldsymbol{45}\right)\\
		\left(\boldsymbol{1},\boldsymbol{3},\boldsymbol{1},\boldsymbol{20'}\right),\,\,\left(\boldsymbol{1},\boldsymbol{3},\boldsymbol{1},\boldsymbol{35}\right),\,\,\left(\boldsymbol{1},\boldsymbol{3},\boldsymbol{1},\boldsymbol{45}\right),\,\,\left(\boldsymbol{1},\boldsymbol{3},\boldsymbol{3},\boldsymbol{15}\right),\,\,\left(\boldsymbol{1},\boldsymbol{3},\boldsymbol{3},\boldsymbol{45}\right)
		\end{array}$\tabularnewline
		\hline 
		$V^{2}\left(\bar{\psi}^{M_{1}}\right)^{2}$ & + & $\begin{array}{c}
		\left(\boldsymbol{1},\boldsymbol{1},\boldsymbol{1},\boldsymbol{15}\right),\,\,\left(\boldsymbol{1},\boldsymbol{1},\boldsymbol{1},\boldsymbol{45}\right),\,\,\left(\boldsymbol{1},\boldsymbol{1},\boldsymbol{3},\boldsymbol{20'}\right),\,\,\left(\boldsymbol{1},\boldsymbol{1},\boldsymbol{3},\boldsymbol{35}\right),\,\,\left(\boldsymbol{1},\boldsymbol{1},\boldsymbol{3},\boldsymbol{45}\right)\\
		\left(\boldsymbol{1},\boldsymbol{3},\boldsymbol{1},\boldsymbol{20'}\right),\,\,\left(\boldsymbol{1},\boldsymbol{3},\boldsymbol{1},\boldsymbol{35}\right),\,\,\left(\boldsymbol{1},\boldsymbol{3},\boldsymbol{1},\boldsymbol{45}\right),\,\,\left(\boldsymbol{1},\boldsymbol{3},\boldsymbol{3},\boldsymbol{15}\right),\,\,\left(\boldsymbol{1},\boldsymbol{3},\boldsymbol{3},\boldsymbol{45}\right)
		\end{array}$\tabularnewline
		\hline 
		$SCV^{3}\bar{\psi}^{M_{1}}$ & - & $\begin{array}{c}
		\left(\boldsymbol{1},\boldsymbol{1},\boldsymbol{1},\boldsymbol{1}\right),\,\,2\left(\boldsymbol{1},\boldsymbol{1},\boldsymbol{1},\boldsymbol{15}\right),\,\,\left(\boldsymbol{1},\boldsymbol{1},\boldsymbol{1},\boldsymbol{20'}\right),\,\,\left(\boldsymbol{1},\boldsymbol{1},\boldsymbol{1},\boldsymbol{35}\right),\,\,2\left(\boldsymbol{1},\boldsymbol{1},\boldsymbol{1},\boldsymbol{45}\right),\\
		\left(\boldsymbol{1},\boldsymbol{1},\boldsymbol{3},\boldsymbol{1}\right),\,\,2\left(\boldsymbol{1},\boldsymbol{1},\boldsymbol{3},\boldsymbol{15}\right),\,\,\left(\boldsymbol{1},\boldsymbol{1},\boldsymbol{3},\boldsymbol{20'}\right),\,\,\left(\boldsymbol{1},\boldsymbol{1},\boldsymbol{3},\boldsymbol{35}\right),\,\,2\left(\boldsymbol{1},\boldsymbol{1},\boldsymbol{3},\boldsymbol{45}\right),\\
		\left(\boldsymbol{1},\boldsymbol{3},\boldsymbol{1},\boldsymbol{1}\right),\,\,2\left(\boldsymbol{1},\boldsymbol{3},\boldsymbol{1},\boldsymbol{15}\right),\,\,\left(\boldsymbol{1},\boldsymbol{3},\boldsymbol{1},\boldsymbol{20'}\right),\,\,\left(\boldsymbol{1},\boldsymbol{3},\boldsymbol{1},\boldsymbol{35}\right),\,\,2\left(\boldsymbol{1},\boldsymbol{3},\boldsymbol{1},\boldsymbol{45}\right),\\
		\left(\boldsymbol{1},\boldsymbol{3},\boldsymbol{3},\boldsymbol{1}\right),\,\,2\left(\boldsymbol{1},\boldsymbol{3},\boldsymbol{3},\boldsymbol{15}\right),\,\,\left(\boldsymbol{1},\boldsymbol{3},\boldsymbol{3},\boldsymbol{20'}\right),\,\,\left(\boldsymbol{1},\boldsymbol{3},\boldsymbol{3},\boldsymbol{35}\right),\,\,2\left(\boldsymbol{1},\boldsymbol{3},\boldsymbol{3},\boldsymbol{45}\right),
		\end{array}$\tabularnewline
		\hline 
		$C^{2}\left(\bar{\psi}^{\Phi}\right)^{2}\bar{\psi}^{M_{2}}$ & - & $\left(\boldsymbol{1},\boldsymbol{3},\boldsymbol{3},\boldsymbol{45}\right)$\tabularnewline
		\hline 
		$C^{2}V^{2}\bar{\psi}^{\Phi}\bar{\psi}^{M_{2}}$ & + & $\left(\boldsymbol{1},\boldsymbol{3},\boldsymbol{1},\boldsymbol{15}\right),\,\,\left(\boldsymbol{1},\boldsymbol{3},\boldsymbol{3},\boldsymbol{20'}\right),\,\,\left(\boldsymbol{1},\boldsymbol{3},\boldsymbol{3},\boldsymbol{35}\right),\,\,\left(\boldsymbol{1},\boldsymbol{3},\boldsymbol{1},\boldsymbol{45}\right),\,\,\left(\boldsymbol{1},\boldsymbol{3},\boldsymbol{3},\boldsymbol{45}\right)$\tabularnewline
		\hline 
		$\left(\bar{\psi}^{\Phi}\right)^{2}\bar{\psi}^{M_{3}}\bar{\psi}^{M_{2}}$ & + & $\left(\boldsymbol{1},\boldsymbol{3},\boldsymbol{3},\boldsymbol{45}\right)$\tabularnewline
		\hline 
		$V^{2}\bar{\psi}^{\Phi}\bar{\psi}^{M_{3}}\bar{\psi}^{M_{2}}$ & - & $\left(\boldsymbol{1},\boldsymbol{3},\boldsymbol{3},\boldsymbol{20'}\right),\,\,\left(\boldsymbol{1},\boldsymbol{3},\boldsymbol{3},\boldsymbol{35}\right),\,\,\left(\boldsymbol{1},\boldsymbol{3},\boldsymbol{3},\boldsymbol{45}\right)$\tabularnewline
		\hline 
		$V^{4}\bar{\psi}^{M_{3}}\bar{\psi}^{M_{2}}$ & + & $\begin{array}{c}
		\left(\boldsymbol{1},\boldsymbol{3},\boldsymbol{3},\boldsymbol{20'}\right),\,\,\left(\boldsymbol{1},\boldsymbol{3},\boldsymbol{3},\boldsymbol{35}\right)\end{array}$\tabularnewline
		\hline 
		$C^{2}V^{4}\bar{\psi}^{M_{2}}$ & - & $\begin{array}{c}
		\left(\boldsymbol{1},\boldsymbol{3},\boldsymbol{3},\boldsymbol{20'}\right),\,\,\left(\boldsymbol{1},\boldsymbol{3},\boldsymbol{3},\boldsymbol{35}\right),\,\,\left(\boldsymbol{1},\boldsymbol{3},\boldsymbol{1},\boldsymbol{15}\right),\,\,\left(\boldsymbol{1},\boldsymbol{3},\boldsymbol{1},\boldsymbol{45}\right),\,\,\left(\boldsymbol{1},\boldsymbol{3},\boldsymbol{3},\boldsymbol{1}\right)\end{array}$\tabularnewline
		\hline 
		$S^{2}\left(\bar{\psi}^{\Phi}\right)^{2}\bar{\psi}^{M_{3}}$ & - & $\left(\boldsymbol{1},\boldsymbol{3},\boldsymbol{3},\boldsymbol{45}\right)$\tabularnewline
		\hline 
		$S^{2}V^{2}\bar{\psi}^{\Phi}\bar{\psi}^{M_{3}}$ & + & $\left(\boldsymbol{1},\boldsymbol{1},\boldsymbol{3},\boldsymbol{15}\right),\,\,\left(\boldsymbol{1},\boldsymbol{3},\boldsymbol{3},\boldsymbol{20'}\right),\,\,\left(\boldsymbol{1},\boldsymbol{3},\boldsymbol{3},\boldsymbol{35}\right),\,\,\left(\boldsymbol{1},\boldsymbol{1},\boldsymbol{3},\boldsymbol{45}\right),\,\,\left(\boldsymbol{1},\boldsymbol{3},\boldsymbol{3},\boldsymbol{45}\right)$\tabularnewline
		\hline 
		$S^{2}V^{4}\bar{\psi}^{M_{3}}$ & - & $\begin{array}{c}
		\left(\boldsymbol{1},\boldsymbol{3},\boldsymbol{3},\boldsymbol{20'}\right),\,\,\left(\boldsymbol{1},\boldsymbol{3},\boldsymbol{3},\boldsymbol{35}\right),\,\,\left(\boldsymbol{1},\boldsymbol{1},\boldsymbol{3},\boldsymbol{15}\right),\,\,\left(\boldsymbol{1},\boldsymbol{1},\boldsymbol{3},\boldsymbol{45}\right),\,\,\left(\boldsymbol{1},\boldsymbol{3},\boldsymbol{3},\boldsymbol{1}\right)\end{array}$\tabularnewline
		\hline 
		$S^{2}C^{2}V^{2}\bar{\psi}^{\Phi}$ & - & $\begin{array}{c}
		2\left(\boldsymbol{1},\boldsymbol{1},\boldsymbol{1},\boldsymbol{20'}\right),\,\,2\left(\boldsymbol{1},\boldsymbol{1},\boldsymbol{1},\boldsymbol{35}\right),\,\,2\left(\boldsymbol{1},\boldsymbol{1},\boldsymbol{1},\boldsymbol{45}\right),\,\,2\left(\boldsymbol{1},\boldsymbol{3},\boldsymbol{3},\boldsymbol{20'}\right),\\
		2\left(\boldsymbol{1},\boldsymbol{3},\boldsymbol{3},\boldsymbol{35}\right),\,\,2\left(\boldsymbol{1},\boldsymbol{3},\boldsymbol{3},\boldsymbol{45}\right),\,\,\left(\boldsymbol{1},\boldsymbol{1},\boldsymbol{1},\boldsymbol{15}\right),\,\,\left(\boldsymbol{1},\boldsymbol{1},\boldsymbol{1},\boldsymbol{45}\right),\\
		2\left(\boldsymbol{1},\boldsymbol{3},\boldsymbol{1},\boldsymbol{15}\right),\,\,2\left(\boldsymbol{1},\boldsymbol{3},\boldsymbol{1},\boldsymbol{45}\right),\,\,2\left(\boldsymbol{1},\boldsymbol{1},\boldsymbol{3},\boldsymbol{15}\right),\,\,2\left(\boldsymbol{1},\boldsymbol{1},\boldsymbol{3},\boldsymbol{45}\right)
		\end{array}$\tabularnewline
		\hline 
		$S^{2}C^{2}\left(\bar{\psi}^{\Phi}\right)^{2}$ & + & $\left(\boldsymbol{1},\boldsymbol{1},\boldsymbol{1},\boldsymbol{45}\right),\,\,\left(\boldsymbol{1},\boldsymbol{3},\boldsymbol{3},\boldsymbol{45}\right)$\tabularnewline
		\hline
	\end{longtable}
\end{center}
Summing (with signs) the above characters, we obtain  
\begin{equation*}
\sum R\left(-1\right)^{F}=-\begin{array}{c}
\left(\boldsymbol{1},\boldsymbol{3},\boldsymbol{3},\boldsymbol{1}\right)\end{array}-\left(\boldsymbol{1},\boldsymbol{3},\boldsymbol{1},\boldsymbol{1}\right)-\left(\boldsymbol{1},\boldsymbol{1},\boldsymbol{3},\boldsymbol{1}\right)-2\begin{array}{c}
\left(\boldsymbol{1},\boldsymbol{1},\boldsymbol{1},\boldsymbol{15}\right)\end{array}-2\left(\boldsymbol{1},\boldsymbol{1},\boldsymbol{1},\boldsymbol{1}\right)
\end{equation*}
\begin{equation}
=-\boldsymbol{15}_{\widetilde{SU}(4)}-2\,\boldsymbol{15}_{SU(4)_v}-2
\end{equation}
which corresponds to the enhancement of the flavor symmetry $SU\left(2\right)^{2}\times SU\left(4\right)\times U\left(1\right)^{2}$ to $SU\left(4\right)^{3}\times U\left(1\right)^{2}$ in the IR. Moreover, the dimension of the conformal manifold vanishes. 

\

Next, in the second part, we examine a model which is similar to the previous one but without the "flipper" field $M_0$; that is, we remove this field and its corresponding term in the superpotential \eqref{AppWSpin8}. We obtain the matter content given in the following table, 
\begin{center}
	\begin{tabular}{|c||c|c|c|c|c|c|c|}
		\hline
		Field  & $Spin\left(8\right)_{g}$ & $SU\left(2\right)_{s}$ & $SU\left(2\right)_{c}$ & $SU\left(4\right)_{v}$ & $U\left(1\right)_{a}$ & $U\left(1\right)_{b}$ & $U\left(1\right)_{\hat r}$\\
		\hline 
		$S$ & $\boldsymbol{8}_{s}$ & $\boldsymbol{2}$ & $\boldsymbol{1}$ & $\boldsymbol{1}$ & -2 & 0 & $\frac{1}{3}$\\
		
		$C$ & $\boldsymbol{8}_{c}$ & $\boldsymbol{1}$ & $\boldsymbol{2}$ & $\boldsymbol{1}$ & 0 & -2 & $\frac{1}{3}$\\
		
		$V$ & $\boldsymbol{8}_{v}$ & $\boldsymbol{1}$ & $\boldsymbol{1}$ & $\boldsymbol{4}$ & 1 & 1 & $\frac{1}{6}$\\
		
		$M_{1}$ & $\boldsymbol{1}$ & $\boldsymbol{2}$ & $\boldsymbol{2}$ & $\boldsymbol{\overline{4}}$ & 1 & 1 & $\frac{7}{6}$\\
		
		$M_{2}$ & $\boldsymbol{1}$ & $\boldsymbol{3}$ & $\boldsymbol{1}$ & $\boldsymbol{1}$ & 4 & 0 & $\frac{4}{3}$\\
		
		$M_{3}$ & $\boldsymbol{1}$ & $\boldsymbol{1}$ & $\boldsymbol{3}$ & $\boldsymbol{1}$ & 0 & 4 & $\frac{4}{3}$\\
		
		$\Phi$ & $\boldsymbol{1}$ & $\boldsymbol{1}$ & $\boldsymbol{1}$ & $\boldsymbol{\overline{10}}$ & -2 & -2 & $\frac{5}{3}$\\
		\hline
	\end{tabular}
\end{center}
The superpotential is now given by 
\begin{equation}
W=M_{1}SCV+M_{2}S^{2}+M_{3}C^{2}+\Phi V^{2}
\end{equation}
and $\hat r$ written in the table is the superconformal R charge. Since the enhancement that we want to identify is less trivial in this case (in particular, there is a nonvanishing positive contribution at order $pq$ coming from marginal operators), we calculate the full index up to order $pq$ instead of focusing only at this order. The analytical calculation at orders $(pq)^{<\,1}$ is done in the same way as the one performed at order $pq$,\footnote{Note that in order to find the contribution to the index of the second model at order $pq$ one should just remove from the table of representations corresponding to order $pq$ in the first model the contributions of all the operators involving $M_0$ and $\bar{\psi}^{M_0}$.} and in addition to that the index was calculated in Mathematica (which yielded the same result). We get 
\begin{equation*}
I=1+\boldsymbol{6}_{SU\left(4\right)_{v}}\left(a^{2}b^{-2}+a^{-2}b^{2}\right)\left(pq\right)^{\frac{1}{2}}+2\,\boldsymbol{2}_{SU\left(2\right)_{s}}\boldsymbol{2}_{SU\left(2\right)_{c}}\boldsymbol{\bar{4}}_{SU\left(4\right)_{v}}ab\left(pq\right)^{\frac{7}{12}}
\end{equation*}
\begin{equation*}
+\left[\left(\boldsymbol{3}_{SU\left(2\right)_{s}}+\boldsymbol{3}_{SU\left(2\right)_{c}}\right)\left(a^{4}+b^{4}\right)+a^{-4}b^{-4}\right]\left(pq\right)^{\frac{2}{3}}+\left[\boldsymbol{\overline{10}}_{SU\left(4\right)_{v}}+\boldsymbol{6}_{SU\left(4\right)_{v}}\right]a^{-2}b^{-2}\left(pq\right)^{\frac{5}{6}}
\end{equation*}
\begin{equation*}
+\boldsymbol{6}_{SU\left(4\right)_{v}}\left(a^{2}b^{-2}+a^{-2}b^{2}\right)\left(p^{\frac{3}{2}}q^{\frac{1}{2}}+p^{\frac{1}{2}}q^{\frac{3}{2}}\right)
\end{equation*}
\begin{equation*}
+\left[\boldsymbol{20'}_{SU\left(4\right)_{v}}\left(a^{4}b^{-4}+a^{-4}b^{4}+1\right)-\boldsymbol{3}_{SU(2)_s}-\boldsymbol{3}_{SU(2)_c}-\boldsymbol{3}_{SU(2)_s}\boldsymbol{3}_{SU(2)_c}-\boldsymbol{15}_{SU\left(4\right)_{v}}-1\right]pq+\ldots\,.
\end{equation*}

We see that this index is consistent with the $U(1)$ corresponding to the fugacity $\frac{a^{2}}{b^{2}}$ enhancing to $SU(2)$ somewhere on the conformal manifold. Defining 
\begin{equation}
\label{U1sSpin8}
U(1)_{e}=\frac{1}{4}\left[U(1)_{a}-U(1)_{b}\right]\,,\,\,\,\,U(1)_{h}=\frac{1}{2}\left[U(1)_{a}+U(1)_{b}\right]
\end{equation}
we get that the representations appearing in the index are consistent with $U(1)_{e}$ enhancing to $SU(2)_{e}$. Moreover, we get that $SU(2)_s\times SU(2)_c$ enhances to $\widetilde{SU}(4)$ as before. Rewriting the index using the representations of $SU(2)_{e}$ and $\widetilde{SU}(4)$, we obtain 
\begin{equation*}
I=1+\boldsymbol{6}_{SU\left(4\right)_{v}}\boldsymbol{2}_{SU\left(2\right)_{e}}\left(pq\right)^{\frac{1}{2}}+2\,\boldsymbol{4}_{\widetilde{SU}(4)}\boldsymbol{\bar{4}}_{SU\left(4\right)_{v}}h\left(pq\right)^{\frac{7}{12}}
\end{equation*}
\begin{equation*}
+\left[\boldsymbol{6}_{\widetilde{SU}(4)}\boldsymbol{2}_{SU\left(2\right)_{e}}h^{2}+h^{-4}\right]\left(pq\right)^{\frac{2}{3}}+\left[\boldsymbol{\overline{10}}_{SU\left(4\right)_{v}}+\boldsymbol{6}_{SU\left(4\right)_{v}}\right]h^{-2}\left(pq\right)^{\frac{5}{6}}
\end{equation*}
\begin{equation*}
+\boldsymbol{6}_{SU\left(4\right)_{v}}\boldsymbol{2}_{SU\left(2\right)_{e}}\left(p^{\frac{3}{2}}q^{\frac{1}{2}}+p^{\frac{1}{2}}q^{\frac{3}{2}}\right)+\left[\boldsymbol{20'}_{SU\left(4\right)_{v}}\boldsymbol{3}_{SU\left(2\right)_{e}}-\boldsymbol{15}_{\widetilde{SU}(4)}-\boldsymbol{15}_{SU\left(4\right)_{v}}-1\right]pq+\ldots\,.
\end{equation*}

As expected, we see that in this case there is a nonvanishing contribution at order $pq$ from marginal operators. Noticing that we have such operators which are of the forms $\left(S^{2}V^{2}\right)\left(C^{2}V^{2}\right)$, $\left(S^{2}V^{2}\right)^2$ and $\left(C^{2}V^{2}\right)^2$, {\it i.e.} (denoting by subscripts the charges under $U(1)_{a}$ and $U(1)_{b}$)
\begin{center}
	\begin{longtable}{|c|c|}
		\hline
		Operator & Representations\\
		\hline \hline
		$S^{2}V^{2}$ & $\left(\boldsymbol{1},\boldsymbol{1},\boldsymbol{1},\boldsymbol{6}\right)_{-2,2}$\\
		\hline 
		$C^{2}V^{2}$ & $\left(\boldsymbol{1},\boldsymbol{1},\boldsymbol{1},\boldsymbol{6}\right)_{2,-2}$\\
		\hline 
		$\left(S^{2}V^{2}\right)\left(C^{2}V^{2}\right)$ & $\left(\boldsymbol{1},\boldsymbol{1},\boldsymbol{1},\boldsymbol{1}\right)_{0,0},\,\,\left(\boldsymbol{1},\boldsymbol{1},\boldsymbol{1},\boldsymbol{15}\right)_{0,0},\,\,\left(\boldsymbol{1},\boldsymbol{1},\boldsymbol{1},\boldsymbol{20'}\right)_{0,0}$\\
		\hline 
		$\left(S^{2}V^{2}\right)^{2}$ & $\left(\boldsymbol{1},\boldsymbol{1},\boldsymbol{1},\boldsymbol{20'}\right)_{-4,4}$\\
		\hline 
		$\left(C^{2}V^{2}\right)^{2}$ & $\left(\boldsymbol{1},\boldsymbol{1},\boldsymbol{1},\boldsymbol{20'}\right)_{4,-4}$\\
		\hline
	\end{longtable}
\end{center}
and that the contribution to the index at order $pq$ comes from marginal operators (with a positive sign) and conserved currents (with a negative sign), we can write the contribution to the index at this order as 
\begin{equation}
\textrm{Marginals}-\textrm{Conserved currents}\,,
\end{equation}
where \footnote{Note that we do not assume that these marginal operator and conserved current contents take place everywhere on the conformal manifold. The statement is that it is consistent to happen somewhere on the manifold.}
\begin{equation}
\label{MarSpin8}
\textrm{Marginals}=\boldsymbol{20'}_{SU\left(4\right)_{v}}\left(a^{4}b^{-4}+a^{-4}b^{4}+1\right)+\boldsymbol{15}_{SU\left(4\right)_{v}}+1
\end{equation}
and
\begin{equation}
\label{ConCurrSpin8}
\textrm{Conserved currents}=\boldsymbol{15}_{\widetilde{SU}(4)}+2\,\boldsymbol{15}_{SU(4)_{v}}+2\,.
\end{equation}
Now, since the representations appearing in the index are consistent with $U(1)_e$ given in \eqref{U1sSpin8} enhancing to $SU(2)_e$ somewhere on the conformal manifold, it is natural to conjecture that this indeed happens; then, at this place where the symmetry enhances two more marginal operators appear that contribute $a^{4}b^{-4}+a^{-4}b^{4}$ to the index at order $pq$, along with two more conserved currents with the same contribution but with the opposite (minus) sign. As a result, at this place the marginal operator and conserved current contents are different from the ones written in Eqs. \eqref{MarSpin8} and \eqref{ConCurrSpin8}. Using representations of $SU(2)_e$, we have instead 
\begin{equation}
\textrm{Marginals}=\left(\boldsymbol{20'}_{SU(4)_{v}}+1\right)\boldsymbol{3}_{SU\left(2\right)_{e}}+\boldsymbol{15}_{SU\left(4\right)_{v}}
\end{equation}
and
\begin{equation}
\textrm{Conserved currents}=\boldsymbol{15}_{\widetilde{SU}(4)}+2\,\boldsymbol{15}_{SU(4)_{v}}+\boldsymbol{3}_{SU\left(2\right)_{e}}+1\,,
\end{equation}
which corresponds to the enhancement of the flavor symmetry $SU\left(2\right)^{2}\times SU\left(4\right)\times U\left(1\right)^{2}$ to $SU\left(4\right)^{3}\times SU(2)\times U\left(1\right)$ in the IR.

\

\subsection*{$\boldsymbol{Spin(9)}$}

\

The matter content is given in the following table, 
\begin{center}
	\begin{tabular}{|c||c|c|c|c|c|}
		\hline
		Field & $Spin\left(9\right)_{g}$ & $SU\left(2\right)_{s}$ & $SU\left(5\right)_{v}$ & $U\left(1\right)_{a}$ & $U\left(1\right)_{r}$\\
		\hline 
		$S$ & $\boldsymbol{16}$ & $\boldsymbol{2}$ & $\boldsymbol{1}$ & -5 & $\frac{1}{2}$\\
		
		$V$ & $\boldsymbol{9}$ & $\boldsymbol{1}$ & $\boldsymbol{5}$ & 4 & 0\\
		
		$M_{0}$ & $\boldsymbol{1}$ & $\boldsymbol{1}$ & $\boldsymbol{\overline{10}}$ & 2 & 1\\
		
		$M_{1}$ & $\boldsymbol{1}$ & $\boldsymbol{3}$ & $\boldsymbol{\overline{5}}$ & 6 & 1\\
		
		$M_{2}$ & $\boldsymbol{1}$ & $\boldsymbol{3}$ & $\boldsymbol{1}$ & 10 & 1\\
		
		$\Phi$ & $\boldsymbol{1}$ & $\boldsymbol{1}$ & $\boldsymbol{\overline{15}}$ & -8 & 2\\
		\hline
	\end{tabular}
\end{center}
The superpotential is given by 
\begin{equation}
W=M_{0}S^{2}V^{2}+M_{1}S^{2}V+M_{2}S^{2}+\Phi V^{2}
\end{equation}
and the superconformal R charge by ${\hat r}=r+0.0376q_{a}$. We see that $U\left(1\right)_{a}$ mixes with $U\left(1\right)_{r}$ in the expression for the IR R symmetry, and so the operators that contribute to the index at order $pq$ should have a vanishing charge under $U\left(1\right)_{a}$. These operators are of the forms
\begin{equation*}
\lambda\lambda,\,\,\,\,\bar{\psi}^{S}S,\,\,\,\,\bar{\psi}^{V}V,\,\,\,\,\bar{\psi}^{M_{0}}M_{0},\,\,\,\,\bar{\psi}^{M_{1}}M_{1},\,\,\,\,\bar{\psi}^{M_{2}}M_{2},\,\,\,\,\bar{\psi}^{\Phi}\Phi,\,\,\,\,M_{0}S^{2}V^{2},
\end{equation*}
\begin{equation*}
M_{1}S^{2}V,\,\,\,\,M_{2}S^{2},\,\,\,\,\Phi V^{2},\,\,\,\,M_{0}V\bar{\psi}^{M_{1}},\,\,\,\,M_{0}\bar{\psi}^{M_{2}}V^{2},\,\,\,\,M_{0}S^{2}\bar{\psi}^{\Phi},\,\,\,\,M_{0}\bar{\psi}^{M_{2}}\bar{\psi}^{\Phi},\,\,\,\,M_{1}\bar{\psi}^{M_{2}}V,
\end{equation*}
\begin{equation*}
S^{4}V^{5},\,\,\,\,S^{2}V^{3}\bar{\psi}^{M_{0}},\,\,\,\,V\left(\bar{\psi}^{M_{0}}\right)^{2},\,\,\,\,V^{2}\bar{\psi}^{M_{1}}\bar{\psi}^{M_{0}},\,\,\,\,\bar{\psi}^{\Phi}\bar{\psi}^{M_{1}}\bar{\psi}^{M_{0}},\,\,\,\,V^{3}\bar{\psi}^{M_{2}}\bar{\psi}^{M_{0}},\,\,\,\,V\bar{\psi}^{\Phi}\bar{\psi}^{M_{2}}\bar{\psi}^{M_{0}},
\end{equation*}
\begin{equation*}
S^{2}V\bar{\psi}^{\Phi}\bar{\psi}^{M_{0}},\,\,\,\,S^{2}V^{4}\bar{\psi}^{M_{1}},\,\,\,\,V^{3}\left(\bar{\psi}^{M_{1}}\right)^{2},\,\,\,\,V\bar{\psi}^{\Phi}\left(\bar{\psi}^{M_{1}}\right)^{2},\,\,\,\,V^{4}\bar{\psi}^{M_{2}}\bar{\psi}^{M_{1}},\,\,\,\,V^{2}\bar{\psi}^{\Phi}\bar{\psi}^{M_{2}}\bar{\psi}^{M_{1}},
\end{equation*}
\begin{equation*}
\left(\bar{\psi}^{\Phi}\right)^{2}\bar{\psi}^{M_{2}}\bar{\psi}^{M_{1}},\,\,\,\,S^{2}V^{2}\bar{\psi}^{\Phi}\bar{\psi}^{M_{1}},\,\,\,\,S^{2}\left(\bar{\psi}^{\Phi}\right)^{2}\bar{\psi}^{M_{1}},\,\,\,\,S^{2}V^{5}\bar{\psi}^{M_{2}},\,\,\,\,V^{5}\left(\bar{\psi}^{M_{2}}\right)^{2},\,\,\,\,V^{3}\bar{\psi}^{\Phi}\left(\bar{\psi}^{M_{2}}\right)^{2},
\end{equation*}
\begin{equation}
V\left(\bar{\psi}^{\Phi}\right)^{2}\left(\bar{\psi}^{M_{2}}\right)^{2},\,\,\,\,S^{2}V^{3}\bar{\psi}^{\Phi}\bar{\psi}^{M_{2}},\,\,\,\,S^{2}V\left(\bar{\psi}^{\Phi}\right)^{2}\bar{\psi}^{M_{2}},\,\,\,\,S^{4}V^{3}\bar{\psi}^{\Phi},\,\,\,\,S^{4}V\left(\bar{\psi}^{\Phi}\right)^{2}.
\label{listSpin9}
\end{equation}

Next, denoting the irreducible representations (and the corresponding characters) of the nonabelian groups in the theory by 
\begin{equation}
\left(Spin\left(9\right)_{g},SU\left(2\right)_{s},SU\left(5\right)_{v}\right),
\end{equation}
we find the following representations of the gauge singlets corresponding
to the operators in \eqref{listSpin9}, 
\begin{center}
	\begin{longtable}{|c|c|c|}
		\hline
		$\mathrm{Operator}$ & $\left(-1\right)^{F}$ & $\mathrm{Representations}\,\,(R)$\\
		\hline \hline 
		$\lambda\lambda$ & + & $\left(\boldsymbol{1},\boldsymbol{1},\boldsymbol{1}\right)$\tabularnewline
		\hline 
		$\bar{\psi}^{S}S$ & - & $\left(\boldsymbol{1},\boldsymbol{1},\boldsymbol{1}\right),\,\,\left(\boldsymbol{1},\boldsymbol{3},\boldsymbol{1}\right)$\tabularnewline
		\hline 
		$\bar{\psi}^{V}V$ & - & $\left(\boldsymbol{1},\boldsymbol{1},\boldsymbol{1}\right),\,\,\left(\boldsymbol{1},\boldsymbol{1},\boldsymbol{24}\right)$\tabularnewline
		\hline 
		$\bar{\psi}^{M_{0}}M_{0}$ & - & $\left(\boldsymbol{1},\boldsymbol{1},\boldsymbol{1}\right),\,\,\left(\boldsymbol{1},\boldsymbol{1},\boldsymbol{24}\right),\,\,\left(\boldsymbol{1},\boldsymbol{1},\boldsymbol{75}\right)$\tabularnewline
		\hline 
		$\bar{\psi}^{M_{1}}M_{1}$ & - & $\left(\boldsymbol{1},\boldsymbol{1},\boldsymbol{1}\right),\,\,\left(\boldsymbol{1},\boldsymbol{1},\boldsymbol{24}\right),\,\,\left(\boldsymbol{1},\boldsymbol{3},\boldsymbol{1}\right),\,\,\left(\boldsymbol{1},\boldsymbol{3},\boldsymbol{24}\right),\,\,\left(\boldsymbol{1},\boldsymbol{5},\boldsymbol{1}\right),\,\,\left(\boldsymbol{1},\boldsymbol{5},\boldsymbol{24}\right)$\tabularnewline
		\hline 
		$\bar{\psi}^{M_{2}}M_{2}$ & - & $\left(\boldsymbol{1},\boldsymbol{1},\boldsymbol{1}\right),\,\,\left(\boldsymbol{1},\boldsymbol{3},\boldsymbol{1}\right),\,\,\left(\boldsymbol{1},\boldsymbol{5},\boldsymbol{1}\right)$\tabularnewline
		\hline 
		$\bar{\psi}^{\Phi}\Phi$ & - & $\left(\boldsymbol{1},\boldsymbol{1},\boldsymbol{1}\right),\,\,\left(\boldsymbol{1},\boldsymbol{1},\boldsymbol{24}\right),\,\,\left(\boldsymbol{1},\boldsymbol{1},\boldsymbol{200}\right)$\tabularnewline
		\hline 
		$M_{0}S^{2}V^{2}$ & + & $\left(\boldsymbol{1},\boldsymbol{3},\boldsymbol{24}\right),\,\,\left(\boldsymbol{1},\boldsymbol{3},\boldsymbol{126}\right),\,\,\left(\boldsymbol{1},\boldsymbol{1},\boldsymbol{1}\right),\,\,\left(\boldsymbol{1},\boldsymbol{1},\boldsymbol{24}\right),\,\,\left(\boldsymbol{1},\boldsymbol{1},\boldsymbol{75}\right)$\tabularnewline
		\hline 
		$M_{1}S^{2}V$ & + & $\left(\boldsymbol{1},\boldsymbol{1},\boldsymbol{1}\right),\,\,\left(\boldsymbol{1},\boldsymbol{1},\boldsymbol{24}\right),\,\,\left(\boldsymbol{1},\boldsymbol{3},\boldsymbol{1}\right),\,\,\left(\boldsymbol{1},\boldsymbol{3},\boldsymbol{24}\right),\,\,\left(\boldsymbol{1},\boldsymbol{5},\boldsymbol{1}\right),\,\,\left(\boldsymbol{1},\boldsymbol{5},\boldsymbol{24}\right)$\tabularnewline
		\hline 
		$M_{2}S^{2}$ & + & $\left(\boldsymbol{1},\boldsymbol{1},\boldsymbol{1}\right),\,\,\left(\boldsymbol{1},\boldsymbol{3},\boldsymbol{1}\right),\,\,\left(\boldsymbol{1},\boldsymbol{5},\boldsymbol{1}\right)$\tabularnewline
		\hline 
		$\Phi V^{2}$ & + & $\left(\boldsymbol{1},\boldsymbol{1},\boldsymbol{1}\right),\,\,\left(\boldsymbol{1},\boldsymbol{1},\boldsymbol{24}\right),\,\,\left(\boldsymbol{1},\boldsymbol{1},\boldsymbol{200}\right)$\tabularnewline
		\hline 
		$M_{0}V\bar{\psi}^{M_{1}}$ & - & $\textrm{No gauge singlets}$\tabularnewline
		\hline 
		$M_{0}\bar{\psi}^{M_{2}}V^{2}$ & - & $\left(\boldsymbol{1},\boldsymbol{3},\boldsymbol{24}\right),\,\,\left(\boldsymbol{1},\boldsymbol{3},\boldsymbol{126}\right)$\tabularnewline
		\hline 
		$M_{0}S^{2}\bar{\psi}^{\Phi}$ & - & $\left(\boldsymbol{1},\boldsymbol{3},\boldsymbol{24}\right),\,\,\left(\boldsymbol{1},\boldsymbol{3},\boldsymbol{126}\right)$\tabularnewline
		\hline 
		$M_{0}\bar{\psi}^{M_{2}}\bar{\psi}^{\Phi}$ & + & $\left(\boldsymbol{1},\boldsymbol{3},\boldsymbol{24}\right),\,\,\left(\boldsymbol{1},\boldsymbol{3},\boldsymbol{126}\right)$\tabularnewline
		\hline 
		$M_{1}\bar{\psi}^{M_{2}}V$ & - & $\textrm{No gauge singlets}$\tabularnewline
		\hline 
		$S^{4}V^{5}$ & + & $\begin{array}{c}
		2\left(\boldsymbol{1},\boldsymbol{1},\boldsymbol{75}\right),\,\,\left(\boldsymbol{1},\boldsymbol{1},\boldsymbol{126'}\right),\,\,\left(\boldsymbol{1},\boldsymbol{1},\boldsymbol{175'}\right),\,\,\left(\boldsymbol{1},\boldsymbol{1},\boldsymbol{224}\right),\\
		2\left(\boldsymbol{1},\boldsymbol{3},\boldsymbol{75}\right),\,\,\left(\boldsymbol{1},\boldsymbol{3},\boldsymbol{126'}\right),\,\,2\left(\boldsymbol{1},\boldsymbol{3},\boldsymbol{175'}\right),\,\,2\left(\boldsymbol{1},\boldsymbol{3},\boldsymbol{224}\right),\\
		\left(\boldsymbol{1},\boldsymbol{5},\boldsymbol{75}\right),\,\,\left(\boldsymbol{1},\boldsymbol{5},\boldsymbol{126'}\right),\,\,\left(\boldsymbol{1},\boldsymbol{5},\boldsymbol{175'}\right),\,\,\left(\boldsymbol{1},\boldsymbol{5},\boldsymbol{224}\right),\\
		2\left(\boldsymbol{1},\boldsymbol{1},\boldsymbol{24}\right),\,\,\left(\boldsymbol{1},\boldsymbol{1},\boldsymbol{126}\right),\,\,3\left(\boldsymbol{1},\boldsymbol{3},\boldsymbol{24}\right),\,\,3\left(\boldsymbol{1},\boldsymbol{3},\boldsymbol{126}\right),\\
		2\left(\boldsymbol{1},\boldsymbol{1},\boldsymbol{1}\right),\,\,2\left(\boldsymbol{1},\boldsymbol{3},\boldsymbol{1}\right),\,\,\left(\boldsymbol{1},\boldsymbol{5},\boldsymbol{1}\right),\,\,\left(\boldsymbol{1},\boldsymbol{5},\boldsymbol{24}\right)
		\end{array}$\tabularnewline
		\hline 
		$S^{2}V^{3}\bar{\psi}^{M_{0}}$ & - & $\begin{array}{c}
		\left(\boldsymbol{1},\boldsymbol{3},\boldsymbol{24}\right),\,\,\left(\boldsymbol{1},\boldsymbol{3},\boldsymbol{75}\right),\,\,2\left(\boldsymbol{1},\boldsymbol{3},\boldsymbol{126}\right),\,\,\left(\boldsymbol{1},\boldsymbol{3},\boldsymbol{175'}\right),\,\,\left(\boldsymbol{1},\boldsymbol{3},\boldsymbol{224}\right),\\
		\left(\boldsymbol{1},\boldsymbol{1},\boldsymbol{1}\right),\,\,\left(\boldsymbol{1},\boldsymbol{1},\boldsymbol{24}\right),\,\,\left(\boldsymbol{1},\boldsymbol{1},\boldsymbol{75}\right)
		\end{array}$\tabularnewline
		\hline 
		$V\left(\bar{\psi}^{M_{0}}\right)^{2}$ & + & $\textrm{No gauge singlets}$\tabularnewline
		\hline 
		$V^{2}\bar{\psi}^{M_{1}}\bar{\psi}^{M_{0}}$ & + & $\left(\boldsymbol{1},\boldsymbol{3},\boldsymbol{24}\right),\,\,\left(\boldsymbol{1},\boldsymbol{3},\boldsymbol{75}\right),\,\,2\left(\boldsymbol{1},\boldsymbol{3},\boldsymbol{126}\right),\,\,\left(\boldsymbol{1},\boldsymbol{3},\boldsymbol{175'}\right),\,\,\left(\boldsymbol{1},\boldsymbol{3},\boldsymbol{224}\right)$\tabularnewline
		\hline 
		$\bar{\psi}^{\Phi}\bar{\psi}^{M_{1}}\bar{\psi}^{M_{0}}$ & - & $\left(\boldsymbol{1},\boldsymbol{3},\boldsymbol{24}\right),\,\,\left(\boldsymbol{1},\boldsymbol{3},\boldsymbol{75}\right),\,\,2\left(\boldsymbol{1},\boldsymbol{3},\boldsymbol{126}\right),\,\,\left(\boldsymbol{1},\boldsymbol{3},\boldsymbol{175'}\right),\,\,\left(\boldsymbol{1},\boldsymbol{3},\boldsymbol{224}\right)$\tabularnewline
		\hline 
		$V^{3}\bar{\psi}^{M_{2}}\bar{\psi}^{M_{0}}$ & + & $\textrm{No gauge singlets}$\tabularnewline
		\hline 
		$V\bar{\psi}^{\Phi}\bar{\psi}^{M_{2}}\bar{\psi}^{M_{0}}$ & - & $\textrm{No gauge singlets}$\tabularnewline
		\hline 
		$S^{2}V\bar{\psi}^{\Phi}\bar{\psi}^{M_{0}}$ & + & $\left(\boldsymbol{1},\boldsymbol{3},\boldsymbol{24}\right),\,\,\left(\boldsymbol{1},\boldsymbol{3},\boldsymbol{75}\right),\,\,2\left(\boldsymbol{1},\boldsymbol{3},\boldsymbol{126}\right),\,\,\left(\boldsymbol{1},\boldsymbol{3},\boldsymbol{175'}\right),\,\,\left(\boldsymbol{1},\boldsymbol{3},\boldsymbol{224}\right)$\tabularnewline
		\hline 
		$S^{2}V^{4}\bar{\psi}^{M_{1}}$ & - & $\begin{array}{c}
		\left(\boldsymbol{1},\boldsymbol{1},\boldsymbol{1}\right),\,\,\left(\boldsymbol{1},\boldsymbol{1},\boldsymbol{24}\right),\,\,\left(\boldsymbol{1},\boldsymbol{1},\boldsymbol{75}\right),\,\,\left(\boldsymbol{1},\boldsymbol{1},\boldsymbol{126'}\right),\,\,\left(\boldsymbol{1},\boldsymbol{1},\boldsymbol{175'}\right),\,\,\left(\boldsymbol{1},\boldsymbol{1},\boldsymbol{224}\right),\\
		\left(\boldsymbol{1},\boldsymbol{3},\boldsymbol{1}\right),\,\,\left(\boldsymbol{1},\boldsymbol{3},\boldsymbol{24}\right),\,\,\left(\boldsymbol{1},\boldsymbol{3},\boldsymbol{75}\right),\,\,\left(\boldsymbol{1},\boldsymbol{3},\boldsymbol{126'}\right),\,\,\left(\boldsymbol{1},\boldsymbol{3},\boldsymbol{175'}\right),\,\,\left(\boldsymbol{1},\boldsymbol{3},\boldsymbol{224}\right),\\
		\left(\boldsymbol{1},\boldsymbol{5},\boldsymbol{1}\right),\,\,\left(\boldsymbol{1},\boldsymbol{5},\boldsymbol{24}\right),\,\,\left(\boldsymbol{1},\boldsymbol{5},\boldsymbol{75}\right),\,\,\left(\boldsymbol{1},\boldsymbol{5},\boldsymbol{126'}\right),\,\,\left(\boldsymbol{1},\boldsymbol{5},\boldsymbol{175'}\right),\,\,\left(\boldsymbol{1},\boldsymbol{5},\boldsymbol{224}\right),\\
		\left(\boldsymbol{1},\boldsymbol{3},\boldsymbol{24}\right),\,\,\left(\boldsymbol{1},\boldsymbol{3},\boldsymbol{75}\right),\,\,2\left(\boldsymbol{1},\boldsymbol{3},\boldsymbol{126}\right),\,\,\left(\boldsymbol{1},\boldsymbol{3},\boldsymbol{175'}\right),\,\,\left(\boldsymbol{1},\boldsymbol{3},\boldsymbol{224}\right)
		\end{array}$\tabularnewline
		\hline 
		$V^{3}\left(\bar{\psi}^{M_{1}}\right)^{2}$ & + & $\textrm{No gauge singlets}$\tabularnewline
		\hline 
		$V\bar{\psi}^{\Phi}\left(\bar{\psi}^{M_{1}}\right)^{2}$ & - & $\textrm{No gauge singlets}$\tabularnewline
		\hline 
		$V^{4}\bar{\psi}^{M_{2}}\bar{\psi}^{M_{1}}$ & + & $\begin{array}{c}
		\left(\boldsymbol{1},\boldsymbol{1},\boldsymbol{75}\right),\,\,\left(\boldsymbol{1},\boldsymbol{1},\boldsymbol{126'}\right),\,\,\left(\boldsymbol{1},\boldsymbol{1},\boldsymbol{175'}\right),\,\,\left(\boldsymbol{1},\boldsymbol{1},\boldsymbol{224}\right),\\
		\left(\boldsymbol{1},\boldsymbol{3},\boldsymbol{75}\right),\,\,\left(\boldsymbol{1},\boldsymbol{3},\boldsymbol{126'}\right),\,\,\left(\boldsymbol{1},\boldsymbol{3},\boldsymbol{175'}\right),\,\,\left(\boldsymbol{1},\boldsymbol{3},\boldsymbol{224}\right),\\
		\left(\boldsymbol{1},\boldsymbol{5},\boldsymbol{75}\right),\,\,\left(\boldsymbol{1},\boldsymbol{5},\boldsymbol{126'}\right),\,\,\left(\boldsymbol{1},\boldsymbol{5},\boldsymbol{175'}\right),\,\,\left(\boldsymbol{1},\boldsymbol{5},\boldsymbol{224}\right)
		\end{array}$\tabularnewline
		\hline 
		$V^{2}\bar{\psi}^{\Phi}\bar{\psi}^{M_{2}}\bar{\psi}^{M_{1}}$ & - & $\begin{array}{c}
		\left(\boldsymbol{1},\boldsymbol{1},\boldsymbol{75}\right),\,\,\left(\boldsymbol{1},\boldsymbol{1},\boldsymbol{126}\right),\,\,\left(\boldsymbol{1},\boldsymbol{1},\boldsymbol{126'}\right),\,\,2\left(\boldsymbol{1},\boldsymbol{1},\boldsymbol{175'}\right),\,\,2\left(\boldsymbol{1},\boldsymbol{1},\boldsymbol{224}\right),\\
		\left(\boldsymbol{1},\boldsymbol{3},\boldsymbol{75}\right),\,\,\left(\boldsymbol{1},\boldsymbol{3},\boldsymbol{126}\right),\,\,\left(\boldsymbol{1},\boldsymbol{3},\boldsymbol{126'}\right),\,\,2\left(\boldsymbol{1},\boldsymbol{3},\boldsymbol{175'}\right),\,\,2\left(\boldsymbol{1},\boldsymbol{3},\boldsymbol{224}\right),\\
		\left(\boldsymbol{1},\boldsymbol{5},\boldsymbol{75}\right),\,\,\left(\boldsymbol{1},\boldsymbol{5},\boldsymbol{126}\right),\,\,\left(\boldsymbol{1},\boldsymbol{5},\boldsymbol{126'}\right),\,\,2\left(\boldsymbol{1},\boldsymbol{5},\boldsymbol{175'}\right),\,\,2\left(\boldsymbol{1},\boldsymbol{5},\boldsymbol{224}\right)
		\end{array}$\tabularnewline
		\hline 
		$\left(\bar{\psi}^{\Phi}\right)^{2}\bar{\psi}^{M_{2}}\bar{\psi}^{M_{1}}$ & + & $\begin{array}{c}
		\left(\boldsymbol{1},\boldsymbol{1},\boldsymbol{126}\right),\,\,\left(\boldsymbol{1},\boldsymbol{1},\boldsymbol{175'}\right),\,\,\left(\boldsymbol{1},\boldsymbol{1},\boldsymbol{224}\right),\\
		\left(\boldsymbol{1},\boldsymbol{3},\boldsymbol{126}\right),\,\,\left(\boldsymbol{1},\boldsymbol{3},\boldsymbol{175'}\right),\,\,\left(\boldsymbol{1},\boldsymbol{3},\boldsymbol{224}\right),\\
		\left(\boldsymbol{1},\boldsymbol{5},\boldsymbol{126}\right),\,\,\left(\boldsymbol{1},\boldsymbol{5},\boldsymbol{175'}\right),\,\,\left(\boldsymbol{1},\boldsymbol{5},\boldsymbol{224}\right)
		\end{array}$\tabularnewline
		\hline 
		$S^{2}V^{2}\bar{\psi}^{\Phi}\bar{\psi}^{M_{1}}$ & + & $\begin{array}{c}
		\left(\boldsymbol{1},\boldsymbol{1},\boldsymbol{75}\right),\,\,\left(\boldsymbol{1},\boldsymbol{1},\boldsymbol{126}\right),\,\,\left(\boldsymbol{1},\boldsymbol{1},\boldsymbol{126'}\right),\,\,2\left(\boldsymbol{1},\boldsymbol{1},\boldsymbol{175'}\right),\,\,2\left(\boldsymbol{1},\boldsymbol{1},\boldsymbol{224}\right),\\
		\left(\boldsymbol{1},\boldsymbol{3},\boldsymbol{75}\right),\,\,\left(\boldsymbol{1},\boldsymbol{3},\boldsymbol{126}\right),\,\,\left(\boldsymbol{1},\boldsymbol{3},\boldsymbol{126'}\right),\,\,2\left(\boldsymbol{1},\boldsymbol{3},\boldsymbol{175'}\right),\,\,2\left(\boldsymbol{1},\boldsymbol{3},\boldsymbol{224}\right),\\
		\left(\boldsymbol{1},\boldsymbol{5},\boldsymbol{75}\right),\,\,\left(\boldsymbol{1},\boldsymbol{5},\boldsymbol{126}\right),\,\,\left(\boldsymbol{1},\boldsymbol{5},\boldsymbol{126'}\right),\,\,2\left(\boldsymbol{1},\boldsymbol{5},\boldsymbol{175'}\right),\,\,2\left(\boldsymbol{1},\boldsymbol{5},\boldsymbol{224}\right)\\
		\left(\boldsymbol{1},\boldsymbol{3},\boldsymbol{24}\right),\,\,\left(\boldsymbol{1},\boldsymbol{3},\boldsymbol{75}\right),\,\,2\left(\boldsymbol{1},\boldsymbol{3},\boldsymbol{126}\right),\,\,\left(\boldsymbol{1},\boldsymbol{3},\boldsymbol{175'}\right),\,\,\left(\boldsymbol{1},\boldsymbol{3},\boldsymbol{224}\right)
		\end{array}$\tabularnewline
		\hline 
		$S^{2}\left(\bar{\psi}^{\Phi}\right)^{2}\bar{\psi}^{M_{1}}$ & - & $\begin{array}{c}
		\left(\boldsymbol{1},\boldsymbol{1},\boldsymbol{126}\right),\,\,\left(\boldsymbol{1},\boldsymbol{1},\boldsymbol{175'}\right),\,\,\left(\boldsymbol{1},\boldsymbol{1},\boldsymbol{224}\right),\\
		\left(\boldsymbol{1},\boldsymbol{3},\boldsymbol{126}\right),\,\,\left(\boldsymbol{1},\boldsymbol{3},\boldsymbol{175'}\right),\,\,\left(\boldsymbol{1},\boldsymbol{3},\boldsymbol{224}\right),\\
		\left(\boldsymbol{1},\boldsymbol{5},\boldsymbol{126}\right),\,\,\left(\boldsymbol{1},\boldsymbol{5},\boldsymbol{175'}\right),\,\,\left(\boldsymbol{1},\boldsymbol{5},\boldsymbol{224}\right)
		\end{array}$\tabularnewline
		\hline 
		$S^{2}V^{5}\bar{\psi}^{M_{2}}$ & - & $\begin{array}{c}
		\left(\boldsymbol{1},\boldsymbol{1},\boldsymbol{1}\right),\,\,\left(\boldsymbol{1},\boldsymbol{1},\boldsymbol{75}\right),\,\,\left(\boldsymbol{1},\boldsymbol{1},\boldsymbol{126'}\right),\,\,\left(\boldsymbol{1},\boldsymbol{1},\boldsymbol{175'}\right),\,\,\left(\boldsymbol{1},\boldsymbol{1},\boldsymbol{224}\right),\\
		\left(\boldsymbol{1},\boldsymbol{3},\boldsymbol{1}\right),\,\,\left(\boldsymbol{1},\boldsymbol{3},\boldsymbol{75}\right),\,\,\left(\boldsymbol{1},\boldsymbol{3},\boldsymbol{126'}\right),\,\,\left(\boldsymbol{1},\boldsymbol{3},\boldsymbol{175'}\right),\,\,\left(\boldsymbol{1},\boldsymbol{3},\boldsymbol{224}\right),\\
		\left(\boldsymbol{1},\boldsymbol{5},\boldsymbol{1}\right),\,\,\left(\boldsymbol{1},\boldsymbol{5},\boldsymbol{75}\right),\,\,\left(\boldsymbol{1},\boldsymbol{5},\boldsymbol{126'}\right),\,\,\left(\boldsymbol{1},\boldsymbol{5},\boldsymbol{175'}\right),\,\,\left(\boldsymbol{1},\boldsymbol{5},\boldsymbol{224}\right),\\
		\begin{array}{c}
		\left(\boldsymbol{1},\boldsymbol{3},\boldsymbol{24}\right),\,\,\left(\boldsymbol{1},\boldsymbol{3},\boldsymbol{126}\right)\end{array}
		\end{array}$\tabularnewline
		\hline 
		$V^{5}\left(\bar{\psi}^{M_{2}}\right)^{2}$ & + & $\textrm{No gauge singlets}$\tabularnewline
		\hline 
		$V^{3}\bar{\psi}^{\Phi}\left(\bar{\psi}^{M_{2}}\right)^{2}$ & - & $\textrm{No gauge singlets}$\tabularnewline
		\hline 
		$V\left(\bar{\psi}^{\Phi}\right)^{2}\left(\bar{\psi}^{M_{2}}\right)^{2}$ & + & $\textrm{No gauge singlets}$\tabularnewline
		\hline
		$S^{2}V^{3}\bar{\psi}^{\Phi}\bar{\psi}^{M_{2}}$ & + & $\begin{array}{c}
		\left(\boldsymbol{1},\boldsymbol{1},\boldsymbol{75}\right),\,\,\left(\boldsymbol{1},\boldsymbol{1},\boldsymbol{126}\right),\,\,\left(\boldsymbol{1},\boldsymbol{1},\boldsymbol{126'}\right),\,\,2\left(\boldsymbol{1},\boldsymbol{1},\boldsymbol{175'}\right),\,\,2\left(\boldsymbol{1},\boldsymbol{1},\boldsymbol{224}\right),\\
		\left(\boldsymbol{1},\boldsymbol{3},\boldsymbol{75}\right),\,\,\left(\boldsymbol{1},\boldsymbol{3},\boldsymbol{126}\right),\,\,\left(\boldsymbol{1},\boldsymbol{3},\boldsymbol{126'}\right),\,\,2\left(\boldsymbol{1},\boldsymbol{3},\boldsymbol{175'}\right),\,\,2\left(\boldsymbol{1},\boldsymbol{3},\boldsymbol{224}\right),\\
		\left(\boldsymbol{1},\boldsymbol{5},\boldsymbol{75}\right),\,\,\left(\boldsymbol{1},\boldsymbol{5},\boldsymbol{126}\right),\,\,\left(\boldsymbol{1},\boldsymbol{5},\boldsymbol{126'}\right),\,\,2\left(\boldsymbol{1},\boldsymbol{5},\boldsymbol{175'}\right),\,\,2\left(\boldsymbol{1},\boldsymbol{5},\boldsymbol{224}\right)\\
		\left(\boldsymbol{1},\boldsymbol{3},\boldsymbol{24}\right),\,\,\left(\boldsymbol{1},\boldsymbol{3},\boldsymbol{126}\right)
		\end{array}$\tabularnewline
		\hline 
		$S^{2}V\left(\bar{\psi}^{\Phi}\right)^{2}\bar{\psi}^{M_{2}}$ & - & $\begin{array}{c}
		\left(\boldsymbol{1},\boldsymbol{1},\boldsymbol{126}\right),\,\,\left(\boldsymbol{1},\boldsymbol{1},\boldsymbol{175'}\right),\,\,\left(\boldsymbol{1},\boldsymbol{1},\boldsymbol{224}\right),\\
		\left(\boldsymbol{1},\boldsymbol{3},\boldsymbol{126}\right),\,\,\left(\boldsymbol{1},\boldsymbol{3},\boldsymbol{175'}\right),\,\,\left(\boldsymbol{1},\boldsymbol{3},\boldsymbol{224}\right),\\
		\left(\boldsymbol{1},\boldsymbol{5},\boldsymbol{126}\right),\,\,\left(\boldsymbol{1},\boldsymbol{5},\boldsymbol{175'}\right),\,\,\left(\boldsymbol{1},\boldsymbol{5},\boldsymbol{224}\right)
		\end{array}$\tabularnewline
		\hline 
		$S^{4}V^{3}\bar{\psi}^{\Phi}$ & - & $\begin{array}{c}
		\left(\boldsymbol{1},\boldsymbol{1},\boldsymbol{75}\right),\,\,\left(\boldsymbol{1},\boldsymbol{1},\boldsymbol{126}\right),\,\,\left(\boldsymbol{1},\boldsymbol{1},\boldsymbol{126'}\right),\,\,2\left(\boldsymbol{1},\boldsymbol{1},\boldsymbol{175'}\right),\,\,2\left(\boldsymbol{1},\boldsymbol{1},\boldsymbol{224}\right),\\
		\left(\boldsymbol{1},\boldsymbol{3},\boldsymbol{75}\right),\,\,\left(\boldsymbol{1},\boldsymbol{3},\boldsymbol{126}\right),\,\,\left(\boldsymbol{1},\boldsymbol{3},\boldsymbol{126'}\right),\,\,2\left(\boldsymbol{1},\boldsymbol{3},\boldsymbol{175'}\right),\,\,2\left(\boldsymbol{1},\boldsymbol{3},\boldsymbol{224}\right),\\
		\left(\boldsymbol{1},\boldsymbol{5},\boldsymbol{75}\right),\,\,\left(\boldsymbol{1},\boldsymbol{5},\boldsymbol{126}\right),\,\,\left(\boldsymbol{1},\boldsymbol{5},\boldsymbol{126'}\right),\,\,2\left(\boldsymbol{1},\boldsymbol{5},\boldsymbol{175'}\right),\,\,2\left(\boldsymbol{1},\boldsymbol{5},\boldsymbol{224}\right)\\
		2\left(\boldsymbol{1},\boldsymbol{3},\boldsymbol{24}\right),\,\,\left(\boldsymbol{1},\boldsymbol{3},\boldsymbol{75}\right),\,\,3\left(\boldsymbol{1},\boldsymbol{3},\boldsymbol{126}\right),\,\,\left(\boldsymbol{1},\boldsymbol{3},\boldsymbol{175'}\right),\,\,\left(\boldsymbol{1},\boldsymbol{3},\boldsymbol{224}\right)\\
		\left(\boldsymbol{1},\boldsymbol{1},\boldsymbol{24}\right),\,\,\left(\boldsymbol{1},\boldsymbol{1},\boldsymbol{126}\right)
		\end{array}$\tabularnewline
		\hline 
		$S^{4}V\left(\bar{\psi}^{\Phi}\right)^{2}$ & + & $\begin{array}{c}
		\left(\boldsymbol{1},\boldsymbol{1},\boldsymbol{126}\right),\,\,\left(\boldsymbol{1},\boldsymbol{1},\boldsymbol{175'}\right),\,\,\left(\boldsymbol{1},\boldsymbol{1},\boldsymbol{224}\right),\\
		\left(\boldsymbol{1},\boldsymbol{3},\boldsymbol{126}\right),\,\,\left(\boldsymbol{1},\boldsymbol{3},\boldsymbol{175'}\right),\,\,\left(\boldsymbol{1},\boldsymbol{3},\boldsymbol{224}\right),\\
		\left(\boldsymbol{1},\boldsymbol{5},\boldsymbol{126}\right),\,\,\left(\boldsymbol{1},\boldsymbol{5},\boldsymbol{175'}\right),\,\,\left(\boldsymbol{1},\boldsymbol{5},\boldsymbol{224}\right)
		\end{array}$\tabularnewline
		\hline
	\end{longtable}
\end{center}
Summing (with signs) the above characters, we obtain 
\begin{equation*}
\sum R\left(-1\right)^{F}=-\left(\boldsymbol{1},\boldsymbol{3},\boldsymbol{1}\right)-\left(\boldsymbol{1},\boldsymbol{5},\boldsymbol{1}\right)-2\begin{array}{c}
\left(\boldsymbol{1},\boldsymbol{1},\boldsymbol{24}\right)\end{array}-2\begin{array}{c}
\left(\boldsymbol{1},\boldsymbol{1},\boldsymbol{1}\right)\end{array}
\end{equation*}
\begin{equation}
=-\boldsymbol{8}_{SU(3)}-2\,\boldsymbol{24}_{SU(5)_v}-2
\end{equation}
which corresponds to the enhancement of the flavor symmetry $SU\left(2\right)\times SU\left(5\right)\times U\left(1\right)$ to $SU\left(3\right)\times SU\left(5\right)^{2}\times U\left(1\right)^{2}$ in the IR. Moreover, the dimension of the conformal manifold vanishes. 

\

\subsection*{$\boldsymbol{Spin(10)}$}

\

The matter content is given in the following table, 
\begin{center}
	\begin{tabular}{|c||c|c|c|c|c|}
		\hline
		Field  & $Spin\left(10\right)_{g}$ & $SU\left(2\right)_{s}$ & $SU\left(6\right)_{v}$ & $U\left(1\right)_{a}$ & $U\left(1\right)_{r}$\\
		\hline
		$S$ & $\boldsymbol{16}$ & $\boldsymbol{2}$ & $\boldsymbol{1}$ & 3 & $\frac{1}{2}$\\
		
		$V$ & $\boldsymbol{10}$ & $\boldsymbol{1}$ & $\boldsymbol{6}$ & -2 & 0\\
		
		$M$ & $\boldsymbol{1}$ & $\boldsymbol{3}$ & $\boldsymbol{\overline{6}}$ & -4 & 1\\
		
		$\Phi$ & $\boldsymbol{1}$ & $\boldsymbol{1}$ & $\boldsymbol{\overline{21}}$ & 4 & 2\\
		\hline
	\end{tabular}
\end{center}
The superpotential is given by 
\begin{equation}
W=MS^{2}V+\Phi V^{2}
\end{equation}
and the superconformal R charge by ${\hat r}=r-0.076q_{a}$. We see that $U\left(1\right)_{a}$ mixes with $U\left(1\right)_{r}$ in the expression for the IR R symmetry, and so the operators that contribute to the index at order $pq$ should have a vanishing charge under $U\left(1\right)_{a}$. These operators are of the forms
\begin{equation*}
\lambda\lambda,\,\,\,\,\bar{\psi}^{S}S,\,\,\,\,\bar{\psi}^{V}V,\,\,\,\,\bar{\psi}^{M}M,\,\,\,\,\bar{\psi}^{\Phi}\Phi,\,\,\,\,MS^{2}V,\,\,\,\,\Phi V^{2},\,\,\,\,S^{4}V^{6},
\end{equation*}
\begin{equation*}
S^{2}V^{5}\bar{\psi}^{M},\,\,\,\,V^{4}\left(\bar{\psi}^{M}\right)^{2},\,\,\,\,V^{2}\bar{\psi}^{\Phi}\left(\bar{\psi}^{M}\right)^{2},\,\,\,\,\left(\bar{\psi}^{\Phi}\right)^{2}\left(\bar{\psi}^{M}\right)^{2},\,\,\,\,S^{2}V^{3}\bar{\psi}^{\Phi}\bar{\psi}^{M},
\end{equation*}
\begin{equation}
S^{2}V\left(\bar{\psi}^{\Phi}\right)^{2}\bar{\psi}^{M},\,\,\,\,S^{4}V^{4}\bar{\psi}^{\Phi},\,\,\,\,S^{4}V^{2}\left(\bar{\psi}^{\Phi}\right)^{2},\,\,\,\,S^{4}\left(\bar{\psi}^{\Phi}\right)^{3}.
\label{listSpin10}
\end{equation}

Next, denoting the irreducible representations (and the corresponding characters) of the nonabelian groups in the theory by 
\begin{equation}
\left(Spin\left(10\right)_{g},SU\left(2\right)_{s},SU\left(6\right)_{v}\right),
\end{equation}
we find the following representations of the gauge singlets corresponding
to the operators in \eqref{listSpin10}, 
\begin{center}
	\begin{longtable}{|c|c|c|}
		\hline
		$\mathrm{Operator}$ & $\left(-1\right)^{F}$ & $\mathrm{Representations}\,\,(R)$\\
		\hline \hline 
		$\lambda\lambda$ & + & $\left(\boldsymbol{1},\boldsymbol{1},\boldsymbol{1}\right)$\tabularnewline
		\hline 
		$\bar{\psi}^{S}S$ & - & $\left(\boldsymbol{1},\boldsymbol{1},\boldsymbol{1}\right),\,\,\left(\boldsymbol{1},\boldsymbol{3},\boldsymbol{1}\right)$\tabularnewline
		\hline 
		$\bar{\psi}^{V}V$ & - & $\left(\boldsymbol{1},\boldsymbol{1},\boldsymbol{1}\right),\,\,\left(\boldsymbol{1},\boldsymbol{1},\boldsymbol{35}\right)$\tabularnewline
		\hline 
		$\bar{\psi}^{M}M$ & - & $\left(\boldsymbol{1},\boldsymbol{1},\boldsymbol{1}\right),\,\,\left(\boldsymbol{1},\boldsymbol{1},\boldsymbol{35}\right),\,\,\left(\boldsymbol{1},\boldsymbol{3},\boldsymbol{1}\right),\,\,\left(\boldsymbol{1},\boldsymbol{3},\boldsymbol{35}\right),\,\,\left(\boldsymbol{1},\boldsymbol{5},\boldsymbol{1}\right),\,\,\left(\boldsymbol{1},\boldsymbol{5},\boldsymbol{35}\right)$\tabularnewline
		\hline 
		$\bar{\psi}^{\Phi}\Phi$ & - & $\left(\boldsymbol{1},\boldsymbol{1},\boldsymbol{1}\right),\,\,\left(\boldsymbol{1},\boldsymbol{1},\boldsymbol{35}\right),\,\,\left(\boldsymbol{1},\boldsymbol{1},\boldsymbol{405}\right)$\tabularnewline
		\hline 
		$\Phi V^{2}$ & + & $\left(\boldsymbol{1},\boldsymbol{1},\boldsymbol{1}\right),\,\,\left(\boldsymbol{1},\boldsymbol{1},\boldsymbol{35}\right),\,\,\left(\boldsymbol{1},\boldsymbol{1},\boldsymbol{405}\right)$\tabularnewline
		\hline 
		$MS^{2}V$ & + & $\left(\boldsymbol{1},\boldsymbol{1},\boldsymbol{1}\right),\,\,\left(\boldsymbol{1},\boldsymbol{1},\boldsymbol{35}\right),\,\,\left(\boldsymbol{1},\boldsymbol{3},\boldsymbol{1}\right),\,\,\left(\boldsymbol{1},\boldsymbol{3},\boldsymbol{35}\right),\,\,\left(\boldsymbol{1},\boldsymbol{5},\boldsymbol{1}\right),\,\,\left(\boldsymbol{1},\boldsymbol{5},\boldsymbol{35}\right)$\tabularnewline
		\hline 
		$S^{4}V^{6}$ & + & $\begin{array}{c}
		\left(\boldsymbol{1},\boldsymbol{1},\boldsymbol{1}\right),\,\,2\left(\boldsymbol{1},\boldsymbol{1},\boldsymbol{35}\right),\,\,\left(\boldsymbol{1},\boldsymbol{1},\boldsymbol{280}\right),\,\,2\left(\boldsymbol{1},\boldsymbol{1},\boldsymbol{462}\right),\\
		3\left(\boldsymbol{1},\boldsymbol{1},\boldsymbol{175}\right),\,\,3\left(\boldsymbol{1},\boldsymbol{1},\boldsymbol{1134'}\right),\,\,\left(\boldsymbol{1},\boldsymbol{1},\boldsymbol{896}\right),\\
		\left(\boldsymbol{1},\boldsymbol{1},\boldsymbol{1050''}\right),\,\,2\left(\boldsymbol{1},\boldsymbol{3},\boldsymbol{189}\right),\,\,\left(\boldsymbol{1},\boldsymbol{3},\boldsymbol{490}\right),\,\,2\left(\boldsymbol{1},\boldsymbol{3},\boldsymbol{840''}\right),\\
		2\left(\boldsymbol{1},\boldsymbol{3},\boldsymbol{896}\right),\,\,\left(\boldsymbol{1},\boldsymbol{3},\boldsymbol{1050''}\right),\,\,\left(\boldsymbol{1},\boldsymbol{5},\boldsymbol{175}\right),\\
		\left(\boldsymbol{1},\boldsymbol{5},\boldsymbol{462}\right),\,\,\left(\boldsymbol{1},\boldsymbol{5},\boldsymbol{896}\right),\,\,\left(\boldsymbol{1},\boldsymbol{5},\boldsymbol{1050''}\right),\,\,2\left(\boldsymbol{1},\boldsymbol{5},\boldsymbol{1134'}\right),\\
		\left(\boldsymbol{1},\boldsymbol{3},\boldsymbol{1}\right),\,\,2\left(\boldsymbol{1},\boldsymbol{3},\boldsymbol{35}\right),\,\,2\left(\boldsymbol{1},\boldsymbol{3},\boldsymbol{280}\right),\,\,\left(\boldsymbol{1},\boldsymbol{5},\boldsymbol{35}\right)
		\end{array}$\tabularnewline
		\hline 
		$S^{2}V^{5}\bar{\psi}^{M}$ & - & $\begin{array}{c}
		\left(\boldsymbol{1},\boldsymbol{1},\boldsymbol{1}\right),\,\,\left(\boldsymbol{1},\boldsymbol{1},\boldsymbol{35}\right),\,\,\left(\boldsymbol{1},\boldsymbol{1},\boldsymbol{175}\right),\,\,\left(\boldsymbol{1},\boldsymbol{1},\boldsymbol{189}\right),\\
		\left(\boldsymbol{1},\boldsymbol{1},\boldsymbol{462}\right),\,\,\left(\boldsymbol{1},\boldsymbol{1},\boldsymbol{490}\right),\,\,\left(\boldsymbol{1},\boldsymbol{1},\boldsymbol{840''}\right),\,\,2\left(\boldsymbol{1},\boldsymbol{1},\boldsymbol{896}\right),\\
		2\left(\boldsymbol{1},\boldsymbol{1},\boldsymbol{1050''}\right),\,\,2\left(\boldsymbol{1},\boldsymbol{1},\boldsymbol{1134'}\right),\,\,\left(\boldsymbol{1},\boldsymbol{3},\boldsymbol{1}\right),\\
		\left(\boldsymbol{1},\boldsymbol{3},\boldsymbol{35}\right),\,\,\left(\boldsymbol{1},\boldsymbol{3},\boldsymbol{175}\right),\,\,\left(\boldsymbol{1},\boldsymbol{3},\boldsymbol{189}\right),\,\,\left(\boldsymbol{1},\boldsymbol{3},\boldsymbol{462}\right),\\
		\left(\boldsymbol{1},\boldsymbol{3},\boldsymbol{490}\right),\,\,\left(\boldsymbol{1},\boldsymbol{3},\boldsymbol{840''}\right),\,\,2\left(\boldsymbol{1},\boldsymbol{3},\boldsymbol{896}\right),\,\,2\left(\boldsymbol{1},\boldsymbol{3},\boldsymbol{1050''}\right),\\
		2\left(\boldsymbol{1},\boldsymbol{3},\boldsymbol{1134'}\right),\,\,\left(\boldsymbol{1},\boldsymbol{5},\boldsymbol{1}\right),\,\,\left(\boldsymbol{1},\boldsymbol{5},\boldsymbol{35}\right),\,\,\left(\boldsymbol{1},\boldsymbol{5},\boldsymbol{175}\right),\\
		\left(\boldsymbol{1},\boldsymbol{5},\boldsymbol{189}\right),\,\,\left(\boldsymbol{1},\boldsymbol{5},\boldsymbol{462}\right),\,\,\left(\boldsymbol{1},\boldsymbol{5},\boldsymbol{490}\right),\,\,\left(\boldsymbol{1},\boldsymbol{5},\boldsymbol{840''}\right),\\
		2\left(\boldsymbol{1},\boldsymbol{5},\boldsymbol{896}\right),\,\,2\left(\boldsymbol{1},\boldsymbol{5},\boldsymbol{1050''}\right),\,\,2\left(\boldsymbol{1},\boldsymbol{5},\boldsymbol{1134'}\right),\,\,\left(\boldsymbol{1},\boldsymbol{3},\boldsymbol{35}\right),\\
		\left(\boldsymbol{1},\boldsymbol{3},\boldsymbol{189}\right),\,\,2\left(\boldsymbol{1},\boldsymbol{3},\boldsymbol{280}\right),\,\,\left(\boldsymbol{1},\boldsymbol{3},\boldsymbol{840''}\right),\,\,\left(\boldsymbol{1},\boldsymbol{3},\boldsymbol{896}\right)
		\end{array}$\tabularnewline
		\hline 
		$V^{4}\left(\bar{\psi}^{M}\right)^{2}$ & + & $\begin{array}{c}
		\left(\boldsymbol{1},\boldsymbol{1},\boldsymbol{189}\right),\,\,\left(\boldsymbol{1},\boldsymbol{1},\boldsymbol{490}\right),\,\,\left(\boldsymbol{1},\boldsymbol{1},\boldsymbol{840''}\right),\\
		\left(\boldsymbol{1},\boldsymbol{1},\boldsymbol{896}\right),\,\,\left(\boldsymbol{1},\boldsymbol{1},\boldsymbol{1050''}\right),\,\,\left(\boldsymbol{1},\boldsymbol{3},\boldsymbol{175}\right),\,\,\left(\boldsymbol{1},\boldsymbol{3},\boldsymbol{462}\right),\\
		\left(\boldsymbol{1},\boldsymbol{3},\boldsymbol{896}\right),\,\,\left(\boldsymbol{1},\boldsymbol{3},\boldsymbol{1050''}\right),\,\,2\left(\boldsymbol{1},\boldsymbol{3},\boldsymbol{1134'}\right),\,\,\left(\boldsymbol{1},\boldsymbol{5},\boldsymbol{189}\right),\\
		\left(\boldsymbol{1},\boldsymbol{5},\boldsymbol{490}\right),\,\,\left(\boldsymbol{1},\boldsymbol{5},\boldsymbol{840''}\right),\,\,\left(\boldsymbol{1},\boldsymbol{5},\boldsymbol{896}\right),\,\,\left(\boldsymbol{1},\boldsymbol{5},\boldsymbol{1050''}\right)
		\end{array}$\tabularnewline
		\hline 
		$V^{2}\bar{\psi}^{\Phi}\left(\bar{\psi}^{M}\right)^{2}$ & - & $\begin{array}{c}
		\left(\boldsymbol{1},\boldsymbol{1},\boldsymbol{189}\right),\,\,\left(\boldsymbol{1},\boldsymbol{1},\boldsymbol{280}\right),\\
		\left(\boldsymbol{1},\boldsymbol{1},\boldsymbol{490}\right),\,\,2\left(\boldsymbol{1},\boldsymbol{1},\boldsymbol{840''}\right),\,\,2\left(\boldsymbol{1},\boldsymbol{1},\boldsymbol{896}\right),\\
		\left(\boldsymbol{1},\boldsymbol{1},\boldsymbol{1050''}\right),\,\,\left(\boldsymbol{1},\boldsymbol{1},\boldsymbol{1134'}\right),\,\,\left(\boldsymbol{1},\boldsymbol{3},\boldsymbol{175}\right),\,\,\left(\boldsymbol{1},\boldsymbol{3},\boldsymbol{462}\right),\\
		\left(\boldsymbol{1},\boldsymbol{3},\boldsymbol{490}\right),\,\,\left(\boldsymbol{1},\boldsymbol{3},\boldsymbol{840''}\right),\,\,2\left(\boldsymbol{1},\boldsymbol{3},\boldsymbol{896}\right),\,\,2\left(\boldsymbol{1},\boldsymbol{3},\boldsymbol{1050''}\right),\\
		3\left(\boldsymbol{1},\boldsymbol{3},\boldsymbol{1134'}\right),\,\,\left(\boldsymbol{1},\boldsymbol{5},\boldsymbol{189}\right),\,\,\left(\boldsymbol{1},\boldsymbol{5},\boldsymbol{280}\right),\,\,\left(\boldsymbol{1},\boldsymbol{5},\boldsymbol{490}\right),\\
		2\left(\boldsymbol{1},\boldsymbol{5},\boldsymbol{840''}\right),\,\,2\left(\boldsymbol{1},\boldsymbol{5},\boldsymbol{896}\right),\,\,\left(\boldsymbol{1},\boldsymbol{5},\boldsymbol{1050''}\right),\,\,\left(\boldsymbol{1},\boldsymbol{5},\boldsymbol{1134'}\right)
		\end{array}$\tabularnewline
		\hline 
		$\left(\bar{\psi}^{\Phi}\right)^{2}\left(\bar{\psi}^{M}\right)^{2}$ & + & $\begin{array}{c}
		\left(\boldsymbol{1},\boldsymbol{1},\boldsymbol{280}\right),\,\,\left(\boldsymbol{1},\boldsymbol{1},\boldsymbol{840''}\right),\,\,\left(\boldsymbol{1},\boldsymbol{1},\boldsymbol{896}\right),\,\,\left(\boldsymbol{1},\boldsymbol{1},\boldsymbol{1134'}\right),\\
		\left(\boldsymbol{1},\boldsymbol{3},\boldsymbol{490}\right),\,\,\left(\boldsymbol{1},\boldsymbol{3},\boldsymbol{840''}\right),\,\,\left(\boldsymbol{1},\boldsymbol{3},\boldsymbol{896}\right),\,\,\left(\boldsymbol{1},\boldsymbol{3},\boldsymbol{1050''}\right),\,\,\left(\boldsymbol{1},\boldsymbol{3},\boldsymbol{1134'}\right),\\
		\left(\boldsymbol{1},\boldsymbol{5},\boldsymbol{280}\right),\,\,\left(\boldsymbol{1},\boldsymbol{5},\boldsymbol{840''}\right),\,\,\left(\boldsymbol{1},\boldsymbol{5},\boldsymbol{896}\right),\,\,\left(\boldsymbol{1},\boldsymbol{5},\boldsymbol{1134'}\right),
		\end{array}$\tabularnewline
		\hline 
		$S^{2}V^{3}\bar{\psi}^{\Phi}\bar{\psi}^{M}$ & + & $\begin{array}{c}
		\left(\boldsymbol{1},\boldsymbol{1},\boldsymbol{175}\right),\,\,\left(\boldsymbol{1},\boldsymbol{1},\boldsymbol{189}\right),\,\,\left(\boldsymbol{1},\boldsymbol{1},\boldsymbol{280}\right),\,\,\left(\boldsymbol{1},\boldsymbol{1},\boldsymbol{462}\right),\,\,2\left(\boldsymbol{1},\boldsymbol{1},\boldsymbol{490}\right),\\
		3\left(\boldsymbol{1},\boldsymbol{1},\boldsymbol{840''}\right),\,\,4\left(\boldsymbol{1},\boldsymbol{1},\boldsymbol{896}\right),\,\,3\left(\boldsymbol{1},\boldsymbol{1},\boldsymbol{1050''}\right),\,\,4\left(\boldsymbol{1},\boldsymbol{1},\boldsymbol{1134'}\right),\\
		\left(\boldsymbol{1},\boldsymbol{3},\boldsymbol{175}\right),\,\,\left(\boldsymbol{1},\boldsymbol{3},\boldsymbol{189}\right),\,\,\left(\boldsymbol{1},\boldsymbol{3},\boldsymbol{280}\right),\,\,\left(\boldsymbol{1},\boldsymbol{3},\boldsymbol{462}\right),\,\,2\left(\boldsymbol{1},\boldsymbol{3},\boldsymbol{490}\right),\\
		3\left(\boldsymbol{1},\boldsymbol{3},\boldsymbol{840''}\right),\,\,4\left(\boldsymbol{1},\boldsymbol{3},\boldsymbol{896}\right),\,\,3\left(\boldsymbol{1},\boldsymbol{3},\boldsymbol{1050''}\right),\,\,4\left(\boldsymbol{1},\boldsymbol{3},\boldsymbol{1134'}\right),\\
		\left(\boldsymbol{1},\boldsymbol{5},\boldsymbol{175}\right),\,\,\left(\boldsymbol{1},\boldsymbol{5},\boldsymbol{189}\right),\,\,\left(\boldsymbol{1},\boldsymbol{5},\boldsymbol{280}\right),\,\,\left(\boldsymbol{1},\boldsymbol{5},\boldsymbol{462}\right),\,\,2\left(\boldsymbol{1},\boldsymbol{5},\boldsymbol{490}\right),\\
		3\left(\boldsymbol{1},\boldsymbol{5},\boldsymbol{840''}\right),\,\,4\left(\boldsymbol{1},\boldsymbol{5},\boldsymbol{896}\right),\,\,3\left(\boldsymbol{1},\boldsymbol{5},\boldsymbol{1050''}\right),\,\,4\left(\boldsymbol{1},\boldsymbol{5},\boldsymbol{1134'}\right),\\
		\left(\boldsymbol{1},\boldsymbol{3},\boldsymbol{35}\right),\,\,\left(\boldsymbol{1},\boldsymbol{3},\boldsymbol{189}\right),\,\,2\left(\boldsymbol{1},\boldsymbol{3},\boldsymbol{280}\right),\,\,\left(\boldsymbol{1},\boldsymbol{3},\boldsymbol{840''}\right),\,\,\left(\boldsymbol{1},\boldsymbol{3},\boldsymbol{896}\right)
		\end{array}$\tabularnewline
		\hline 
		$S^{2}V\left(\bar{\psi}^{\Phi}\right)^{2}\bar{\psi}^{M}$ & - & $\begin{array}{c}
		\left(\boldsymbol{1},\boldsymbol{1},\boldsymbol{280}\right),\,\,\left(\boldsymbol{1},\boldsymbol{1},\boldsymbol{490}\right),\,\,2\left(\boldsymbol{1},\boldsymbol{1},\boldsymbol{840''}\right),\\
		2\left(\boldsymbol{1},\boldsymbol{1},\boldsymbol{896}\right),\,\,\left(\boldsymbol{1},\boldsymbol{1},\boldsymbol{1050''}\right),\,\,2\left(\boldsymbol{1},\boldsymbol{1},\boldsymbol{1134'}\right),\\
		\left(\boldsymbol{1},\boldsymbol{3},\boldsymbol{280}\right),\,\,\left(\boldsymbol{1},\boldsymbol{3},\boldsymbol{490}\right),\,\,2\left(\boldsymbol{1},\boldsymbol{3},\boldsymbol{840''}\right),\,\,2\left(\boldsymbol{1},\boldsymbol{3},\boldsymbol{896}\right),\\
		\left(\boldsymbol{1},\boldsymbol{3},\boldsymbol{1050''}\right),\,\,2\left(\boldsymbol{1},\boldsymbol{3},\boldsymbol{1134'}\right),\,\,\left(\boldsymbol{1},\boldsymbol{5},\boldsymbol{280}\right),\,\,\left(\boldsymbol{1},\boldsymbol{5},\boldsymbol{490}\right),\\
		2\left(\boldsymbol{1},\boldsymbol{5},\boldsymbol{840''}\right),\,\,2\left(\boldsymbol{1},\boldsymbol{5},\boldsymbol{896}\right),\,\,\left(\boldsymbol{1},\boldsymbol{5},\boldsymbol{1050''}\right),\,\,2\left(\boldsymbol{1},\boldsymbol{5},\boldsymbol{1134'}\right)
		\end{array}$\tabularnewline
		\hline 
		$S^{4}V^{4}\bar{\psi}^{\Phi}$ & - & $\begin{array}{c}
		\left(\boldsymbol{1},\boldsymbol{1},\boldsymbol{35}\right),\,\,2\left(\boldsymbol{1},\boldsymbol{1},\boldsymbol{175}\right),\,\,\left(\boldsymbol{1},\boldsymbol{1},\boldsymbol{280}\right),\,\,2\left(\boldsymbol{1},\boldsymbol{1},\boldsymbol{462}\right),\\
		\left(\boldsymbol{1},\boldsymbol{1},\boldsymbol{490}\right),\,\,\left(\boldsymbol{1},\boldsymbol{1},\boldsymbol{840''}\right),\,\,3\left(\boldsymbol{1},\boldsymbol{1},\boldsymbol{896}\right),\,\,3\left(\boldsymbol{1},\boldsymbol{1},\boldsymbol{1050''}\right),\\
		5\left(\boldsymbol{1},\boldsymbol{1},\boldsymbol{1134'}\right),\,\,\left(\boldsymbol{1},\boldsymbol{3},\boldsymbol{35}\right),\,\,2\left(\boldsymbol{1},\boldsymbol{3},\boldsymbol{189}\right),\,\,3\left(\boldsymbol{1},\boldsymbol{3},\boldsymbol{280}\right),\\
		\left(\boldsymbol{1},\boldsymbol{3},\boldsymbol{490}\right),\,\,3\left(\boldsymbol{1},\boldsymbol{3},\boldsymbol{840''}\right),\,\,3\left(\boldsymbol{1},\boldsymbol{3},\boldsymbol{896}\right),\,\,\left(\boldsymbol{1},\boldsymbol{3},\boldsymbol{1050''}\right),\\
		\left(\boldsymbol{1},\boldsymbol{3},\boldsymbol{1134'}\right),\,\,\left(\boldsymbol{1},\boldsymbol{5},\boldsymbol{175}\right),\,\,\left(\boldsymbol{1},\boldsymbol{5},\boldsymbol{462}\right),\,\,\left(\boldsymbol{1},\boldsymbol{5},\boldsymbol{490}\right),\\
		\left(\boldsymbol{1},\boldsymbol{5},\boldsymbol{840''}\right),\,\,2\left(\boldsymbol{1},\boldsymbol{5},\boldsymbol{896}\right),\,\,2\left(\boldsymbol{1},\boldsymbol{5},\boldsymbol{1050''}\right),\,\,3\left(\boldsymbol{1},\boldsymbol{5},\boldsymbol{1134'}\right)
		\end{array}$\tabularnewline
		\hline 
		$S^{4}V^{2}\left(\bar{\psi}^{\Phi}\right)^{2}$ & + & $\begin{array}{c}
		2\left(\boldsymbol{1},\boldsymbol{1},\boldsymbol{490}\right),\,\,2\left(\boldsymbol{1},\boldsymbol{1},\boldsymbol{840''}\right),\,\,2\left(\boldsymbol{1},\boldsymbol{1},\boldsymbol{896}\right),\\
		2\left(\boldsymbol{1},\boldsymbol{1},\boldsymbol{1050''}\right),\,\,2\left(\boldsymbol{1},\boldsymbol{1},\boldsymbol{1134'}\right),\,\,\left(\boldsymbol{1},\boldsymbol{3},\boldsymbol{280}\right),\\
		\left(\boldsymbol{1},\boldsymbol{3},\boldsymbol{840''}\right),\,\,\left(\boldsymbol{1},\boldsymbol{3},\boldsymbol{896}\right),\,\,\left(\boldsymbol{1},\boldsymbol{3},\boldsymbol{1134'}\right),\,\,\left(\boldsymbol{1},\boldsymbol{5},\boldsymbol{490}\right),\\
		\left(\boldsymbol{1},\boldsymbol{5},\boldsymbol{840''}\right),\,\,\left(\boldsymbol{1},\boldsymbol{5},\boldsymbol{896}\right),\,\,\left(\boldsymbol{1},\boldsymbol{5},\boldsymbol{1050''}\right),\,\,\left(\boldsymbol{1},\boldsymbol{5},\boldsymbol{1134'}\right)
		\end{array}$\tabularnewline
		\hline 
		$S^{4}\left(\bar{\psi}^{\Phi}\right)^{3}$ & - & $\left(\boldsymbol{1},\boldsymbol{1},\boldsymbol{490}\right),\,\,\left(\boldsymbol{1},\boldsymbol{1},\boldsymbol{840''}\right)$\tabularnewline
		\hline
	\end{longtable}
\end{center}
Summing (with signs) the above characters, we obtain 
\begin{equation}
\sum R\left(-1\right)^{F}=\left(\boldsymbol{1},\boldsymbol{1},\boldsymbol{175}\right)-\left(\boldsymbol{1},\boldsymbol{1},\boldsymbol{35}\right)-\left(\boldsymbol{1},\boldsymbol{3},\boldsymbol{1}\right)-\begin{array}{c}
\left(\boldsymbol{1},\boldsymbol{5},\boldsymbol{1}\right)\end{array}-\begin{array}{c}
\left(\boldsymbol{1},\boldsymbol{1},\boldsymbol{1}\right)\end{array}.
\end{equation}
We see that in this case there is a nonvanishing contribution from marginal operators. Noticing that we have such an operator which is of the form $\left(S^{2}V^{3}\right)^{2}$, {\it i.e.}\footnote{Note that the representations $\left(\boldsymbol{1},\boldsymbol{3},\boldsymbol{56}\right)$
	and $\left(\boldsymbol{1},\boldsymbol{3},\boldsymbol{70}\right)$
	corresponding to an operator of the form $S^{2}V^{3}$ (that appear in the table above)
	are not considered since they are obtained from the following product
	of operators: $\left(S^{2}V\right)\left(V^{2}\right)$ (which are
	flipped).}
\begin{center}
	\begin{longtable}{|c|c|}
		\hline
		Operator & Representations\\
		\hline \hline
		$S^{2}V^{3}$ & $\left(\boldsymbol{1},\boldsymbol{1},\boldsymbol{20}\right)$\tabularnewline
		\hline 
		$\left(S^{2}V^{3}\right)^{2}$ & $\left(\boldsymbol{1},\boldsymbol{1},\boldsymbol{35}\right),\,\,\left(\boldsymbol{1},\boldsymbol{1},\boldsymbol{175}\right)$\tabularnewline
		\hline
	\end{longtable}
\end{center}
and that the contribution to the index at order $pq$ comes from marginal operators (with a positive sign) and conserved currents (with a negative sign), we can write this sum as 
\begin{equation}
\sum R\left(-1\right)^{F}=\textrm{Marginals}-\textrm{Conserved currents},
\end{equation}
where \footnote{Note that as in the $Spin(8)$ case, we do not assume that these marginal operator and conserved current contents take place everywhere on the conformal manifold. The statement is that it is consistent to happen somewhere on the manifold.}
\begin{equation}
\textrm{Marginals}=\boldsymbol{175}_{SU(6)_v}+\boldsymbol{35}_{SU(6)_v}
\end{equation}
and
\begin{equation}
\textrm{Conserved currents}=2\,\boldsymbol{35}_{SU(6)_v}+\boldsymbol{8}_{SU(3)}+1,
\end{equation}
which corresponds to the enhancement of the flavor symmetry $SU\left(2\right)\times SU\left(6\right)\times U\left(1\right)$ to $SU\left(3\right)\times SU\left(6\right)^{2}\times U\left(1\right)$ in the IR.

\

\subsection*{$\boldsymbol{Spin(11)}$}

\

The matter content is given in the following table, 
\begin{center}
	\begin{tabular}{|c||c|c|c|c|}
		\hline
		Field  & $Spin\left(11\right)_{g}$ & $SU\left(7\right)_{v}$ & $U\left(1\right)_{a}$ & $U\left(1\right)_{r}$\\
		\hline 
		$S$ & $\boldsymbol{32}$ & $\boldsymbol{1}$ & -7 & $\frac{1}{2}$\\
		
		$V$ & $\boldsymbol{11}$ & $\boldsymbol{7}$ & 4 & 0\\
		
		$M_{0}$ & $\boldsymbol{1}$ & $\boldsymbol{\overline{21}}$ & 6 & 1\\
		
		$M_{1}$ & $\boldsymbol{1}$ & $\boldsymbol{\overline{7}}$ & 10 & 1\\
		
		$\Phi$ & $\boldsymbol{1}$ & $\boldsymbol{\overline{28}}$ & -8 & 2\\
		\hline
	\end{tabular}
\end{center}
The superpotential is given by 
\begin{equation}
W=M_{0}S^{2}V^{2}+M_{1}S^{2}V+\Phi V^{2}
\end{equation}
and the superconformal R charge by ${\hat r}=r+0.030q_{a}$. We see that $U\left(1\right)_{a}$ mixes with $U\left(1\right)_{r}$ in the expression for the IR R symmetry, and so the operators that contribute to the index at order $pq$ should have a vanishing charge under $U\left(1\right)_{a}$. These operators are of the forms
\begin{equation*}
\lambda\lambda,\,\,\,\,\bar{\psi}^{S}S,\,\,\,\,\bar{\psi}^{V}V,\,\,\,\,\bar{\psi}^{M_{0}}M_{0},\,\,\,\,\bar{\psi}^{M_{1}}M_{1},\,\,\,\,\bar{\psi}^{\Phi}\Phi,\,\,\,\,S^{2}V^{2}M_{0},\,\,\,\,S^{2}VM_{1},
\end{equation*}
\begin{equation*}
\Phi V^{2},\,\,\,\,VM_{0}\bar{\psi}^{M_{1}},\,\,\,\,S^{2}M_{0}\bar{\psi}^{\Phi},\,\,\,\,S^{4}V^{7},\,\,\,\,S^{2}V^{5}\bar{\psi}^{M_{0}},\,\,\,\,V^{3}\left(\bar{\psi}^{M_{0}}\right)^{2},\,\,\,\,V\bar{\psi}^{\Phi}\left(\bar{\psi}^{M_{0}}\right)^{2},
\end{equation*}
\begin{equation*}
V^{4}\bar{\psi}^{M_{0}}\bar{\psi}^{M_{1}},\,\,\,\,V^{2}\bar{\psi}^{\Phi}\bar{\psi}^{M_{0}}\bar{\psi}^{M_{1}},\,\,\,\,\left(\bar{\psi}^{\Phi}\right)^{2}\bar{\psi}^{M_{0}}\bar{\psi}^{M_{1}},\,\,\,\,S^{2}V^{3}\bar{\psi}^{\Phi}\bar{\psi}^{M_{0}},\,\,\,\,S^{2}V\left(\bar{\psi}^{\Phi}\right)^{2}\bar{\psi}^{M_{0}},
\end{equation*}
\begin{equation*}
S^{2}V^{6}\bar{\psi}^{M_{1}},\,\,\,\,V^{5}\left(\bar{\psi}^{M_{1}}\right)^{2},\,\,\,\,V^{3}\bar{\psi}^{\Phi}\left(\bar{\psi}^{M_{1}}\right)^{2},\,\,\,\,V\left(\bar{\psi}^{\Phi}\right)^{2}\left(\bar{\psi}^{M_{1}}\right)^{2},\,\,\,\,S^{2}V^{4}\bar{\psi}^{\Phi}\bar{\psi}^{M_{1}},
\end{equation*}
\begin{equation}
S^{2}V^{2}\left(\bar{\psi}^{\Phi}\right)^{2}\bar{\psi}^{M_{1}},\,\,\,\,S^{2}\left(\bar{\psi}^{\Phi}\right)^{3}\bar{\psi}^{M_{1}},\,\,\,\,S^{4}V^{5}\bar{\psi}^{\Phi},\,\,\,\,S^{4}V^{3}\left(\bar{\psi}^{\Phi}\right)^{2},\,\,\,\,S^{4}V\left(\bar{\psi}^{\Phi}\right)^{3}.
\label{listSpin11}
\end{equation}

Next, denoting the irreducible representations (and the corresponding characters) of the nonabelian groups in the theory by 
\begin{equation}
\left(Spin\left(11\right)_{g},SU\left(7\right)_{v}\right),
\end{equation}
we find the following representations of the gauge singlets corresponding
to the operators in \eqref{listSpin11},
\begin{longtable}{|c|c|c|}
		\hline
		$\mathrm{Operator}$ & $\left(-1\right)^{F}$ & $\mathrm{Representations}\,\,(R)$\\
		\hline \hline 
		$\lambda\lambda$ & + & $\left(\boldsymbol{1},\boldsymbol{1}\right)$\tabularnewline
		\hline 
		$\bar{\psi}^{S}S$ & - & $\left(\boldsymbol{1},\boldsymbol{1}\right)$\tabularnewline
		\hline 
		$\bar{\psi}^{V}V$ & - & $\left(\boldsymbol{1},\boldsymbol{1}\right),\,\,\left(\boldsymbol{1},\boldsymbol{48}\right)$\tabularnewline
		\hline 
		$\bar{\psi}^{M_{0}}M_{0}$ & - & $\left(\boldsymbol{1},\boldsymbol{1}\right),\,\,\left(\boldsymbol{1},\boldsymbol{48}\right),\,\,\left(\boldsymbol{1},\boldsymbol{392}\right)$\tabularnewline
		\hline 
		$\bar{\psi}^{M_{1}}M_{1}$ & - & $\left(\boldsymbol{1},\boldsymbol{1}\right),\,\,\left(\boldsymbol{1},\boldsymbol{48}\right)$\tabularnewline
		\hline 
		$\bar{\psi}^{\Phi}\Phi$ & - & $\left(\boldsymbol{1},\boldsymbol{1}\right),\,\,\left(\boldsymbol{1},\boldsymbol{48}\right),\,\,\left(\boldsymbol{1},\boldsymbol{735'}\right)$\tabularnewline
		\hline 
		$\Phi V^{2}$ & + & $\left(\boldsymbol{1},\boldsymbol{1}\right),\,\,\left(\boldsymbol{1},\boldsymbol{48}\right),\,\,\left(\boldsymbol{1},\boldsymbol{735'}\right)$\tabularnewline
		\hline 
		$S^{2}VM_{1}$ & + & $\left(\boldsymbol{1},\boldsymbol{1}\right),\,\,\left(\boldsymbol{1},\boldsymbol{48}\right)$\tabularnewline
		\hline 
		$S^{2}V^{2}M_{0}$ & + & $\left(\boldsymbol{1},\boldsymbol{1}\right),\,\,\left(\boldsymbol{1},\boldsymbol{48}\right),\,\,\left(\boldsymbol{1},\boldsymbol{392}\right)$\tabularnewline
		\hline 
		$VM_{0}\bar{\psi}^{M_{1}}$ & - & $\textrm{No gauge singlets}$\tabularnewline
		\hline 
		$S^{2}M_{0}\bar{\psi}^{\Phi}$ & - & $\textrm{No gauge singlets}$\tabularnewline
		\hline 
		$S^{4}V^{7}$ & + & $\begin{array}{c}
		2\left(\boldsymbol{1},\boldsymbol{392}\right),\,\,\left(\boldsymbol{1},\boldsymbol{2400}\right),\,\,2\left(\boldsymbol{1},\boldsymbol{2940}\right),\,\,2\left(\boldsymbol{1},\boldsymbol{3528}\right),\,\,2\left(\boldsymbol{1},\boldsymbol{4950}\right),\\
		\left(\boldsymbol{1},\boldsymbol{1}\right),\,\,\left(\boldsymbol{1},\boldsymbol{784}\right),\,\,\left(\boldsymbol{1},\boldsymbol{2646}\right),\:\,\left(\boldsymbol{1},\boldsymbol{4704''}\right),\,\,\left(\boldsymbol{1},\boldsymbol{4752}\right),\\
		\left(\boldsymbol{1},\boldsymbol{6468'}\right),\,\,2\left(\boldsymbol{1},\boldsymbol{7350}\right),\,\,2\left(\boldsymbol{1},\boldsymbol{48}\right),\,\,\left(\boldsymbol{1},\boldsymbol{540}\right)
		\end{array}$\tabularnewline
		\hline 
		$S^{2}V^{5}\bar{\psi}^{M_{0}}$ & - & $\begin{array}{c}
		\left(\boldsymbol{1},\boldsymbol{4752}\right),\,\,2\left(\boldsymbol{1},\boldsymbol{4950}\right),\,\,2\left(\boldsymbol{1},\boldsymbol{392}\right),\,\,\left(\boldsymbol{1},\boldsymbol{784}\right),\,\,\left(\boldsymbol{1},\boldsymbol{2646}\right),\\
		2\left(\boldsymbol{1},\boldsymbol{2940}\right),\,\,2\left(\boldsymbol{1},\boldsymbol{3528}\right),\,\,\left(\boldsymbol{1},\boldsymbol{4704''}\right),\,\,2\left(\boldsymbol{1},\boldsymbol{7350}\right),\\
		\left(\boldsymbol{1},\boldsymbol{2400}\right),\,\,\left(\boldsymbol{1},\boldsymbol{6468'}\right),\,\,\left(\boldsymbol{1},\boldsymbol{1}\right),\,\,\left(\boldsymbol{1},\boldsymbol{48}\right)
		\end{array}$\tabularnewline
		\hline 
		$V^{3}\left(\bar{\psi}^{M_{0}}\right)^{2}$ & + & $\textrm{No gauge singlets}$\tabularnewline
		\hline 
		$V\bar{\psi}^{\Phi}\left(\bar{\psi}^{M_{0}}\right)^{2}$ & - & $\textrm{No gauge singlets}$\tabularnewline
		\hline 
		$V^{4}\bar{\psi}^{M_{0}}\bar{\psi}^{M_{1}}$ & + & $\begin{array}{c}
		\left(\boldsymbol{1},\boldsymbol{392}\right),\,\,\left(\boldsymbol{1},\boldsymbol{784}\right),\,\,\left(\boldsymbol{1},\boldsymbol{2400}\right),\,\,\left(\boldsymbol{1},\boldsymbol{2646}\right),\,\,2\left(\boldsymbol{1},\boldsymbol{2940}\right),\,\,2\left(\boldsymbol{1},\boldsymbol{3528}\right),\\
		\left(\boldsymbol{1},\boldsymbol{4704''}\right),\,\,\left(\boldsymbol{1},\boldsymbol{4752}\right),\,\,2\left(\boldsymbol{1},\boldsymbol{4950}\right),\,\,\left(\boldsymbol{1},\boldsymbol{6468'}\right),\,\,2\left(\boldsymbol{1},\boldsymbol{7350}\right)
		\end{array}$\tabularnewline
		\hline 
		$V^{2}\bar{\psi}^{\Phi}\bar{\psi}^{M_{0}}\bar{\psi}^{M_{1}}$ & - & $\begin{array}{c}
		\left(\boldsymbol{1},\boldsymbol{392}\right),\,\,\left(\boldsymbol{1},\boldsymbol{540}\right),\,\,\left(\boldsymbol{1},\boldsymbol{784}\right),\,\,3\left(\boldsymbol{1},\boldsymbol{2400}\right),\,\,2\left(\boldsymbol{1},\boldsymbol{2646}\right),\,\,4\left(\boldsymbol{1},\boldsymbol{2940}\right),\\
		3\left(\boldsymbol{1},\boldsymbol{3528}\right),\,\,2\left(\boldsymbol{1},\boldsymbol{4704''}\right),\,\,\left(\boldsymbol{1},\boldsymbol{4752}\right),\,\,3\left(\boldsymbol{1},\boldsymbol{4950}\right),\,\,2\left(\boldsymbol{1},\boldsymbol{6468'}\right),\,\,5\left(\boldsymbol{1},\boldsymbol{7350}\right)
		\end{array}$\tabularnewline
		\hline 
		$\left(\bar{\psi}^{\Phi}\right)^{2}\bar{\psi}^{M_{0}}\bar{\psi}^{M_{1}}$ & + & $\begin{array}{c}
		\left(\boldsymbol{1},\boldsymbol{540}\right),\,\,2\left(\boldsymbol{1},\boldsymbol{2400}\right),\,\,\left(\boldsymbol{1},\boldsymbol{2646}\right),\,\,2\left(\boldsymbol{1},\boldsymbol{2940}\right),\,\,\left(\boldsymbol{1},\boldsymbol{3528}\right),\\
		\left(\boldsymbol{1},\boldsymbol{4704''}\right),\,\,\left(\boldsymbol{1},\boldsymbol{4950}\right),\,\,\left(\boldsymbol{1},\boldsymbol{6468'}\right),\,\,3\left(\boldsymbol{1},\boldsymbol{7350}\right)
		\end{array}$\tabularnewline
		\hline 
		$S^{2}V^{3}\bar{\psi}^{\Phi}\bar{\psi}^{M_{0}}$ & + & $\begin{array}{c}
		\left(\boldsymbol{1},\boldsymbol{392}\right),\,\,\left(\boldsymbol{1},\boldsymbol{540}\right),\,\,\left(\boldsymbol{1},\boldsymbol{784}\right),\,\,3\left(\boldsymbol{1},\boldsymbol{2400}\right),\,\,2\left(\boldsymbol{1},\boldsymbol{2646}\right),\,\,4\left(\boldsymbol{1},\boldsymbol{2940}\right),\\
		3\left(\boldsymbol{1},\boldsymbol{3528}\right),\,\,2\left(\boldsymbol{1},\boldsymbol{4704''}\right),\,\,\left(\boldsymbol{1},\boldsymbol{4752}\right),\,\,3\left(\boldsymbol{1},\boldsymbol{4950}\right),\,\,2\left(\boldsymbol{1},\boldsymbol{6468'}\right),\,\,5\left(\boldsymbol{1},\boldsymbol{7350}\right)
		\end{array}$\tabularnewline
		\hline 
		$S^{2}V\left(\bar{\psi}^{\Phi}\right)^{2}\bar{\psi}^{M_{0}}$ & - & $\begin{array}{c}
		\left(\boldsymbol{1},\boldsymbol{540}\right),\,\,2\left(\boldsymbol{1},\boldsymbol{2400}\right),\,\,\left(\boldsymbol{1},\boldsymbol{2646}\right),\,\,2\left(\boldsymbol{1},\boldsymbol{2940}\right),\,\,\left(\boldsymbol{1},\boldsymbol{3528}\right),\\
		\left(\boldsymbol{1},\boldsymbol{4704''}\right),\,\,\left(\boldsymbol{1},\boldsymbol{4950}\right),\,\,\left(\boldsymbol{1},\boldsymbol{6468'}\right),\,\,3\left(\boldsymbol{1},\boldsymbol{7350}\right)
		\end{array}$\tabularnewline
		\hline 
		$S^{2}V^{6}\bar{\psi}^{M_{1}}$ & - & $\begin{array}{c}
		\left(\boldsymbol{1},\boldsymbol{1}\right),\,\,\left(\boldsymbol{1},\boldsymbol{48}\right),\,\,\left(\boldsymbol{1},\boldsymbol{392}\right),\,\,\left(\boldsymbol{1},\boldsymbol{784}\right),\,\,\left(\boldsymbol{1},\boldsymbol{2400}\right),\\
		\left(\boldsymbol{1},\boldsymbol{2646}\right),\,\,2\left(\boldsymbol{1},\boldsymbol{2940}\right),\,\,2\left(\boldsymbol{1},\boldsymbol{3528}\right),\,\,\left(\boldsymbol{1},\boldsymbol{4704''}\right),\\
		\left(\boldsymbol{1},\boldsymbol{4752}\right),\,\,2\left(\boldsymbol{1},\boldsymbol{4950}\right),\,\,\left(\boldsymbol{1},\boldsymbol{6468'}\right),\,\,2\left(\boldsymbol{1},\boldsymbol{7350}\right)
		\end{array}$\tabularnewline
		\hline 
		$V^{5}\left(\bar{\psi}^{M_{1}}\right)^{2}$ & + & $\textrm{No gauge singlets}$\tabularnewline
		\hline 
		$V^{3}\bar{\psi}^{\Phi}\left(\bar{\psi}^{M_{1}}\right)^{2}$ & - & $\textrm{No gauge singlets}$\tabularnewline
		\hline 
		$V\left(\bar{\psi}^{\Phi}\right)^{2}\left(\bar{\psi}^{M_{1}}\right)^{2}$ & + & $\textrm{No gauge singlets}$\tabularnewline
		\hline 
		$S^{2}V^{4}\bar{\psi}^{\Phi}\bar{\psi}^{M_{1}}$ & + & $\begin{array}{c}
		\left(\boldsymbol{1},\boldsymbol{392}\right),\,\,\left(\boldsymbol{1},\boldsymbol{540}\right),\,\,\left(\boldsymbol{1},\boldsymbol{784}\right),\,\,3\left(\boldsymbol{1},\boldsymbol{2400}\right),\,\,2\left(\boldsymbol{1},\boldsymbol{2646}\right),\,\,4\left(\boldsymbol{1},\boldsymbol{2940}\right),\\
		3\left(\boldsymbol{1},\boldsymbol{3528}\right),\,\,2\left(\boldsymbol{1},\boldsymbol{4704''}\right),\,\,\left(\boldsymbol{1},\boldsymbol{4752}\right),\,\,3\left(\boldsymbol{1},\boldsymbol{4950}\right),\,\,2\left(\boldsymbol{1},\boldsymbol{6468'}\right),\,\,5\left(\boldsymbol{1},\boldsymbol{7350}\right)
		\end{array}$\tabularnewline
		\hline 
		$S^{2}V^{2}\left(\bar{\psi}^{\Phi}\right)^{2}\bar{\psi}^{M_{1}}$ & - & $\begin{array}{c}
		\left(\boldsymbol{1},\boldsymbol{540}\right),\,\,2\left(\boldsymbol{1},\boldsymbol{2400}\right),\,\,\left(\boldsymbol{1},\boldsymbol{2646}\right),\,\,2\left(\boldsymbol{1},\boldsymbol{2940}\right),\,\,\left(\boldsymbol{1},\boldsymbol{3528}\right),\\
		\left(\boldsymbol{1},\boldsymbol{4704''}\right),\,\,\left(\boldsymbol{1},\boldsymbol{4950}\right),\,\,\left(\boldsymbol{1},\boldsymbol{6468'}\right),\,\,3\left(\boldsymbol{1},\boldsymbol{7350}\right)
		\end{array}$\tabularnewline
		\hline 
		$S^{2}\left(\bar{\psi}^{\Phi}\right)^{3}\bar{\psi}^{M_{1}}$ & + & $\textrm{No gauge singlets}$\tabularnewline
		\hline 
		$S^{4}V^{5}\bar{\psi}^{\Phi}$ & - & $\begin{array}{c}
		\left(\boldsymbol{1},\boldsymbol{48}\right),\,\,\left(\boldsymbol{1},\boldsymbol{392}\right),\,\,2\left(\boldsymbol{1},\boldsymbol{540}\right),\,\,\left(\boldsymbol{1},\boldsymbol{784}\right),\,\,3\left(\boldsymbol{1},\boldsymbol{2400}\right),\\
		2\left(\boldsymbol{1},\boldsymbol{2646}\right),\,\,4\left(\boldsymbol{1},\boldsymbol{2940}\right),\,\,3\left(\boldsymbol{1},\boldsymbol{3528}\right),\,\,2\left(\boldsymbol{1},\boldsymbol{4704''}\right),\\
		\left(\boldsymbol{1},\boldsymbol{4752}\right),\,\,3\left(\boldsymbol{1},\boldsymbol{4950}\right),\,\,2\left(\boldsymbol{1},\boldsymbol{6468'}\right),\,\,5\left(\boldsymbol{1},\boldsymbol{7350}\right)
		\end{array}$\tabularnewline
		\hline 
		$S^{4}V^{3}\left(\bar{\psi}^{\Phi}\right)^{2}$ & + & $\begin{array}{c}
		\left(\boldsymbol{1},\boldsymbol{540}\right),\,\,2\left(\boldsymbol{1},\boldsymbol{2400}\right),\,\,\left(\boldsymbol{1},\boldsymbol{2646}\right),\,\,2\left(\boldsymbol{1},\boldsymbol{2940}\right),\,\,\left(\boldsymbol{1},\boldsymbol{3528}\right),\\
		\left(\boldsymbol{1},\boldsymbol{4704''}\right),\,\,\left(\boldsymbol{1},\boldsymbol{4950}\right),\,\,\left(\boldsymbol{1},\boldsymbol{6468'}\right),\,\,3\left(\boldsymbol{1},\boldsymbol{7350}\right)
		\end{array}$\tabularnewline
		\hline 
		$S^{4}V\left(\bar{\psi}^{\Phi}\right)^{3}$ & - & $\textrm{No gauge singlets}$\tabularnewline
		\hline
	\end{longtable}
Summing (with signs) the above characters, we obtain 
\begin{equation}
\sum R\left(-1\right)^{F}=-2\left(\boldsymbol{1},\boldsymbol{48}\right)-2\left(\boldsymbol{1},\boldsymbol{1}\right)=-2\,\boldsymbol{48}_{SU(7)_v}-2
\end{equation}
which corresponds to the enhancement of the flavor symmetry $SU\left(7\right)\times U\left(1\right)$ to $SU\left(7\right)^{2}\times U\left(1\right)^{2}$ in the IR. Moreover, the dimension of the conformal manifold vanishes.

\

\subsection*{$\boldsymbol{Spin(12)}$}

\

The matter content is given in the following table, 
\begin{center}
	\begin{tabular}{|c||c|c|c|c|}
		\hline
		Field  & $Spin\left(12\right)_{g}$ & $SU\left(8\right)_{v}$ & $U\left(1\right)_{a}$ & $U\left(1\right)_{r}$\\
		\hline 
		$S$ & $\boldsymbol{\overline{32}}$ & $\boldsymbol{1}$ & 2 & $\frac{1}{2}$\\
		
		$V$ & $\boldsymbol{12}$ & $\boldsymbol{8}$ & -1 & 0\\
		
		$M$ & $\boldsymbol{1}$ & $\boldsymbol{\overline{28}}$ & -2 & 1\\
		
		$\Phi$ & $\boldsymbol{1}$ & $\boldsymbol{\overline{36}}$ & 2 & 2\\
		\hline
	\end{tabular}
\end{center}
The superpotential is given by 
\begin{equation}
W=MS^{2}V^{2}+\Phi V^{2}
\end{equation}
and the superconformal R charge by ${\hat r}=r-0.110q_{a}$. We see that $U\left(1\right)_{a}$ mixes with $U\left(1\right)_{r}$ in the expression for the IR R symmetry, and so the operators that contribute to the index at order $pq$ should have a vanishing charge under $U\left(1\right)_{a}$. These operators are of the forms
\begin{equation*}
\lambda\lambda,\,\,\,\,\bar{\psi}^{S}S,\,\,\,\,\bar{\psi}^{V}V,\,\,\,\,\bar{\psi}^{M}M,\,\,\,\,\bar{\psi}^{\Phi}\Phi,\,\,\,\,S^{2}V^{2}M,\,\,\,\,\Phi V^{2},
\end{equation*}
\begin{equation*}
S^{2}\bar{\psi}^{\Phi}M,\,\,\,\,S^{4}V^{8},\,\,\,\,S^{2}V^{6}\bar{\psi}^{M},\,\,\,\,V^{4}\left(\bar{\psi}^{M}\right)^{2},\,\,\,\,S^{2}V^{4}\bar{\psi}^{\Phi}\bar{\psi}^{M},
\end{equation*}
\begin{equation*}
S^{2}V^{2}\left(\bar{\psi}^{\Phi}\right)^{2}\bar{\psi}^{M},\,\,\,\,S^{2}\left(\bar{\psi}^{\Phi}\right)^{3}\bar{\psi}^{M},\,\,\,\,V^{2}\bar{\psi}^{\Phi}\left(\bar{\psi}^{M}\right)^{2},\,\,\,\,\left(\bar{\psi}^{\Phi}\right)^{2}\left(\bar{\psi}^{M}\right)^{2},
\end{equation*}
\begin{equation}
S^{4}V^{6}\bar{\psi}^{\Phi},\,\,\,\,S^{4}V^{4}\left(\bar{\psi}^{\Phi}\right)^{2},\,\,\,\,S^{4}V^{2}\left(\bar{\psi}^{\Phi}\right)^{3},\,\,\,\,S^{4}\left(\bar{\psi}^{\Phi}\right)^{4}.
\label{listSpin12}
\end{equation}

Next, denoting the irreducible representations (and the corresponding characters) of the nonabelian groups in the theory by 
\begin{equation}
\left(Spin\left(12\right)_{g},SU\left(8\right)_{v}\right),
\end{equation}
we find the following representations of the gauge singlets corresponding
to the operators in \eqref{listSpin12},
\begin{center}
	\begin{longtable}{|c|c|c|}
		\hline
		$\mathrm{Operator}$ & $\left(-1\right)^{F}$ & $\mathrm{Representations}\,\,(R)$\\
		\hline \hline 
		$\lambda\lambda$ & + & $\left(\boldsymbol{1},\boldsymbol{1}\right)$\tabularnewline
		\hline 
		$\bar{\psi}^{S}S$ & - & $\left(\boldsymbol{1},\boldsymbol{1}\right)$\tabularnewline
		\hline 
		$\bar{\psi}^{V}V$ & - & $\left(\boldsymbol{1},\boldsymbol{1}\right),\,\,\left(\boldsymbol{1},\boldsymbol{63}\right)$\tabularnewline
		\hline 
		$\bar{\psi}^{M}M$ & - & $\left(\boldsymbol{1},\boldsymbol{1}\right),\,\,\left(\boldsymbol{1},\boldsymbol{63}\right),\,\,\left(\boldsymbol{1},\boldsymbol{720}\right)$\tabularnewline
		\hline 
		$\bar{\psi}^{\Phi}\Phi$ & - & $\left(\boldsymbol{1},\boldsymbol{1}\right),\,\,\left(\boldsymbol{1},\boldsymbol{63}\right),\,\,\left(\boldsymbol{1},\boldsymbol{1232}\right)$\tabularnewline
		\hline 
		$S^{2}V^{2}M$ & + & $\left(\boldsymbol{1},\boldsymbol{1}\right),\,\,\left(\boldsymbol{1},\boldsymbol{63}\right),\,\,\left(\boldsymbol{1},\boldsymbol{720}\right)$\tabularnewline
		\hline 
		$\Phi V^{2}$ & + & $\left(\boldsymbol{1},\boldsymbol{1}\right),\,\,\left(\boldsymbol{1},\boldsymbol{63}\right),\,\,\left(\boldsymbol{1},\boldsymbol{1232}\right)$\tabularnewline
		\hline 
		$S^{2}\bar{\psi}^{\Phi}M$ & - & $\textrm{No gauge singlets}$\tabularnewline
		\hline 
		$S^{4}V^{8}$ & + & $\begin{array}{c}
		\left(\boldsymbol{1},\boldsymbol{6435}\right),\,\,2\left(\boldsymbol{1},\boldsymbol{34320'}\right),\,\,2\left(\boldsymbol{1},\boldsymbol{13860}\right),\,\,3\left(\boldsymbol{1},\boldsymbol{25872}\right),\,\,2\left(\boldsymbol{1},\boldsymbol{1764}\right),\\
		\left(\boldsymbol{1},\boldsymbol{63}\right),\,\,\left(\boldsymbol{1},\boldsymbol{945}\right),\,\,\left(\boldsymbol{1},\boldsymbol{5775}\right),\,\,\left(\boldsymbol{1},\boldsymbol{17325}\right),\,\,2\left(\boldsymbol{1},\boldsymbol{720}\right),\\
		\left(\boldsymbol{1},\boldsymbol{7680}\right),\,\,2\left(\boldsymbol{1},\boldsymbol{15120'}\right),\,\,\left(\boldsymbol{1},\boldsymbol{1}\right),\,\,\left(\boldsymbol{1},\boldsymbol{14700}\right),\,\,\left(\boldsymbol{1},\boldsymbol{41580}\right),\,\,\left(\boldsymbol{1},\boldsymbol{50688}\right)
		\end{array}$\tabularnewline
		\hline 
		$S^{2}V^{6}\bar{\psi}^{M}$ & - & $\begin{array}{c}
		\left(\boldsymbol{1},\boldsymbol{5775}\right),\,\,2\left(\boldsymbol{1},\boldsymbol{17325}\right),\,\,2\left(\boldsymbol{1},\boldsymbol{25872}\right),\,\,2\left(\boldsymbol{1},\boldsymbol{29700}\right),\,\,2\left(\boldsymbol{1},\boldsymbol{50688}\right),\\
		\left(\boldsymbol{1},\boldsymbol{27027}\right),\,\,\left(\boldsymbol{1},\boldsymbol{34320'}\right),\,\,2\left(\boldsymbol{1},\boldsymbol{720}\right),\,\,\left(\boldsymbol{1},\boldsymbol{1764}\right),\,\,\left(\boldsymbol{1},\boldsymbol{2352}\right),\\
		2\left(\boldsymbol{1},\boldsymbol{7680}\right),\,\,2\left(\boldsymbol{1},\boldsymbol{14700}\right),\,\,3\left(\boldsymbol{1},\boldsymbol{15120'}\right),\,\,\left(\boldsymbol{1},\boldsymbol{13860}\right),\,\,2\left(\boldsymbol{1},\boldsymbol{41580}\right),\\
		\left(\boldsymbol{1},\boldsymbol{1}\right),\,\,\left(\boldsymbol{1},\boldsymbol{63}\right),\,\,\left(\boldsymbol{1},\boldsymbol{15876}\right)
		\end{array}$\tabularnewline
		\hline 
		$V^{4}\left(\bar{\psi}^{M}\right)^{2}$ & + & $\begin{array}{c}
		\left(\boldsymbol{1},\boldsymbol{17325}\right),\,\,\left(\boldsymbol{1},\boldsymbol{27027}\right),\,\,\left(\boldsymbol{1},\boldsymbol{29700}\right),\\
		\left(\boldsymbol{1},\boldsymbol{50688}\right),\,\,\left(\boldsymbol{1},\boldsymbol{2352}\right),\,\,\left(\boldsymbol{1},\boldsymbol{7680}\right),\,\,\left(\boldsymbol{1},\boldsymbol{14700}\right),\\
		\left(\boldsymbol{1},\boldsymbol{15120'}\right),\,\,\left(\boldsymbol{1},\boldsymbol{15876}\right),\,\,\left(\boldsymbol{1},\boldsymbol{29700}\right),\,\,\left(\boldsymbol{1},\boldsymbol{41580}\right)
		\end{array}$\tabularnewline
		\hline 
		$S^{2}V^{4}\bar{\psi}^{\Phi}\bar{\psi}^{M}$ & + & $\begin{array}{c}
		\left(\boldsymbol{1},\boldsymbol{720}\right),\,\,\left(\boldsymbol{1},\boldsymbol{945}\right),\,\,\left(\boldsymbol{1},\boldsymbol{1764}\right),\\
		\left(\boldsymbol{1},\boldsymbol{2352}\right),\,\,3\left(\boldsymbol{1},\boldsymbol{5775}\right),\,\,4\left(\boldsymbol{1},\boldsymbol{7680}\right),\\
		\left(\boldsymbol{1},\boldsymbol{13860}\right),\,\,4\left(\boldsymbol{1},\boldsymbol{14700}\right),\,\,4\left(\boldsymbol{1},\boldsymbol{15120'}\right),\,\,2\left(\boldsymbol{1},\boldsymbol{15876}\right),\\
		3\left(\boldsymbol{1},\boldsymbol{17325}\right),\,\,4\left(\boldsymbol{1},\boldsymbol{25872}\right),\,\,\left(\boldsymbol{1},\boldsymbol{27027}\right),\,\,6\left(\boldsymbol{1},\boldsymbol{29700}\right),\\
		\left(\boldsymbol{1},\boldsymbol{33264}\right),\,\,\left(\boldsymbol{1},\boldsymbol{34320'}\right),\,\,4\left(\boldsymbol{1},\boldsymbol{41580}\right),\,\,4\left(\boldsymbol{1},\boldsymbol{50688}\right)
		\end{array}$\tabularnewline
		\hline 
		$S^{2}V^{2}\left(\bar{\psi}^{\Phi}\right)^{2}\bar{\psi}^{M}$ & - & $\begin{array}{c}
		\left(\boldsymbol{1},\boldsymbol{945}\right),\,\,2\left(\boldsymbol{1},\boldsymbol{5775}\right),\,\,2\left(\boldsymbol{1},\boldsymbol{7680}\right),\,\,2\left(\boldsymbol{1},\boldsymbol{14700}\right),\\
		\left(\boldsymbol{1},\boldsymbol{15120'}\right),\,\,\left(\boldsymbol{1},\boldsymbol{15876}\right),\,\,\left(\boldsymbol{1},\boldsymbol{17325}\right),\,\,2\left(\boldsymbol{1},\boldsymbol{25872}\right),\\
		4\left(\boldsymbol{1},\boldsymbol{29700}\right),\,\,\left(\boldsymbol{1},\boldsymbol{33264}\right),\,\,2\left(\boldsymbol{1},\boldsymbol{41580}\right),\,\,2\left(\boldsymbol{1},\boldsymbol{50688}\right)
		\end{array}$\tabularnewline
		\hline 
		$S^{2}\left(\bar{\psi}^{\Phi}\right)^{3}\bar{\psi}^{M}$ & + & $\textrm{No gauge singlets}$\tabularnewline
		\hline 
		$V^{2}\bar{\psi}^{\Phi}\left(\bar{\psi}^{M}\right)^{2}$ & - & $\begin{array}{c}
		\left(\boldsymbol{1},\boldsymbol{2352}\right),\,\,\left(\boldsymbol{1},\boldsymbol{5775}\right),\,\,2\left(\boldsymbol{1},\boldsymbol{7680}\right),\,\,2\left(\boldsymbol{1},\boldsymbol{14700}\right),\\
		2\left(\boldsymbol{1},\boldsymbol{15120'}\right),\,\,\left(\boldsymbol{1},\boldsymbol{15876}\right),\,\,2\left(\boldsymbol{1},\boldsymbol{17325}\right),\,\,\left(\boldsymbol{1},\boldsymbol{25872}\right),\\
		\left(\boldsymbol{1},\boldsymbol{27027}\right),\,\,4\left(\boldsymbol{1},\boldsymbol{29700}\right),\,\,2\left(\boldsymbol{1},\boldsymbol{41580}\right),\,\,2\left(\boldsymbol{1},\boldsymbol{50688}\right)
		\end{array}$\tabularnewline
		\hline 
		$\left(\bar{\psi}^{\Phi}\right)^{2}\left(\bar{\psi}^{M}\right)^{2}$ & + & $\begin{array}{c}
		\left(\boldsymbol{1},\boldsymbol{5775}\right),\,\,\left(\boldsymbol{1},\boldsymbol{7680}\right),\,\,\left(\boldsymbol{1},\boldsymbol{14700}\right),\,\,\left(\boldsymbol{1},\boldsymbol{15120'}\right),\,\,\left(\boldsymbol{1},\boldsymbol{17325}\right),\\
		\left(\boldsymbol{1},\boldsymbol{25872}\right),\,\,2\left(\boldsymbol{1},\boldsymbol{29700}\right),\,\,\left(\boldsymbol{1},\boldsymbol{41580}\right),\,\,\left(\boldsymbol{1},\boldsymbol{50688}\right)
		\end{array}$\tabularnewline
		\hline 
		$S^{4}V^{6}\bar{\psi}^{\Phi}$ & - & $\begin{array}{c}
		\left(\boldsymbol{1},\boldsymbol{6435}\right),\,\,\left(\boldsymbol{1},\boldsymbol{21021}\right),\,\,3\left(\boldsymbol{1},\boldsymbol{34320'}\right),\,\,2\left(\boldsymbol{1},\boldsymbol{13860}\right),\,\,5\left(\boldsymbol{1},\boldsymbol{25872}\right),\\
		2\left(\boldsymbol{1},\boldsymbol{33264}\right),\,\,3\left(\boldsymbol{1},\boldsymbol{41580}\right),\,\,3\left(\boldsymbol{1},\boldsymbol{50688}\right),\,\,2\left(\boldsymbol{1},\boldsymbol{1764}\right),\,\,3\left(\boldsymbol{1},\boldsymbol{14700}\right),\\
		2\left(\boldsymbol{1},\boldsymbol{15120'}\right),\,\,\left(\boldsymbol{1},\boldsymbol{15876}\right),\,\,2\left(\boldsymbol{1},\boldsymbol{29700}\right),\,\,2\left(\boldsymbol{1},\boldsymbol{5775}\right),\,\,2\left(\boldsymbol{1},\boldsymbol{7680}\right),\\
		\left(\boldsymbol{1},\boldsymbol{17325}\right),\,\,\left(\boldsymbol{1},\boldsymbol{720}\right),\,\,2\left(\boldsymbol{1},\boldsymbol{945}\right),\,\,\left(\boldsymbol{1},\boldsymbol{63}\right)
		\end{array}$\tabularnewline
		\hline 
		$S^{4}V^{4}\left(\bar{\psi}^{\Phi}\right)^{2}$ & + & $\begin{array}{c}
		\left(\boldsymbol{1},\boldsymbol{945}\right),\,\,\left(\boldsymbol{1},\boldsymbol{5775}\right),\,\,\left(\boldsymbol{1},\boldsymbol{7680}\right),\,\,2\left(\boldsymbol{1},\boldsymbol{14700}\right),\,\,2\left(\boldsymbol{1},\boldsymbol{15876}\right),\\
		\left(\boldsymbol{1},\boldsymbol{21021}\right),\,\,2\left(\boldsymbol{1},\boldsymbol{25872}\right),\,\,\left(\boldsymbol{1},\boldsymbol{27027}\right),\,\,3\left(\boldsymbol{1},\boldsymbol{29700}\right),\,\,3\left(\boldsymbol{1},\boldsymbol{33264}\right),\\
		\left(\boldsymbol{1},\boldsymbol{34320'}\right),\,\,3\left(\boldsymbol{1},\boldsymbol{41580}\right),\,\,3\left(\boldsymbol{1},\boldsymbol{50688}\right)
		\end{array}$\tabularnewline
		\hline 
		$S^{4}V^{2}\left(\bar{\psi}^{\Phi}\right)^{3}$ & - & $\begin{array}{c}
		\left(\boldsymbol{1},\boldsymbol{15876}\right),\,\,\left(\boldsymbol{1},\boldsymbol{17325}\right),\,\,\left(\boldsymbol{1},\boldsymbol{27027}\right),\,\,\left(\boldsymbol{1},\boldsymbol{29700}\right),\,\,\left(\boldsymbol{1},\boldsymbol{33264}\right),\\
		2\left(\boldsymbol{1},\boldsymbol{41580}\right),\,\,\left(\boldsymbol{1},\boldsymbol{50688}\right),
		\end{array}$\tabularnewline
		\hline 
		$S^{4}\left(\bar{\psi}^{\Phi}\right)^{4}$ & + & $\left(\boldsymbol{1},\boldsymbol{41580}\right),\,\,\left(\boldsymbol{1},\boldsymbol{17325}\right)$\tabularnewline
		\hline
	\end{longtable}
\end{center}
Summing (with signs) the above characters, we obtain 
\begin{equation}
\sum R\left(-1\right)^{F}=-2\left(\boldsymbol{1},\boldsymbol{63}\right)-\left(\boldsymbol{1},\boldsymbol{1}\right)=-2\,\boldsymbol{63}_{SU(8)_v}-1
\end{equation}
which corresponds to the enhancement of the flavor symmetry $SU\left(8\right)\times U\left(1\right)$ to $SU\left(8\right)^{2}\times U\left(1\right)$ in the IR. Moreover, the dimension of the conformal manifold vanishes.

\

\section{Symmetry enhancement in the $\boldsymbol{SU(6)}$ and $\boldsymbol{SU(8)}$ gauge theories}

We present the calculation of the superconformal index at order $pq$ corresponding to the $SU(6)$ and $SU(8)$ gauge theories of section \ref{secsuen} using the same method as in the previous section. Notice that, as before, the superpotential is a relevant deformation. 

\

\subsection*{$\boldsymbol{SU(6)}$}

\

The matter content is given in the following table, 
\begin{center}
	\begin{tabular}{|c||c|c|c|c|c|c|c|}
		\hline
		Field & $SU\left(6\right)_{g}$ & $SU\left(2\right)_{1}$ & $SU\left(2\right)_{2}$ & $SU\left(6\right)$ & $U\left(1\right)_{a}$ & $U\left(1\right)_{b}$ & $U\left(1\right)_{r}$ \\
		\hline 
		$A$ & $\boldsymbol{15}$ & $\boldsymbol{2}$ & $\boldsymbol{1}$ & $\boldsymbol{1}$ & -3 & -1 & 0\\
		
		$\bar{Q}$ & $\boldsymbol{\overline{6}}$ & $\boldsymbol{1}$ & $\boldsymbol{1}$ & $\boldsymbol{6}$ & 4 & 0 & $\frac{1}{2}$\\
		
		$F$ & $\boldsymbol{6}$ & $\boldsymbol{1}$ & $\boldsymbol{2}$ & $\boldsymbol{1}$ & 0 & 4 & $\frac{1}{2}$\\
		
		$M_{0}$ & $\boldsymbol{1}$ & $\boldsymbol{1}$ & $\boldsymbol{2}$ & $\boldsymbol{\overline{6}}$ & -4 & -4 & 1\\
		
		$M_{1}$ & $\boldsymbol{1}$ & $\boldsymbol{2}$ & $\boldsymbol{1}$ & $\boldsymbol{\overline{15}}$ & -5 & 1 & 1\\
		
		$\Phi$ & $\boldsymbol{1}$ & $\boldsymbol{4}$ & $\boldsymbol{1}$ & $\boldsymbol{1}$ & 9 & 3 & 2\\
		\hline
	\end{tabular}
\end{center}
The superpotential is given by 
\begin{equation}
W=M_{0}F\bar{Q}+M_{1}A\bar{Q}^{2}+\Phi A^{3}
\end{equation}
and the superconformal R charge by $\hat{r}=r-0.0175q_{a}-0.0481q_{b}$. We see that $U\left(1\right)_{b}$ and $U\left(1\right)_{a}$ mix with $U\left(1\right)_{r}$ in the expression for the IR R symmetry, and so the operators that contribute to the index at order $pq$ should have vanishing charges under $U\left(1\right)_{b}$ and $U\left(1\right)_{a}$. These operators are of the forms
\begin{equation*}
\lambda\lambda,\,\,\,\,\bar{\psi}^{A}A,\,\,\,\,\bar{\psi}^{\bar{Q}}\bar{Q},\,\,\,\,\bar{\psi}^{F}F,\,\,\,\,\bar{\psi}^{M_{0}}M_{0},\,\,\,\,\bar{\psi}^{M_{1}}M_{1},\,\,\,\,\bar{\psi}^{\Phi}\Phi,
\end{equation*}
\begin{equation*}
M_{0}F\bar{Q},\,\,\,\,M_{1}A\bar{Q}^{2},\,\,\,\,\Phi A^{3},\,\,\,\,A^{4}\bar{Q}^{3}F,\,\,\,\,A^{4}\bar{Q}^{2}\bar{\psi}^{M_{0}},\,\,\,\,A\bar{Q}^{2}\bar{\psi}^{\Phi}\bar{\psi}^{M_{0}},
\end{equation*}
\begin{equation}
\bar{\psi}^{M_{1}}\bar{\psi}^{\Phi}\bar{\psi}^{M_{0}},\,\,\,\,A^{3}\bar{\psi}^{M_{1}}\bar{\psi}^{M_{0}},\,\,\,\,A^{3}\bar{Q}F\bar{\psi}^{M_{1}},\,\,\,\,\bar{Q}F\bar{\psi}^{\Phi}\bar{\psi}^{M_{1}},\,\,\,\,A\bar{Q}^{3}F\bar{\psi}^{\Phi}.
\label{listSU6}
\end{equation}

Next, denoting the irreducible representations (and the corresponding characters) of the nonabelian groups in the theory by 
\begin{equation}
\left(SU\left(6\right)_{g},SU\left(2\right)_{1},SU\left(2\right)_{2},SU\left(6\right)\right),
\end{equation}
we find the following representations of the gauge singlets corresponding
to the operators in \eqref{listSU6},
\begin{center}
	\begin{longtable}{|c|c|c|}
		\hline
		$\mathrm{Operator}$ & $\left(-1\right)^{F}$ & $\mathrm{Representations}\,\,(R)$\\
		\hline \hline 
		$\lambda\lambda$ & + & $\left(\boldsymbol{1},\boldsymbol{1},\boldsymbol{1},\boldsymbol{1}\right)$\tabularnewline
		\hline 
		$\bar{\psi}^{A}A$ & - & $\left(\boldsymbol{1},\boldsymbol{1},\boldsymbol{1},\boldsymbol{1}\right),\,\,\left(\boldsymbol{1},\boldsymbol{3},\boldsymbol{1},\boldsymbol{1}\right)$\tabularnewline
		\hline 
		$\bar{\psi}^{\bar{Q}}\bar{Q}$ & - & $\left(\boldsymbol{1},\boldsymbol{1},\boldsymbol{1},\boldsymbol{1}\right),\,\,\left(\boldsymbol{1},\boldsymbol{1},\boldsymbol{1},\boldsymbol{35}\right)$\tabularnewline
		\hline 
		$\bar{\psi}^{F}F$ & - & $\left(\boldsymbol{1},\boldsymbol{1},\boldsymbol{1},\boldsymbol{1}\right),\,\,\left(\boldsymbol{1},\boldsymbol{1},\boldsymbol{3},\boldsymbol{1}\right)$\tabularnewline
		\hline 
		$\bar{\psi}^{M_{0}}M_{0}$ & - & $\left(\boldsymbol{1},\boldsymbol{1},\boldsymbol{1},\boldsymbol{1}\right),\,\,\left(\boldsymbol{1},\boldsymbol{1},\boldsymbol{1},\boldsymbol{35}\right),\,\,\left(\boldsymbol{1},\boldsymbol{1},\boldsymbol{3},\boldsymbol{1}\right),\,\,\left(\boldsymbol{1},\boldsymbol{1},\boldsymbol{3},\boldsymbol{35}\right)$\tabularnewline
		\hline 
		$\bar{\psi}^{M_{1}}M_{1}$ & - & $\begin{array}{c}
		\left(\boldsymbol{1},\boldsymbol{1},\boldsymbol{1},\boldsymbol{1}\right),\,\,\left(\boldsymbol{1},\boldsymbol{1},\boldsymbol{1},\boldsymbol{35}\right),\,\,\left(\boldsymbol{1},\boldsymbol{1},\boldsymbol{1},\boldsymbol{189}\right),\\
		\left(\boldsymbol{1},\boldsymbol{3},\boldsymbol{1},\boldsymbol{1}\right),\,\,\left(\boldsymbol{1},\boldsymbol{3},\boldsymbol{1},\boldsymbol{35}\right),\,\,\left(\boldsymbol{1},\boldsymbol{3},\boldsymbol{1},\boldsymbol{189}\right)
		\end{array}$\tabularnewline
		\hline 
		$\bar{\psi}^{\Phi}\Phi$ & - & $\left(\boldsymbol{1},\boldsymbol{1},\boldsymbol{1},\boldsymbol{1}\right),\,\,\left(\boldsymbol{1},\boldsymbol{3},\boldsymbol{1},\boldsymbol{1}\right),\,\,\left(\boldsymbol{1},\boldsymbol{5},\boldsymbol{1},\boldsymbol{1}\right),\,\,\left(\boldsymbol{1},\boldsymbol{7},\boldsymbol{1},\boldsymbol{1}\right)$\tabularnewline
		\hline 
		$M_{0}F\bar{Q}$ & + & $\left(\boldsymbol{1},\boldsymbol{1},\boldsymbol{1},\boldsymbol{1}\right),\,\,\left(\boldsymbol{1},\boldsymbol{1},\boldsymbol{1},\boldsymbol{35}\right),\,\,\left(\boldsymbol{1},\boldsymbol{1},\boldsymbol{3},\boldsymbol{1}\right),\,\,\left(\boldsymbol{1},\boldsymbol{1},\boldsymbol{3},\boldsymbol{35}\right)$\tabularnewline
		\hline 
		$M_{1}A\bar{Q}^{2}$ & + & $\begin{array}{c}
		\left(\boldsymbol{1},\boldsymbol{1},\boldsymbol{1},\boldsymbol{1}\right),\,\,\left(\boldsymbol{1},\boldsymbol{1},\boldsymbol{1},\boldsymbol{35}\right),\,\,\left(\boldsymbol{1},\boldsymbol{1},\boldsymbol{1},\boldsymbol{189}\right),\\
		\left(\boldsymbol{1},\boldsymbol{3},\boldsymbol{1},\boldsymbol{1}\right),\,\,\left(\boldsymbol{1},\boldsymbol{3},\boldsymbol{1},\boldsymbol{35}\right),\,\,\left(\boldsymbol{1},\boldsymbol{3},\boldsymbol{1},\boldsymbol{189}\right)
		\end{array}$\tabularnewline
		\hline 
		$\Phi A^{3}$ & + & $\left(\boldsymbol{1},\boldsymbol{1},\boldsymbol{1},\boldsymbol{1}\right),\,\,\left(\boldsymbol{1},\boldsymbol{3},\boldsymbol{1},\boldsymbol{1}\right),\,\,\left(\boldsymbol{1},\boldsymbol{5},\boldsymbol{1},\boldsymbol{1}\right),\,\,\left(\boldsymbol{1},\boldsymbol{7},\boldsymbol{1},\boldsymbol{1}\right)$\tabularnewline
		\hline 
		$A^{4}\bar{Q}^{3}F$ & + & $\begin{array}{c}
		\left(\boldsymbol{1},\boldsymbol{1},\boldsymbol{2},\boldsymbol{20}\right),\,\,\left(\boldsymbol{1},\boldsymbol{3},\boldsymbol{2},\boldsymbol{20}\right),\,\,\left(\boldsymbol{1},\boldsymbol{5},\boldsymbol{2},\boldsymbol{20}\right),\\
		\left(\boldsymbol{1},\boldsymbol{1},\boldsymbol{2},\boldsymbol{70}\right),\,\,\left(\boldsymbol{1},\boldsymbol{3},\boldsymbol{2},\boldsymbol{70}\right),\,\,\left(\boldsymbol{1},\boldsymbol{5},\boldsymbol{2},\boldsymbol{70}\right),\\
		\left(\boldsymbol{1},\boldsymbol{3},\boldsymbol{2},\boldsymbol{20}\right),\,\,\left(\boldsymbol{1},\boldsymbol{1},\boldsymbol{2},\boldsymbol{70}\right),\,\,\left(\boldsymbol{1},\boldsymbol{3},\boldsymbol{2},\boldsymbol{70}\right)
		\end{array}$\tabularnewline
		\hline 
		$A^{4}\bar{Q}^{2}\bar{\psi}^{M_{0}}$ & - & $\begin{array}{c}
		\begin{array}{c}
		\left(\boldsymbol{1},\boldsymbol{1},\boldsymbol{2},\boldsymbol{20}\right),\,\,\left(\boldsymbol{1},\boldsymbol{3},\boldsymbol{2},\boldsymbol{20}\right),\,\,\left(\boldsymbol{1},\boldsymbol{5},\boldsymbol{2},\boldsymbol{20}\right)\end{array}\\
		\begin{array}{c}
		\left(\boldsymbol{1},\boldsymbol{1},\boldsymbol{2},\boldsymbol{70}\right),\,\,\left(\boldsymbol{1},\boldsymbol{3},\boldsymbol{2},\boldsymbol{70}\right),\,\,\left(\boldsymbol{1},\boldsymbol{5},\boldsymbol{2},\boldsymbol{70}\right)\end{array}
		\end{array}$\tabularnewline
		\hline 
		$A\bar{Q}^{2}\bar{\psi}^{\Phi}\bar{\psi}^{M_{0}}$ & + & $\left(\boldsymbol{1},\boldsymbol{3},\boldsymbol{2},\boldsymbol{20}\right),\,\,\left(\boldsymbol{1},\boldsymbol{3},\boldsymbol{2},\boldsymbol{70}\right),\,\,\left(\boldsymbol{1},\boldsymbol{5},\boldsymbol{2},\boldsymbol{20}\right),\,\,\left(\boldsymbol{1},\boldsymbol{5},\boldsymbol{2},\boldsymbol{70}\right)$\tabularnewline
		\hline 
		$\bar{\psi}^{M_{1}}\bar{\psi}^{\Phi}\bar{\psi}^{M_{0}}$ & - & $\left(\boldsymbol{1},\boldsymbol{3},\boldsymbol{2},\boldsymbol{20}\right),\,\,\left(\boldsymbol{1},\boldsymbol{3},\boldsymbol{2},\boldsymbol{70}\right),\,\,\left(\boldsymbol{1},\boldsymbol{5},\boldsymbol{2},\boldsymbol{20}\right),\,\,\left(\boldsymbol{1},\boldsymbol{5},\boldsymbol{2},\boldsymbol{70}\right)$\tabularnewline
		\hline 
		$A^{3}\bar{\psi}^{M_{1}}\bar{\psi}^{M_{0}}$ & + & $\left(\boldsymbol{1},\boldsymbol{3},\boldsymbol{2},\boldsymbol{20}\right),\,\,\left(\boldsymbol{1},\boldsymbol{3},\boldsymbol{2},\boldsymbol{70}\right),\,\,\left(\boldsymbol{1},\boldsymbol{5},\boldsymbol{2},\boldsymbol{20}\right),\,\,\left(\boldsymbol{1},\boldsymbol{5},\boldsymbol{2},\boldsymbol{70}\right)$\tabularnewline
		\hline 
		$A^{3}\bar{Q}F\bar{\psi}^{M_{1}}$ & - & $\begin{array}{c}
		\left(\boldsymbol{1},\boldsymbol{1},\boldsymbol{2},\boldsymbol{20}\right),\,\,\left(\boldsymbol{1},\boldsymbol{1},\boldsymbol{2},\boldsymbol{70}\right),\,\,\left(\boldsymbol{1},\boldsymbol{3},\boldsymbol{2},\boldsymbol{20}\right),\,\,\left(\boldsymbol{1},\boldsymbol{3},\boldsymbol{2},\boldsymbol{70}\right)\\
		\left(\boldsymbol{1},\boldsymbol{3},\boldsymbol{2},\boldsymbol{20}\right),\,\,\left(\boldsymbol{1},\boldsymbol{3},\boldsymbol{2},\boldsymbol{70}\right),\,\,\left(\boldsymbol{1},\boldsymbol{5},\boldsymbol{2},\boldsymbol{20}\right),\,\,\left(\boldsymbol{1},\boldsymbol{5},\boldsymbol{2},\boldsymbol{70}\right)
		\end{array}$\tabularnewline
		\hline 
		$\bar{Q}F\bar{\psi}^{\Phi}\bar{\psi}^{M_{1}}$ & + & $\left(\boldsymbol{1},\boldsymbol{3},\boldsymbol{2},\boldsymbol{20}\right),\,\,\left(\boldsymbol{1},\boldsymbol{3},\boldsymbol{2},\boldsymbol{70}\right),\,\,\left(\boldsymbol{1},\boldsymbol{5},\boldsymbol{2},\boldsymbol{20}\right),\,\,\left(\boldsymbol{1},\boldsymbol{5},\boldsymbol{2},\boldsymbol{70}\right)$\tabularnewline
		\hline 
		$A\bar{Q}^{3}F\bar{\psi}^{\Phi}$ & - & $\left(\boldsymbol{1},\boldsymbol{3},\boldsymbol{2},\boldsymbol{20}\right),\,\,\left(\boldsymbol{1},\boldsymbol{3},\boldsymbol{2},\boldsymbol{70}\right),\,\,\left(\boldsymbol{1},\boldsymbol{5},\boldsymbol{2},\boldsymbol{20}\right),\,\,\left(\boldsymbol{1},\boldsymbol{5},\boldsymbol{2},\boldsymbol{70}\right)$\tabularnewline
		\hline
	\end{longtable}
\end{center}
Summing (with signs) the above characters, we obtain 
\begin{equation*}
\sum R\left(-1\right)^{F}=-\left(\boldsymbol{1},\boldsymbol{3},\boldsymbol{1},\boldsymbol{1}\right)-\left(\boldsymbol{1},\boldsymbol{1},\boldsymbol{3},\boldsymbol{1}\right)-\left(\boldsymbol{1},\boldsymbol{1},\boldsymbol{1},\boldsymbol{35}\right)-\begin{array}{c}
\left(\boldsymbol{1},\boldsymbol{1},\boldsymbol{2},\boldsymbol{20}\right)\end{array}-2\left(\boldsymbol{1},\boldsymbol{1},\boldsymbol{1},\boldsymbol{1}\right)
\end{equation*}
\begin{equation}
=-\boldsymbol{3}_{SU(2)_1}-\boldsymbol{78}_{E_6}-2
\end{equation}
which corresponds to the enhancement of the flavor symmetry $SU(2)^2 \times SU(6) \times U(1)^2$ to $SU(2) \times E_6 \times U(1)^2$ in the IR. Moreover, the dimension of the conformal manifold vanishes.

\

\subsection*{$\boldsymbol{SU(8)}$}

\

The matter content is given in the following table, 
\begin{center}
	\begin{tabular}{|c||c|c|c|c|c|}
		\hline
		Field & $SU\left(8\right)_{g}$ & $SU\left(2\right)$ & $SU\left(8\right)$ & $U\left(1\right)_{a}$ & $U\left(1\right)_{r}$\\
		\hline 
		$A$ & $\boldsymbol{28}$ & $\boldsymbol{2}$ & $\boldsymbol{1}$ & 2 & 0\\
		
		$\bar{Q}$ & $\boldsymbol{\overline{8}}$ & $\boldsymbol{1}$ & $\boldsymbol{8}$ & -3 & $\frac{1}{2}$\\
		
		$M$ & $\boldsymbol{1}$ & $\boldsymbol{2}$ & $\boldsymbol{\overline{28}}$ & -4 & 1\\
		
		$\Phi$ & $\boldsymbol{1}$ & $\boldsymbol{5}$ & $\boldsymbol{1}$ & -8 & 2\\
		\hline
	\end{tabular}
\end{center}
The superpotential is given by 
\begin{equation}
W=MA^{5}\bar{Q}^{2}+\Phi A^{4}
\end{equation}
and the superconformal R charge by $\hat{r}=r+0.0693q_{a}$. We see that $U\left(1\right)_{a}$ mixes with $U\left(1\right)_{r}$ in the expression for the IR R symmetry, and so the operators that contribute to the index at order $pq$ should have a vanishing charge under $U\left(1\right)_{a}$. These operators are of the forms
\begin{equation*}
\lambda\lambda,\,\,\,\,\bar{\psi}^{A}A,\,\,\,\,\bar{\psi}^{\bar{Q}}\bar{Q},\,\,\,\,\bar{\psi}^{M}M,\,\,\,\,\bar{\psi}^{\Phi}\Phi,\,\,\,\,\bar{Q}^{4}A^{6},\,\,\,\,\bar{Q}^{4}A^{2}\bar{\psi}^{\Phi},
\end{equation*}
\begin{equation}
\bar{Q}^{2}A\bar{\psi}^{M},\,\,\,\,\bar{Q}^{2}A^{5}M,\,\,\,\,\bar{Q}^{2}A\bar{\psi}^{\Phi}M,\,\,\,\,M^{2}A^{4},\,\,\,\,M^{2}\bar{\psi}^{\Phi},\,\,\,\,\Phi A^{4}.
\label{listSU8}
\end{equation}

Next, denoting the irreducible representations (and the corresponding characters) of the nonabelian groups in the theory by 
\begin{equation}
\left(SU\left(8\right)_{g},SU\left(2\right),SU\left(8\right)\right),
\end{equation}
we find the following representations of the gauge singlets corresponding
to the operators in \eqref{listSU8},

\

\

\

\begin{center}
	\begin{longtable}{|c|c|c|}
		\hline
		$\mathrm{Operator}$ & $\left(-1\right)^{F}$ & $\mathrm{Representations}\,\,(R)$\\
		\hline \hline 
		$\lambda\lambda$ & + & $\left(\boldsymbol{1},\boldsymbol{1},\boldsymbol{1}\right)$\tabularnewline
		\hline 
		$\bar{\psi}^{A}A$ & - & $\left(\boldsymbol{1},\boldsymbol{3},\boldsymbol{1}\right),\,\,\left(\boldsymbol{1},\boldsymbol{1},\boldsymbol{1}\right)$\tabularnewline
		\hline 
		$\bar{\psi}^{\bar{Q}}\bar{Q}$ & - & $\left(\boldsymbol{1},\boldsymbol{1},\boldsymbol{63}\right),\,\,\left(\boldsymbol{1},\boldsymbol{1},\boldsymbol{1}\right)$\tabularnewline
		\hline 
		$\bar{\psi}^{M}M$ & - & $\left(\boldsymbol{1},\boldsymbol{1},\boldsymbol{1}\right),\,\,\left(\boldsymbol{1},\boldsymbol{1},\boldsymbol{63}\right),\,\,\left(\boldsymbol{1},\boldsymbol{1},\boldsymbol{720}\right),\,\,\left(\boldsymbol{1},\boldsymbol{3},\boldsymbol{1}\right),\,\,\left(\boldsymbol{1},\boldsymbol{3},\boldsymbol{63}\right),\,\,\left(\boldsymbol{1},\boldsymbol{3},\boldsymbol{720}\right)$\tabularnewline
		\hline 
		$\bar{\psi}^{\Phi}\Phi$ & - & $\left(\boldsymbol{1},\boldsymbol{1},\boldsymbol{1}\right),\,\,\left(\boldsymbol{1},\boldsymbol{3},\boldsymbol{1}\right),\,\,\left(\boldsymbol{1},\boldsymbol{5},\boldsymbol{1}\right),\,\,\left(\boldsymbol{1},\boldsymbol{7},\boldsymbol{1}\right),\,\,\left(\boldsymbol{1},\boldsymbol{9},\boldsymbol{1}\right)$\tabularnewline
		\hline 
		$\bar{Q}^{4}A^{6}$ & + & $\begin{array}{c}
		2\left(\boldsymbol{1},\boldsymbol{3},\boldsymbol{70}\right),\,\,\left(\boldsymbol{1},\boldsymbol{5},\boldsymbol{70}\right),\,\,\left(\boldsymbol{1},\boldsymbol{7},\boldsymbol{70}\right),\,\,\left(\boldsymbol{1},\boldsymbol{1},\boldsymbol{336}\right),\,\,2\left(\boldsymbol{1},\boldsymbol{3},\boldsymbol{336}\right),\\
		\left(\boldsymbol{1},\boldsymbol{5},\boldsymbol{336}\right),\,\,\left(\boldsymbol{1},\boldsymbol{7},\boldsymbol{336}\right),\,\,\left(\boldsymbol{1},\boldsymbol{1},\boldsymbol{378}\right),\,\,\left(\boldsymbol{1},\boldsymbol{3},\boldsymbol{378}\right),\,\,\left(\boldsymbol{1},\boldsymbol{5},\boldsymbol{378}\right)
		\end{array}$\tabularnewline
		\hline 
		$\bar{Q}^{4}A^{2}\bar{\psi}^{\Phi}$ & - & $\begin{array}{c}
		\left(\boldsymbol{1},\boldsymbol{3},\boldsymbol{70}\right),\,\,\left(\boldsymbol{1},\boldsymbol{5},\boldsymbol{70}\right),\,\,\left(\boldsymbol{1},\boldsymbol{7},\boldsymbol{70}\right),\,\,\left(\boldsymbol{1},\boldsymbol{3},\boldsymbol{336}\right),\\
		\left(\boldsymbol{1},\boldsymbol{5},\boldsymbol{336}\right),\,\,\left(\boldsymbol{1},\boldsymbol{7},\boldsymbol{336}\right),\,\,\left(\boldsymbol{1},\boldsymbol{5},\boldsymbol{378}\right)
		\end{array}$\tabularnewline
		\hline 
		$\bar{Q}^{2}A\bar{\psi}^{M}$ & - & $\left(\boldsymbol{1},\boldsymbol{1},\boldsymbol{70}\right),\,\,\left(\boldsymbol{1},\boldsymbol{3},\boldsymbol{70}\right),\,\,\left(\boldsymbol{1},\boldsymbol{1},\boldsymbol{336}\right),\,\,\left(\boldsymbol{1},\boldsymbol{3},\boldsymbol{336}\right),\,\,\left(\boldsymbol{1},\boldsymbol{1},\boldsymbol{378}\right),\,\,\left(\boldsymbol{1},\boldsymbol{3},\boldsymbol{378}\right)$\tabularnewline
		\hline 
		$\bar{Q}^{2}A^{5}M$ & + & $\begin{array}{c}
		\left(\boldsymbol{1},\boldsymbol{1},\boldsymbol{1}\right),\,\,\left(\boldsymbol{1},\boldsymbol{1},\boldsymbol{63}\right),\,\,\left(\boldsymbol{1},\boldsymbol{1},\boldsymbol{720}\right),\,\,2\left(\boldsymbol{1},\boldsymbol{3},\boldsymbol{1}\right),\,\,2\left(\boldsymbol{1},\boldsymbol{3},\boldsymbol{63}\right),\,\,2\left(\boldsymbol{1},\boldsymbol{3},\boldsymbol{720}\right),\\
		2\left(\boldsymbol{1},\boldsymbol{5},\boldsymbol{1}\right),\,\,2\left(\boldsymbol{1},\boldsymbol{5},\boldsymbol{63}\right),\,\,2\left(\boldsymbol{1},\boldsymbol{5},\boldsymbol{720}\right),\,\,\left(\boldsymbol{1},\boldsymbol{7},\boldsymbol{1}\right),\,\,\left(\boldsymbol{1},\boldsymbol{7},\boldsymbol{63}\right),\,\,\left(\boldsymbol{1},\boldsymbol{7},\boldsymbol{720}\right)
		\end{array}$\tabularnewline
		\hline 
		$\bar{Q}^{2}A\bar{\psi}^{\Phi}M$ & - & $\begin{array}{c}
		\left(\boldsymbol{1},\boldsymbol{3},\boldsymbol{1}\right),\,\,\left(\boldsymbol{1},\boldsymbol{3},\boldsymbol{63}\right),\,\,\left(\boldsymbol{1},\boldsymbol{3},\boldsymbol{720}\right),\,\,2\left(\boldsymbol{1},\boldsymbol{5},\boldsymbol{1}\right),\,\,2\left(\boldsymbol{1},\boldsymbol{5},\boldsymbol{63}\right),\\
		2\left(\boldsymbol{1},\boldsymbol{5},\boldsymbol{720}\right),\,\,\left(\boldsymbol{1},\boldsymbol{7},\boldsymbol{1}\right),\,\,\left(\boldsymbol{1},\boldsymbol{7},\boldsymbol{63}\right),\,\,\left(\boldsymbol{1},\boldsymbol{7},\boldsymbol{720}\right)
		\end{array}$\tabularnewline
		\hline 
		$M^{2}A^{4}$ & + & $\begin{array}{c}
		\left(\boldsymbol{1},\boldsymbol{3},\boldsymbol{70}\right),\,\,\left(\boldsymbol{1},\boldsymbol{3},\boldsymbol{\overline{336}}\right),\,\,\left(\boldsymbol{1},\boldsymbol{3},\boldsymbol{\overline{378}}\right),\,\,2\left(\boldsymbol{1},\boldsymbol{5},\boldsymbol{70}\right),\,\,2\left(\boldsymbol{1},\boldsymbol{5},\boldsymbol{\overline{336}}\right),\\
		2\left(\boldsymbol{1},\boldsymbol{5},\boldsymbol{\overline{378}}\right),\,\,\left(\boldsymbol{1},\boldsymbol{7},\boldsymbol{70}\right),\,\,\left(\boldsymbol{1},\boldsymbol{7},\boldsymbol{\overline{336}}\right),\,\,\left(\boldsymbol{1},\boldsymbol{7},\boldsymbol{\overline{378}}\right)
		\end{array}$\tabularnewline
		\hline 
		$M^{2}\bar{\psi}^{\Phi}$ & - & $\begin{array}{c}
		\left(\boldsymbol{1},\boldsymbol{3},\boldsymbol{70}\right),\,\,\left(\boldsymbol{1},\boldsymbol{3},\boldsymbol{\overline{336}}\right),\,\,\left(\boldsymbol{1},\boldsymbol{3},\boldsymbol{\overline{378}}\right),\,\,2\left(\boldsymbol{1},\boldsymbol{5},\boldsymbol{70}\right),\,\,2\left(\boldsymbol{1},\boldsymbol{5},\boldsymbol{\overline{336}}\right),\\
		2\left(\boldsymbol{1},\boldsymbol{5},\boldsymbol{\overline{378}}\right),\,\,\left(\boldsymbol{1},\boldsymbol{7},\boldsymbol{70}\right),\,\,\left(\boldsymbol{1},\boldsymbol{7},\boldsymbol{\overline{336}}\right),\,\,\left(\boldsymbol{1},\boldsymbol{7},\boldsymbol{\overline{378}}\right)
		\end{array}$\tabularnewline
		\hline 
		$\Phi A^{4}$ & + & $\left(\boldsymbol{1},\boldsymbol{1},\boldsymbol{1}\right),\,\,\left(\boldsymbol{1},\boldsymbol{3},\boldsymbol{1}\right),\,\,\left(\boldsymbol{1},\boldsymbol{5},\boldsymbol{1}\right),\,\,\left(\boldsymbol{1},\boldsymbol{7},\boldsymbol{1}\right),\,\,\left(\boldsymbol{1},\boldsymbol{9},\boldsymbol{1}\right)$\tabularnewline
		\hline
	\end{longtable}
\end{center}
Summing (with signs) the above characters, we obtain 
\begin{equation*}
\sum R\left(-1\right)^{F}=-\left(\boldsymbol{1},\boldsymbol{1},\boldsymbol{70}\right)-\left(\boldsymbol{1},\boldsymbol{1},\boldsymbol{63}\right)-\left(\boldsymbol{1},\boldsymbol{3},\boldsymbol{1}\right)-\left(\boldsymbol{1},\boldsymbol{1},\boldsymbol{1}\right)
\end{equation*}
\begin{equation}
=-\boldsymbol{133}_{E_{7}}-\boldsymbol{3}_{SU\left(2\right)}-1
\end{equation}
which corresponds to the enhancement of the flavor symmetry $SU(8) \times SU(2) \times U(1)$ to $E_7 \times SU(2) \times U(1)$ in the IR. Moreover, the dimension of the conformal manifold vanishes.

\

\section{Self-dualities and symmetry enhancement in $\boldsymbol{SU(2N)}$ gauge theories}

In \cite{Razamat:2017wsk} two models with $SU(2)$ and $SU(4)$ gauge groups were considered, and the relation between their self-duality and symmetry enhancement properties was discussed. In particular, it was shown explicitly in the $SU(4)$ model how the enhanced symmetry corresponds to a multitude of self-dualities, and it was further used to find a new self-duality. It was also mentioned that these two models are the first two entries in a sequence of $SU(2N)$ gauge theories having such self-dual descriptions. In this section we extend the results obtained for the $SU(4)$ group to this sequence of $SU(2N)$ ($N>2$) theories, and find for each $N$ a new self-dual frame in addition to the ones considered in \cite{Csaki:1997cu,Spiridonov:2009za}. Furthermore, we use these self-dualities to construct for each $N$ a model with symmetry enhancement. We begin with presenting the dualities and continue with the description of the related model with symmetry enhancement. 

\

\subsection*{Dualities}

\

We consider the gauge group $SU(2N)$ and the following matter content:
\begin{center}
	\begin{tabular}{|c||c|c|c|c|c|c|c|}
		\hline
		Field & $SU\left(2N\right)_{g}$ & $SU\left(4\right)_{L}$ & $SU\left(4\right)_{R}$ & $U\left(1\right)_{s}$ & $U\left(1\right)_{b}$ & $U\left(1\right)_{a}$ & $U\left(1\right)_{r}$\\
		\hline 
		$Q$ & $\boldsymbol{2N}$ & $\boldsymbol{4}$ & \textbf{1} & 0 & 1 & $2N-2$ & $\frac{1}{2}$\\
		 
		$\overline{Q}$ & $\overline{\boldsymbol{2N}}$ & \textbf{1} & $\boldsymbol{4}$ & 0 & -1 & $2N-2$ & $\frac{1}{2}$\\
		 
		$A$ & $\boldsymbol{N\left(2N-1\right)}$ & \textbf{1} & \textbf{1} & 1 & 0 & -4 & 0\\
		 
		$\overline{A}$ & $\boldsymbol{\overline{N\left(2N-1\right)}}$ & \textbf{1} & \textbf{1} & -1 & 0 & -4 & 0\\
		\hline
	\end{tabular}
\end{center}

In order to define the various duality frames, we decompose 
\begin{equation*}
SU(4)_{L}\rightarrow SU(2)_{L}\times\widetilde{SU}(2)_{L}\times U(1)_{L}\,\,,\,\,\,\,SU(4)_{R}\rightarrow SU(2)_{R}\times\widetilde{SU}(2)_{R}\times U(1)_{R}
\end{equation*}
and correspondingly define $Q\rightarrow\left(Q_{-},Q_{+}\right)$ and $\overline{Q}\rightarrow\left(\overline{Q}_{-},\overline{Q}_{+}\right)$. Moreover, we omit the $U(1)_r$ column for clarity. In this notation, the matter content is given by: 
\small
\begin{center}
	\begin{tabular}{|c||c|c|c|c|c|c|c|c|c|c|}
		\hline
		Field & $SU\left(2N\right)_{g}$ & $SU\left(2\right)_{L}$ & $\widetilde{SU}\left(2\right)_{L}$ & $U\left(1\right)_{L}$ & $SU\left(2\right)_{R}$ & $\widetilde{SU}\left(2\right)_{R}$ & $U\left(1\right)_{R}$ & $U\left(1\right)_{s}$ & $U\left(1\right)_{b}$ & $U\left(1\right)_{a}$\\
		\hline 
		$Q_{-}$ & $\boldsymbol{2N}$ & $\boldsymbol{2}$ & \textbf{1} & -1 & \textbf{1} & \textbf{1} & 0 & 0 & 1 & $2N-2$\\
		 
		$Q_{+}$ & $\boldsymbol{2N}$ & \textbf{1} & $\boldsymbol{2}$ & 1 & \textbf{1} & \textbf{1} & 0 & 0 & 1 & $2N-2$\\
		 
		$\overline{Q}_{-}$ & $\overline{\boldsymbol{2N}}$ & \textbf{1} & \textbf{1} & 0 & $\boldsymbol{2}$ & \textbf{1} & -1 & 0 & -1 & $2N-2$\\
		 
		$\overline{Q}_{+}$ & $\overline{\boldsymbol{2N}}$ & \textbf{1} & \textbf{1} & 0 & \textbf{1} & $\boldsymbol{2}$ & 1 & 0 & -1 & $2N-2$\\
		 
		$A$ & $\boldsymbol{N\left(2N-1\right)}$ & \textbf{1} & \textbf{1} & 0 & \textbf{1} & \textbf{1} & 0 & 1 & 0 & -4\\
		 
		$\overline{A}$ & $\boldsymbol{\overline{N\left(2N-1\right)}}$ & \textbf{1} & \textbf{1} & 0 & \textbf{1} & \textbf{1} & 0 & -1 & 0 & -4\\
		\hline
	\end{tabular}
\end{center}
\normalsize

The first dual frame is obtained by introducing gauge singlet mesons $M_{k}^{\pm\pm}$, $k=0,\ldots,N-1$ which couple to gauge composite operators through the following superpotential,
\begin{equation}
W=\sum_{k=0}^{N-1}\left(M_{k}^{-+}q_{+}\overline{q}_{-}+M_{k}^{--}q_{+}\overline{q}_{+}+M_{k}^{++}q_{-}\overline{q}_{-}+M_{k}^{+-}q_{-}\overline{q}_{+}\right)\left(a\overline{a}\right)^{N-1-k}\,.
\end{equation}
The matter content is given by: 

	\hskip-2.7cm\begin{tabular}{|c||c|c|c|c|c|c|c|c|c|c|}
		\hline
		Field & $SU\left(2N\right)_{g}$ & $SU\left(2\right)_{L}$ & $\widetilde{SU}\left(2\right)_{L}$ & $U\left(1\right)_{L}$ & $SU\left(2\right)_{R}$ & $\widetilde{SU}\left(2\right)_{R}$ & $U\left(1\right)_{R}$ & $U\left(1\right)_{s}$ & $U\left(1\right)_{b}$ & $U\left(1\right)_{a}$\\
		\hline 
		$q_{-}$ & $\boldsymbol{2N}$ & $\boldsymbol{2}$ & \textbf{1} & 1 & \textbf{1} & \textbf{1} & 0 & 0 & 1 & $2N-2$\\
		 
		$q_{+}$ & $\boldsymbol{2N}$ & \textbf{1} & $\boldsymbol{2}$ & -1 & \textbf{1} & \textbf{1} & 0 & 0 & 1 & $2N-2$\\
		 
		$\overline{q}_{-}$ & $\overline{\boldsymbol{2N}}$ & \textbf{1} & \textbf{1} & 0 & $\boldsymbol{2}$ & \textbf{1} & 1 & 0 & -1 & $2N-2$\\
		 
		$\overline{q}_{+}$ & $\overline{\boldsymbol{2N}}$ & \textbf{1} & \textbf{1} & 0 & \textbf{1} & $\boldsymbol{2}$ & -1 & 0 & -1 & $2N-2$\\
		 
		$a$ & $\boldsymbol{N\left(2N-1\right)}$ & \textbf{1} & \textbf{1} & 0 & \textbf{1} & \textbf{1} & 0 & 1 & 0 & -4\\
		 
		$\overline{a}$ & $\boldsymbol{\overline{N\left(2N-1\right)}}$ & \textbf{1} & \textbf{1} & 0 & \textbf{1} & \textbf{1} & 0 & -1 & 0 & -4\\
		 
		$M_{k}^{+-}$ & \textbf{1} & $\boldsymbol{2}$ & \textbf{1} & -1 & \textbf{1} & $\boldsymbol{2}$ & 1 & 0 & 0 & $4N-4-8k$\\
		 
		$M_{k}^{-+}$ & \textbf{1} & \textbf{1} & $\boldsymbol{2}$ & 1 & $\boldsymbol{2}$ & \textbf{1} & -1 & 0 & 0 & $4N-4-8k$\\
		 
		$M_{k}^{++}$ & \textbf{1} & $\boldsymbol{2}$ & \textbf{1} & -1 & $\boldsymbol{2}$ & \textbf{1} & -1 & 0 & 0 & $4N-4-8k$\\
		 
		$M_{k}^{--}$ & \textbf{1} & \textbf{1} & $\boldsymbol{2}$ & 1 & \textbf{1} & $\boldsymbol{2}$ & 1 & 0 & 0 & $4N-4-8k$\\
		\hline
	\end{tabular}

The second dual frame is obtained by introducing gauge singlet baryons $G^{\pm\pm}$, $\overline{G}^{\pm\pm}$, $H_{m}^{\pm\pm}$, $\overline{H}_{m}^{\pm\pm}$, $m=0,\ldots,N-2$ which couple to gauge composite operators through the following superpotential: 
\begin{equation*}
W=\sum_{m=0}^{N-2}\left[\left(H_{m}^{-+}q_{+}q_{-}+H_{m}^{--}q_{+}q_{+}+H_{m}^{++}q_{-}q_{-}\right)\overline{a}+\left(\overline{H}_{m}^{-+}\overline{q}_{+}\overline{q}_{-}+\overline{H}_{m}^{--}\overline{q}_{+}\overline{q}_{+}+\overline{H}_{m}^{++}\overline{q}_{-}\overline{q}_{-}\right)a\right]\left(a\overline{a}\right)^{N-2-m}
\end{equation*}
\begin{equation*}
+\left(G^{-+}q_{+}q_{-}+G^{--}q_{+}q_{+}+G^{++}q_{-}q_{-}\right)a^{N-1}+\left(\overline{G}^{-+}\overline{q}_{+}\overline{q}_{-}+\overline{G}^{--}\overline{q}_{+}\overline{q}_{+}+\overline{G}^{++}\overline{q}_{-}\overline{q}_{-}\right)\overline{a}^{N-1}\,.
\end{equation*}
The matter content of the gauge charged objects is given by:

	\hskip-2.7cm\begin{tabular}{|c||c|c|c|c|c|c|c|c|c|c|}
		\hline
		Field & $SU\left(2N\right)_{g}$ & $SU\left(2\right)_{L}$ & $\widetilde{SU}\left(2\right)_{L}$ & $U\left(1\right)_{L}$ & $SU\left(2\right)_{R}$ & $\widetilde{SU}\left(2\right)_{R}$ & $U\left(1\right)_{R}$ & $U\left(1\right)_{s}$ & $U\left(1\right)_{b}$ & $U\left(1\right)_{a}$\\
		\hline 
		$q_{-}$ & $\boldsymbol{2N}$ & $\boldsymbol{2}$ & \textbf{1} & -1 & \textbf{1} & \textbf{1} & 0 & 0 & -1 & $2N-2$\\
		 
		$q_{+}$ & $\boldsymbol{2N}$ & \textbf{1} & $\boldsymbol{2}$ & 1 & \textbf{1} & \textbf{1} & 0 & 0 & -1 & $2N-2$\\
		 
		$\overline{q}_{-}$ & $\overline{\boldsymbol{2N}}$ & \textbf{1} & \textbf{1} & 0 & $\boldsymbol{2}$ & \textbf{1} & -1 & 0 & 1 & $2N-2$\\
		 
		$\overline{q}_{+}$ & $\overline{\boldsymbol{2N}}$ & \textbf{1} & \textbf{1} & 0 & \textbf{1} & $\boldsymbol{2}$ & 1 & 0 & 1 & $2N-2$\\
		 
		$a$ & $\boldsymbol{N\left(2N-1\right)}$ & \textbf{1} & \textbf{1} & 0 & \textbf{1} & \textbf{1} & 0 & 1 & 0 & -4\\
		 
		$\overline{a}$ & $\boldsymbol{\overline{N\left(2N-1\right)}}$ & \textbf{1} & \textbf{1} & 0 & \textbf{1} & \textbf{1} & 0 & -1 & 0 & -4\\
		\hline
	\end{tabular}		
		
The matter content of the gauge singlet objects is given by:\footnote{Note that the $U(1)_{s}$ charges of the $G$s and the $H$s have opposite signs with respect to the ones appearing in \cite{Spiridonov:2009za}.} 

\
		
 \hskip-2.7cm\begin{tabular}{|c||c|c|c|c|c|c|c|c|c|c|}
		\hline
		Field & $SU\left(2N\right)_{g}$ & $SU\left(2\right)_{L}$ & $\widetilde{SU}\left(2\right)_{L}$ & $U\left(1\right)_{L}$ & $SU\left(2\right)_{R}$ & $\widetilde{SU}\left(2\right)_{R}$ & $U\left(1\right)_{R}$ & $U\left(1\right)_{s}$ & $U\left(1\right)_{b}$ & $U\left(1\right)_{a}$\\
		\hline 		 
		$H_{m}^{-+}$ & \textbf{1} & $\boldsymbol{2}$ & $\boldsymbol{2}$ & 0 & \textbf{1} & \textbf{1} & 0 & 1 & 2 & $4N-8-8m$\\
		 
		$H_{m}^{++}$ & \textbf{1} & \textbf{1} & \textbf{1} & 2 & \textbf{1} & \textbf{1} & 0 & 1 & 2 & $4N-8-8m$\\
		 
		$H_{m}^{--}$ & \textbf{1} & \textbf{1} & \textbf{1} & -2 & \textbf{1} & \textbf{1} & 0 & 1 & 2 & $4N-8-8m$\\
		 
		$\overline{H}_{m}^{-+}$ & \textbf{1} & \textbf{1} & \textbf{1} & 0 & $\boldsymbol{2}$ & $\boldsymbol{2}$ & 0 & -1 & -2 & $4N-8-8m$\\
		 
		$\overline{H}_{m}^{++}$ & \textbf{1} & \textbf{1} & \textbf{1} & 0 & \textbf{1} & \textbf{1} & 2 & -1 & -2 & $4N-8-8m$\\
		 
		$\overline{H}_{m}^{--}$ & \textbf{1} & \textbf{1} & \textbf{1} & 0 & \textbf{1} & \textbf{1} & -2 & -1 & -2 & $4N-8-8m$\\
		 
		$G^{-+}$ & \textbf{1} & $\boldsymbol{2}$ & $\boldsymbol{2}$ & 0 & \textbf{1} & \textbf{1} & 0 & $-N+1$ & 2 & 0\\
		 
		$G^{++}$ & \textbf{1} & \textbf{1} & \textbf{1} & 2 & \textbf{1} & \textbf{1} & 0 & $-N+1$ & 2 & 0\\
		 
		$G^{--}$ & \textbf{1} & \textbf{1} & \textbf{1} & -2 & \textbf{1} & \textbf{1} & 0 & $-N+1$ & 2 & 0\\
		 
		$\overline{G}^{-+}$ & \textbf{1} & \textbf{1} & \textbf{1} & 0 & $\boldsymbol{2}$ & $\boldsymbol{2}$ & 0 & $N-1$ & -2 & 0\\
		 
		$\overline{G}^{++}$ & \textbf{1} & \textbf{1} & \textbf{1} & 0 & \textbf{1} & \textbf{1} & 2 & $N-1$ & -2 & 0\\
		 
		$\overline{G}^{--}$ & \textbf{1} & \textbf{1} & \textbf{1} & 0 & \textbf{1} & \textbf{1} & -2 & $N-1$ & -2 & 0\\
		\hline
	\end{tabular}

The third and new dual frame is obtained by introducing the gauge singlet mesons $M_{k}^{-+}$, $M_{k}^{+-}$, $k=0,\ldots,N-1$ and the gauge singlet baryons $G^{--}$, $G^{++}$, $\overline{G}^{--}$, $\overline{G}^{++}$, $H_{m}^{--}$, $H_{m}^{++}$, $\overline{H}_{m}^{--}$, $\overline{H}_{m}^{++}$, $m=0,\ldots,N-2$, along with the superpotential 
\begin{equation*}
W=\sum_{k=0}^{N-1}\left(M_{k}^{-+}q_{+}\overline{q}_{-}+M_{k}^{+-}q_{-}\overline{q}_{+}\right)\left(a\overline{a}\right)^{N-1-k}
\end{equation*}
\begin{equation*}
+\sum_{m=0}^{N-2}\left[\left(H_{m}^{--}q_{+}q_{+}+H_{m}^{++}q_{-}q_{-}\right)\overline{a}+\left(\overline{H}_{m}^{--}\overline{q}_{+}\overline{q}_{+}+\overline{H}_{m}^{++}\overline{q}_{-}\overline{q}_{-}\right)a\right]\left(a\overline{a}\right)^{N-2-m}
\end{equation*}
\begin{equation*}
+\left(G^{--}q_{+}q_{+}+G^{++}q_{-}q_{-}\right)a^{N-1}+\left(\overline{G}^{--}\overline{q}_{+}\overline{q}_{+}+\overline{G}^{++}\overline{q}_{-}\overline{q}_{-}\right)\overline{a}^{N-1}\,.
\end{equation*}

The matter content of the gauge charged objects is given by: 

	\hskip-2.7cm\begin{tabular}{|c||c|c|c|c|c|c|c|c|c|c|}
		\hline
		Field & $SU\left(2N\right)_{g}$ & $SU\left(2\right)_{L}$ & $\widetilde{SU}\left(2\right)_{L}$ & $U\left(1\right)_{L}$ & $SU\left(2\right)_{R}$ & $\widetilde{SU}\left(2\right)_{R}$ & $U\left(1\right)_{R}$ & $U\left(1\right)_{s}$ & $U\left(1\right)_{b}$ & $U\left(1\right)_{a}$\\
		\hline 
		$q_{-}$ & $\boldsymbol{2N}$ & $\boldsymbol{2}$ & \textbf{1} & 0 & \textbf{1} & \textbf{1} & -1 & 0 & 1 & $2N-2$\\
		 
		$q_{+}$ & $\boldsymbol{2N}$ & \textbf{1} & $\boldsymbol{2}$ & 0 & \textbf{1} & \textbf{1} & 1 & 0 & 1 & $2N-2$\\
		 
		$\overline{q}_{-}$ & $\overline{\boldsymbol{2N}}$ & \textbf{1} & \textbf{1} & -1 & $\boldsymbol{2}$ & \textbf{1} & 0 & 0 & -1 & $2N-2$\\
		 
		$\overline{q}_{+}$ & $\overline{\boldsymbol{2N}}$ & \textbf{1} & \textbf{1} & 1 & \textbf{1} & $\boldsymbol{2}$ & 0 & 0 & -1 & $2N-2$\\
		 
		$a$ & $\boldsymbol{N\left(2N-1\right)}$ & \textbf{1} & \textbf{1} & 0 & \textbf{1} & \textbf{1} & 0 & 1 & 0 & -4\\
		 
		$\overline{a}$ & $\boldsymbol{\overline{N\left(2N-1\right)}}$ & \textbf{1} & \textbf{1} & 0 & \textbf{1} & \textbf{1} & 0 & -1 & 0 & -4\\
		\hline
	\end{tabular}		

The matter content of the gauge singlet objects is given by: 		
	
	\hskip-2.7cm\begin{tabular}{|c||c|c|c|c|c|c|c|c|c|c|}
		\hline
		Field & $SU\left(2N\right)_{g}$ & $SU\left(2\right)_{L}$ & $\widetilde{SU}\left(2\right)_{L}$ & $U\left(1\right)_{L}$ & $SU\left(2\right)_{R}$ & $\widetilde{SU}\left(2\right)_{R}$ & $U\left(1\right)_{R}$ & $U\left(1\right)_{s}$ & $U\left(1\right)_{b}$ & $U\left(1\right)_{a}$\\
		\hline 	 
		$M_{k}^{+-}$ & \textbf{1} & $\boldsymbol{2}$ & \textbf{1} & -1 & \textbf{1} & $\boldsymbol{2}$ & 1 & 0 & 0 & $4N-4-8k$\\
		 
		$M_{k}^{-+}$ & \textbf{1} & \textbf{1} & $\boldsymbol{2}$ & 1 & $\boldsymbol{2}$ & \textbf{1} & -1 & 0 & 0 & $4N-4-8k$\\
		 
		$H_{m}^{++}$ & \textbf{1} & \textbf{1} & \textbf{1} & 0 & \textbf{1} & \textbf{1} & 2 & 1 & -2 & $4N-8-8m$\\
		 
		$H_{m}^{--}$ & \textbf{1} & \textbf{1} & \textbf{1} & 0 & \textbf{1} & \textbf{1} & -2 & 1 & -2 & $4N-8-8m$\\
		 
		$\overline{H}_{m}^{++}$ & \textbf{1} & \textbf{1} & \textbf{1} & 2 & \textbf{1} & \textbf{1} & 0 & -1 & 2 & $4N-8-8m$\\
		 
		$\overline{H}_{m}^{--}$ & \textbf{1} & \textbf{1} & \textbf{1} & -2 & \textbf{1} & \textbf{1} & 0 & -1 & 2 & $4N-8-8m$\\
		 
		$G^{++}$ & \textbf{1} & \textbf{1} & \textbf{1} & 0 & \textbf{1} & \textbf{1} & 2 & $-N+1$ & -2 & 0\\
		 
		$G^{--}$ & \textbf{1} & \textbf{1} & \textbf{1} & 0 & \textbf{1} & \textbf{1} & -2 & $-N+1$ & -2 & 0\\
		 
		$\overline{G}^{++}$ & \textbf{1} & \textbf{1} & \textbf{1} & 2 & \textbf{1} & \textbf{1} & 0 & $N-1$ & 2 & 0\\
		 
		$\overline{G}^{--}$ & \textbf{1} & \textbf{1} & \textbf{1} & -2 & \textbf{1} & \textbf{1} & 0 & $N-1$ & 2 & 0\\
		\hline
	\end{tabular}

\

Overall, in the first duality frame we take the matter to be in the conjugate representation of the nonabelian symmetry ($SU\left(4\right)_{L}\times SU\left(4\right)_{R}$), in the second we change the signs of the charges under the baryonic
symmetry, and in the third we interchange $U\left(1\right)_{L}$ and $U\left(1\right)_{R}$ (obtained from the decomposition of $SU\left(4\right)_{L}\times SU\left(4\right)_{R}$). The 't Hooft anomalies of all the dual frames match.

\

\subsection*{Enhancement of symmetry}

\

We now turn to use the self-dualities described above in order to construct a model which is mapped into itself under these dualities; as a result, this model will have a symmetry enhancement in the IR. To do that, we flip as usual half of the composite operators involved in the dualities where $N$ is even; however, in the case where $N$ is odd, there is a composite operator the square of which is marginal, and we flip half of the remaining operators (obtained after ignoring the one that squares to a marginal operator). 

This construction is analogous to the one used for the $SU(4)$ theory in \cite{Razamat:2017wsk} (or alternatively for the $Spin(6)$ theory in section \ref{fourd}) and the nonabelian part of the UV symmetry $SU(4)_{L}\times SU(4)_{R}$, viewed as $SO(6)_{L}\times SO(6)_{R}$, is expected to enhance to $SO(12)$ in the IR. This was explicitly verified in the case $N=3$. 

Turning to the details of the model, the matter content is as follows,  
\begin{center}
	\begin{tabular}{|c||c|c|c|c|c|c|c|}
		\hline
		Field & $SU\left(2N\right)_{g}$ & $SU\left(4\right)_{L}$ & $SU\left(4\right)_{R}$ & $U\left(1\right)_{s}$ & $U\left(1\right)_{b}$ & $U\left(1\right)_{a}$ & $U\left(1\right)_{r}$\\
		\hline 
		$Q$ & $\boldsymbol{2N}$ & $\boldsymbol{4}$ & \textbf{1} & 0 & 1 & $2N-2$ & $\frac{1}{2}$\\
		
		$\overline{Q}$ & $\overline{\boldsymbol{2N}}$ & \textbf{1} & $\boldsymbol{4}$ & 0 & -1 & $2N-2$ & $\frac{1}{2}$\\
		
		$A$ & $\boldsymbol{N\left(2N-1\right)}$ & \textbf{1} & \textbf{1} & 1 & 0 & -4 & 0\\
		
		$\overline{A}$ & $\boldsymbol{\overline{N\left(2N-1\right)}}$ & \textbf{1} & \textbf{1} & -1 & 0 & -4 & 0\\
		
		$M_{l}$ & \textbf{1} & $\boldsymbol{\overline{4}}$ & $\boldsymbol{\overline{4}}$ & 0 & 0 & $4N-4-8l$ & 1\\
		
		$H_{m}$ & \textbf{1} & $\boldsymbol{6}$ & \textbf{1} & 1 & -2 & $4N-8-8m$ & 1\\
		
		$G$ & \textbf{1} & $\boldsymbol{6}$ & \textbf{1} & $-N+1$ & -2 & 0 & 1\\
		\hline
	\end{tabular}
\end{center}
In this table, $m=0,\ldots,N-2$ as before and $l=n(N),\ldots,N-1$, where $n(N)=\frac{N}{2}$ for $N$ even and $n(N)=\frac{N+1}{2}$ for $N$ odd. The superpotential is given by\footnote{Note that these superpotential and matter content do not include extra gauge singlet fields that one should add to the model so that all the operators will be above the unitarity bound. These fields flip the operators that violate the bound and their number and charges generally depend on $N$ (and therefore they are not written explicitly). Moreover, the operators which are flipped are composed only of $A$ and $\overline{A}$ and as a result the flipping fields are singlets of the $SU(4)_{L}\times SU(4)_{R}$ part of the global symmetry  that enhances to $SO(12)$. Therefore, these extra fields do not effect the enhancement.}
\begin{equation}
W=\sum_{l=n(N)}^{N-1}M_{l}Q\overline{Q}\left(A\overline{A}\right)^{N-1-l}+\sum_{m=0}^{N-2}H_{m}Q^{2}\overline{A}\left(A\overline{A}\right)^{N-2-m}+GQ^{2}A^{N-1}\,.
\end{equation}

As mentioned above, the UV symmetry $SU(4)_{L}\times SU(4)_{R}\times U(1)_{s}\times U(1)_{b}\times U(1)_{a}$ enhances to $SO(12)\times U(1)_{s}\times U(1)_{b}\times U(1)_{a}$ in the IR. 
The dualities act as the Weyl symmetry of the enhanced group.

\

\bibliographystyle{./aug/ytphys}
\bibliography{./aug/refs}

\end{document}